\documentclass{pas}

\usepackage{graphicx}	
\usepackage{algorithm,algcompatible}
\usepackage{braket}
\usepackage{xcolor}
\usepackage{float, makecell, array}
\usepackage{xspace, bm}
\usepackage{pdflscape}
\usepackage{longtable, makecell}
\usepackage{subfigure}
\usepackage{multirow, multicol}
\usepackage{fontawesome} 
\usepackage[dvipsnames]{xcolor}
\usepackage{scalerel}
\usepackage{enumerate}
\usepackage{chngcntr}

\usepackage{lipsum}
\usepackage{makecell}

\newcommand{\gold}{\textcolor[HTML]{AB8F25}{gold}\xspace}
\newcommand{\silver}{\textcolor[HTML]{808080}{silver}\xspace}
\newcommand{\bronze}{\textcolor{Bittersweet}{bronze}\xspace}

\newcommand{\hbeta}{H$\mathrm{\beta}$\xspace}
\newcommand{\mgii}{Mg\textsc{ii}\xspace}
\newcommand{\civ}{C\textsc{iv}\xspace}
\newcommand{\feii}{Fe\textsc{ii}\xspace}

\newcommand{\hbetamath}{\text{H}\beta}
\newcommand{\mgiimath}{\text{Mg}\textsc{ii}}

\newcommand{\ev}{\mathcal{Z}}

\newcommand{\beq}{\begin{equation}}
\newcommand{\eeq}{\end{equation}}
\newcommand{\bea}{\begin{eqnarray}}
\newcommand{\eea}{\end{eqnarray}}
\newcommand{\multilinecomment}[1]{}

\newcommand{\qm}[1]{``#1''}
\newcommand{\sqm}[1]{`#1'}

\newcommand{\abs}[1]{\stretchleftright{|}{#1}{|}}
\newcommand{\BF}[1]{\mathrm{BF}_\mathrm{#1}}
\newcommand{\Lrat}[1]{\mathrm{LR}_\mathrm{#1}}
\newcommand{\pone}{$16$\textsuperscript{th}}
\newcommand{\ptwo}{$84$\textsuperscript{th}}

\newcommand{\dayu}{\xspace\mathrm{ \ d}}
\newcommand{\angstrom}{\xspace\mathrm{ \ \AA}}

\newcommand{\red}{\textcolor{red}}

\newcommand{\des}{DES\xspace}
\newcommand{\decam}{DECam\xspace}
\newcommand{\ozdes}{OzDES\xspace}
\newcommand{\sdss}{SDSS\xspace}

\newcommand{\javelin}{\texttt{JAVELIN}\xspace}

\newcommand{\PyCCF}{\texttt{PyCCF}\xspace}
\newcommand{\pyroa}{\texttt{PyROA}\xspace}

\newcommand{\cream}{\texttt{CREAM}\xspace}
\newcommand{\mica}{\texttt{MICA}\xspace}
\newcommand{\litmus}{\texttt{LITMUS}\xspace}
\newcommand{\python}{\texttt{python}\xspace}

\newcommand{\numpyro}{\texttt{numpyro}\xspace}
\newcommand{\jax}{\texttt{jax}\xspace}

\newcommand{\jaxns}{\texttt{JAXNS}\xspace}

\usepackage{natbib}
\usepackage{orcidlink}

\DeclareUnicodeCharacter{2009}{ }
\DeclareUnicodeCharacter{1E3A}{L} 
\DeclareUnicodeCharacter{223C}{$\sim$}

\newcommand{\newlags}{$151$\space}
\newcommand{\newhbeta}{$28$\space}
\newcommand{\newmgii}{$54$\space}
\newcommand{\newciv}{$69$\space}

\begin{document}
\lefttitle{Publications of the Astronomical Society of Australia}
\righttitle{McDougall et al.}

\jnlPage{1}{4}
\jnlDoiYr{2026}
\doival{10.1017/pasa.xxxx.xx}

\articletitt{Research Paper}

\title{
AGN Reverberation Mapping with \litmus: Fundamental Limits on Lag Recovery Rates
}

\author{
\gn{Hugh} \sn{McDougall}\orcidlink{0009-0008-5846-1543}$^{1}$,
\gn{Tamara M.} \sn{Davis}\orcidlink{0000-0002-4213-8783}$^{1}$, 
\gn{Benjamin~J.~S. } \sn{Pope}\orcidlink{0000-0003-2595-9114}$^{2}$,  
\gn{Paul} \sn{Martini}\orcidlink{0000-0002-4279-4182}$^{3,4}$, 
\gn{Zhefu} \sn{Yu}\orcidlink{0000-0003-0644-9282}$^{5}$, 
\gn{Chris} \sn{Lidman}\orcidlink{0000-0003-1731-0497}$^{6,7}$, 
\gn{Geraint~F.} \sn{Lewis}\orcidlink{0000-0003-3081-9319}$^{8}$
}

\affil{
$^{1}$ School of Mathematics and Physics, University of Queensland, St Lucia, QLD 4072, Australia, 
$^{2}$ School of Mathematical \& Physical Sciences, 12 Wally's Walk, Macquarie University, Macquarie Park, NSW 2113
$^{3}$ Center for Cosmology and Astro-Particle Physics, The Ohio State University, Columbus, OH 43210, USA, 
$^{4}$ Department of Astronomy, The Ohio State University, Columbus, OH 43210, USA,
$^{5}$ Kavli Institute for Particle Astrophysics and Cosmology, Stanford University, 452 Lomita Mall, Stanford, CA 94305, USA
$^{6}$Centre for Gravitational Astrophysics, College of Science, Australian National University, ACT 2601, Australia, 
$^{7}$Research School of Astronomy and Astrophysics, Australian National University, ACT 2601, Australia, 
$^{8}$Sydney Institute for Astronomy, School of Physics, A28, The University of Sydney, NSW 2006, Australia,
}

\corresp{Hugh McDougall, Email: hughmcdougallemail@gmail.com}

\citeauth{McDougall et al., \red{PLACEHOLDER} {\it Publications of the Astronomical Society of Australia} {\bf 00}, 1--12. https://doi.org/10.1017/pasa.xxxx.xx}

\history{(Received xx xx xxxx; revised xx xx xxxx; accepted xx xx xxxx)}

\begin{abstract}
Reverberation mapping (RM) of Active galactic nuclei (AGN) provides one of the most direct probes of the geometry and kinematics of the broad-line region (BLR) of AGN by measuring time delays between continuum and line variability. However, modern RM surveys frequently suffer difficulties in recovering and gauging the reliability of lag measurements due to the compounding effects of poor signal to noise and the \qm{aliasing} problem, whereby multimodal lag posterior distributions arise due to seasonal gaps in our data. These challenge the reliability of commonly used fitting tools such as \javelin, which can return a high rate of false positive detections. To address this limitation, we implement a new lag measurement package, \litmus, and introduce a new framework that uses Bayesian evidence to identify false positive lag measurements, as well as examine the question of how many AGN present detectable lags in high redshift \qm{industrial scale} surveys like \ozdes and \sdss. Our analysis differs from previous RM studies in six key respects: (i) our inference is explicitly robust to the previously under-diagnosed numerical issues of multimodal posteriors, (ii) the entire sample is analysed using a single consistent methodology, (iii) uncertainty in the underlying AGN variability is fully marginalised, (iv) lag significance is assessed via Bayesian model comparison rather than heuristic metrics, (v) false-positive rates are quantified through comparison against random-chance recoveries and (vi) we use marginal likelihoods to distinguish between sources where a lag is not detectable in our data and sources that show no evidence of reverberation. Using our new pipeline to re-analyse the full \ozdes reverberation mapping sample, we have two major findings: (i) that previous reverberation mapping studies are likely to have overestimated the confidence of recovered lags, and (ii) we find a stark contrast between a very low reverberation percentage for the \mgii line ($f\approx3-28\%$ depending on assumptions) and much higher percentages in the \civ and especially the \hbeta line, which is consistent with $100\%$. We also present a re-analysed set of lags from the \ozdes sample with better-quantified false positive rates.
\end{abstract}


\maketitle

\section{Introduction}
\label{sec: intro}
Supermassive black holes (SMBH) form in the cores of all massive galaxies early in their history, where they subsequently regulate the evolution of their host galaxy and its environment \citep{Kormendy_2013}. In addition to quenching star formation, the demographics of these galactic nuclei and their evolution over cosmic time offer insights about galactic mergers \citep{Volonteri_2003_mergerref, Berti_2008_mergerref}, mechanisms of SMBH origin and growth channels \citep{Volonteri_2021}, and the nano-hertz gravitational wave background \citep{Rosado_2015_PTAref, Goncharov_2025_PTAref}. For the nearest galaxies we can estimate SMBH masses from their dynamical influence on their local environment \citep[e.g.][]{Schodel_2002} or from scaling relationships \citep[e.g.]{Ferrarese_2000,Gebhardt_2000}, but for more distant sources such methods become observationally difficult and untenable for large sample sizes.

For SMBH that are undergoing active accretion (Active Galactic Nuclei, or AGN) we can use the technique of reverberation mapping (RM) \citep{Blandford_McKee_1982, Peterson_1993}. Under the unified model \citep{Urry_1995}, an AGN consists of a number of distinct geometric components with similar distinct radiation signatures. In RM the ionised accretion disk formed by in-falling matter, often brighter than the stellar output of the host galaxy, acts as a central engine whose broad-band continuum radiation powers the emission of the other components. One such component is the Broad Line Region (BLR), a cloud light-weeks to light months removed from the central SMBH. Continuum light from the engine excites ions in the BLR which then fluoresce, resulting in line profiles that are heavily Doppler broadened by the high orbital speed required to maintain the cloud in orbit around the SMBH.\footnote{In this work we are concerned with BLR RM, but for a thorough review of several types of RM see \citet{Cackett_2021}.}

The disk is known to exhibit stochastic fluctuations in its brightness, with the variability known to be approximated by a Damped Random Walk \citep[DRW;][]{Kelly_2009, Kozlowski_2010a, MacLeod_2010}, a red noise process with constrained long-term variability. The DRW model is an example of a Gaussian Process (GP) model \citep[see][for a review]{Aigrain_2023_GPReview}, in which the continuum signal's structure is defined by the shape of its autocorrelation function. Because the disk powers the BLR emission, these fluctuations are echoed in the BLR emission strength, but with a lag time arising from the light travel time for the engine light to reach the cloud. For a characteristic lag, $\Delta t$, this implies a characteristic BLR radius $R_\mathrm{BLR} \approx c\cdot \Delta t$, where $c$ is the speed of light. With repeat photometric and spectroscopic observations of an AGN, we can construct independent light curves for the accretion disk engine and the broad line region. If these are sampled well enough to track the variability and for a long enough temporal baseline to observe the echo, these reverberation lags can be used to estimate the physical scale of the BLR. In concert with Doppler velocities, this offers an estimate of the central SMBH mass:
\begin{equation}    
    M_\mathrm{ BH} = f\frac{ R_\mathrm{ BLR} \braket{\sigma_v^2} }{G}, 
    \label{eq: RM_mass}
\end{equation}
where  $\sigma_v$ is a measure of the broad line width, $G$ is the gravitational constant, and the virial factor $f$ is an empirically derived factor to take into account the geometry and kinematics of the reverberating region \citep{Grier_2013b, Woo_2015, Villafana_2023_dynamical, SDSS-Shen_2023, Shen_2024_dynamical}.

\begin{figure*}
    \centering
    \includegraphics[width=0.9\linewidth]{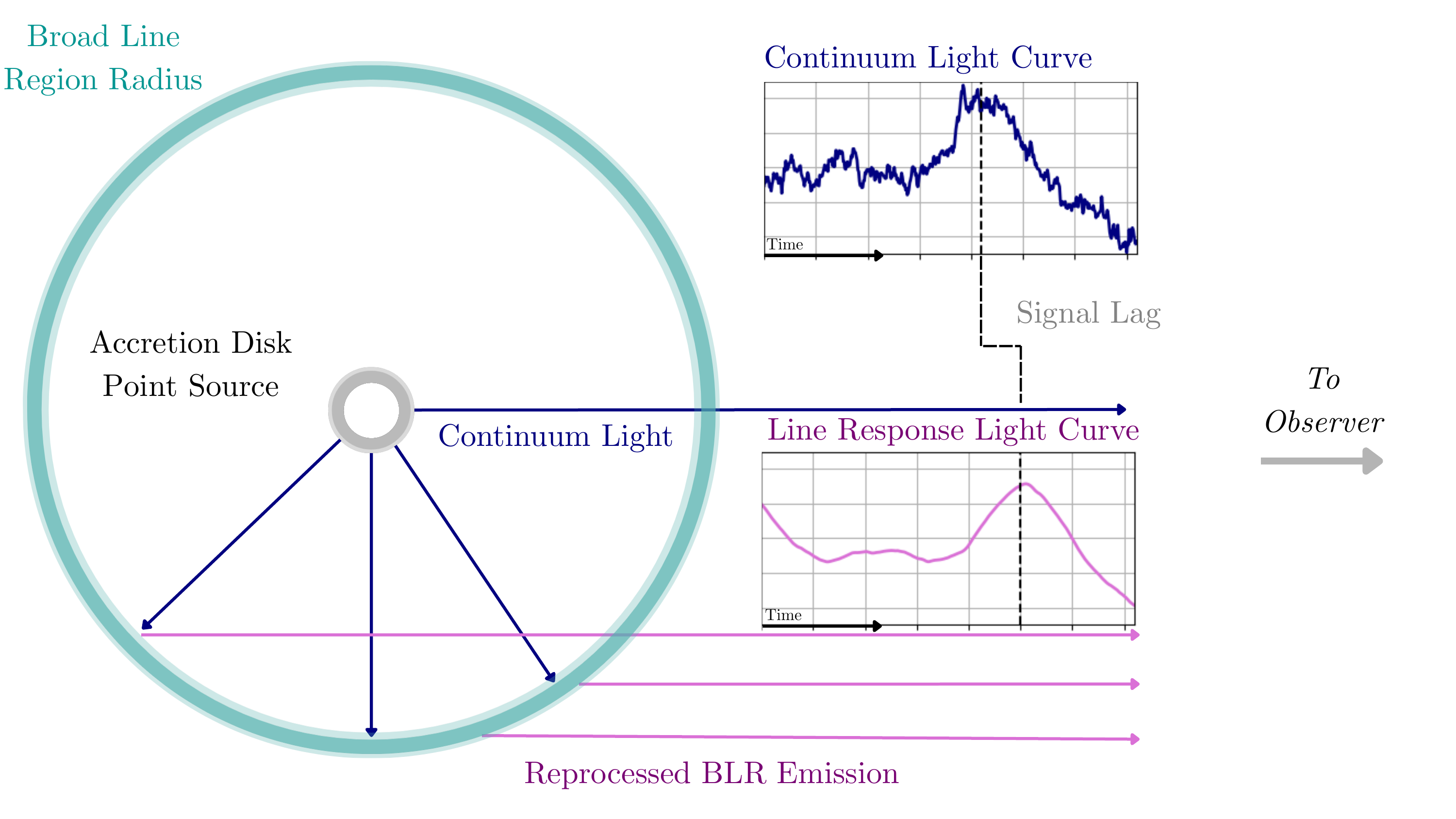}
    \caption{A diagram showing the geometry of BLR reverberation mapping in a simplified form, with different light travel paths shown for light directly from the continuum (navy) and the BLR re-processed light (orchid). In the simplest possible form, BLR RM assumes a geometrically thin BLR of some uniform characteristic radius and an accretion disk of comparatively small radial extent such that it can be approximated as a point source.}
    \label{fig: RM_diagram}
\end{figure*}

The previous decade saw the advent of so called \qm{industrial scale} RM, in which multi-year wide footprint surveys like the Australia Dark Energy Survey \citep[\ozdes;][]{OzDES-DR0-Yuan_2015} and the Sloan Digital Sky Survey \citep[\sdss;][]{SDSS-Shen_2015} harvested optical light curves for hundreds to thousands of AGN out to deep redshifts, with the aim being to construct a statistically significant picture of the AGN population and its evolution over cosmic time, and potentially use AGN as standard rulers for cosmology. One such intended product of these surveys is to constrain the power-law scaling relationship between an AGN's luminosity and its BLR radius as measured through its rest-frame lag, the so-called \qm{radius-luminosity} ($R-L$) relationship, empirically observed for low redshift AGN \citep{Kaspi_2000}. This $R-L$ relationship is described as a power law between the rest-frame lag and the luminosity with Gaussian Scatter in the log-axis,
\begin{equation}
    \log_{10}(\Delta t_\text{R-L}) = \alpha \log_{10}(\lambda L_{\lambda})+ \beta + \sigma\times\epsilon,
    \label{eq: R-L_powerlaw}
\end{equation}
where $\epsilon$ is a random Gaussian variable such that $\sigma$ is the scatter about the $R-L$ relation. Different $R-L$ relations are needed for lags observed with different emission lines from the BLR. The scaling of BLR radius with luminosity is the basis for the technique of \qm{stacking} \citep{Fine_2012_stacking, Fine_2013_stacking, Li_2017_stacking}, in which the the lag constraints from multiple AGN of similar luminosity are combined to achieve a measurement where a single source lacks the SNR to resolve a lag independently \citep[e.g.][]{OzDES-Malik_2024}. Similar relations have been established for other components of the AGN geometry, in particular the dusty torus \citep[e.g. recently]{Sun_2025_torusRL} and the accretion disk \citep[e.g.][]{Wang_2023_RLdisk}. The $R-L$ relation offers a means of estimating the radius, and by Equation~\ref{eq: RM_mass} the mass, of AGN with only a single spectroscopic observation. These \qm{single epoch} mass estimates allow for masses to be estimated for large samples of AGN without arduous time-domain observational campaigns \citep[as was done for the \ozdes sample in][]{OzDES-McDougall_2025}, and also act as the basis for estimating the mass of systems at extremely high redshift \citep[e.g.][]{Xihan_2025_highzAGNJ})

As we move through sources of increasing redshift, rest-frame optical wavelengths can appear redshifted beyond the visible range while rest-frame ultraviolet emission lines become visible. \ozdes and \sdss are both optical surveys, and so use a range of emission lines to perform RM across their sample. For the nearest sources ($z\in \sim[0,0.6]$), the $4861 \angstrom$ \hbeta line is used, the most well studied in the local universe from small-scale surveys \citep[e.g.][]{HBETA_Bentz_2009, HBETA_Bentz_2013, HBETA_Pei_2014, HBETA_Bentz_2014, HBETA_Du_2016, HBETA_Lu_2016, HBETA_Fausnaugh_2017, HBETA_Du_2018, HBETA_Rakshit_2019, HBETA_Zhang_2019, HBETA_Bentz_2016a, HBETA_Bentz_2016b, HBETA_Li_2021, HBETA_Bentz_2023, HBETA_Chen_2025}, and was studied in the sample \ozdes survey by \citet{OzDES-Malik_2023}. At higher redshifts ($z\in \sim[1.6,1.8]$), the $2798 \angstrom$ \mgii line is used, with \ozdes \citep{OzDES-Yu_2021, OzDES-Yu_2023} and \sdss \citep{SDSS-Shen_2023} being the source of the vast majority of measurements for this line, but the overall sample of \mgii lags is complemented by a handful of measurements from smaller surveys that widen the range of AGN luminosities for which lags are available \citep{MgII_Metzroth_2006, Kaspi_2007, MgII_Lira_2018, MgII_Czerny_2019,MgII_Zajacek_2020, MgII_Zajacek_2021}. The \mgii emission line is flanked by \feii lines, and contamination from these can confound RM if not properly identified and removed before constructing the response light curves \citep{OzDES-Yu_2021, OzDES-Yu_2023}.  The most distant RM sources ($z\gtrsim1.8$) use the $1549 \angstrom$ \civ line \citep[e.g.][]{CIV_Rosa_2015, CIV_Peterson_2005, MgII_Metzroth_2006}, believed to be emitted from $\approx2-4$ times smaller a radius than \hbeta or \mgii within the BLR \citep{Kaspi_2007, MgII_Lira_2018}. \civ lags are among the longest in the observer-frame due to cosmic time dilation and being necessarily drawn from the most distant (and most luminous) sources, and limitations of survey length appear to impose a ceiling on the longest of these lags that can be observed within a survey \citep{OzDES-Penton_2021, OzDES-Penton_2025, OzDES-McDougall_2025}.

A major limiting factor in deriving RM results and $R-L$ relations from modern surveys has been the difficulty in finding a reliable and consistent means of constraining and validating lags. The statistical models and numerical approaches / software have varied between surveys or even within a survey for different reverberating lines or as the adopted best-practices have evolved.  Difficulties arise on two fronts: firstly that many AGN light curves have poor signal to noise ratio (SNR) such that lags can be difficult to infer. Secondly, the half-year seasonal gaps that are unavoidable in multi-year ground-based surveys give rise to the problem of \qm{aliasing} \citep{OzDES-Yu_2021,OzDES-Penton_2021, OzDES-Malik_2022}, wherein many lag fitting methods lead to spurious but seemingly convincing lag constraints at $n+\frac{1}{2}$ yearly intervals, (e.g. $180 \dayu$, $540 \dayu$ etc.). These false positives, if not accounted for, contaminate mass measurements and the fitting of the $R-L$ relationship. In past works, both \ozdes and \sdss have adopted various ad-hoc measures of the reliability of a lag recovery\footnote{see \citet{OzDES-McDougall_2025} Section~$4.2$, and Appendix~A for a summary of these fitters and their underlying principles}, often combining lag constraints and reliability measures from a number of lag fitting programs. In this regime, what we refer to as a \qm{cut and constrain} framework, only the most reliable of lags are retained for constraining R-L relations. In \citet{OzDES-McDougall_2025} we demonstrated that apparent differences in the results of \ozdes and \sdss's results can largely be explained by differences in choices made in their analyses, namely their lag constraint and quality cut methods, rather than from differences in their physical sample.

Lag recovery methods can be coarsely divided into GP-based methods and non-GP based. GP-based fitters \citep[e.g. \javelin, \cream and \mica;][]{JAVELIN-Zu_2010, Starkey_2015_CREAM, Li_2016_MICA} make full use of our understanding of the AGN variability to constrain the lag \citep[along with other properties of interest such as the timescale of variability e.g.][]{Lewis_2023} with a Bayesian forward model. They are, in our view, the most principled and complete means of constraining lags, but do not provide direct metrics of the confidence in a lag recovery. Non-GP methods \citep[e.g. \PyCCF and \pyroa;]{PyCCF-Mouyan_2018, pyroa-ferus_2021} make less use of our understanding of the AGN light curve structure, but provide nominal measures of the significance of a lag. \javelin in particular is a widely used tool for RM lag constraint at many scales (e.g. \citet{SDSS-Grier_2017, Homayouni_2019_SDSS_AccDisk, Guo_2022_contRM_ZTF} and more recently \citet{Sun_2025_JAVELINuser, Catalina_2025_JAVELINuser, Mandal_2026_JAVELINuser}), and was the primary method in \ozdes and earlier \sdss data releases. In \citet{McDougall_2025_LITMUS}, we identified that \javelin's choice of Bayesian Sampler, \texttt{emcee} \citep{emcee-Foreman_Mackey_2013}, fails in the multimodal lag posterior distributions that arise from seasonal surveys, meaning that any lags measured with \javelin or similar GP-based fitters would be contaminated by non-convergence that exaggerate the aliasing problem.

In this paper, we re-analyse the entire \ozdes sample in a cohesive framework using \litmus \citep{McDougall_2025_LITMUS}, a new suite of lag estimation tools focused on fitting lags for quasar-like light curves. Relevant to this work, \litmus properly handles the numerical issues of aliasing to give accurate lag posteriors. It also, via Bayesian evidence integrals, gives measures of confidence for whether a signal does or does not exhibit a lag in the same consistent statistical framework as the lag constraint, something not yet done in the field. Because these significance measures can distinguish between the absence of a clear lag and the clear absence of a lag, it also empowers us to ask questions about how many sources in our sample contain lags to be recovered. 

Our analysis differs from previous RM papers firstly in that our results accurately reflect the posterior distribution for our sources, second that we measure the entire \ozdes sample with a single consistent approach, third that we take the heretofore under-examined step of quantifying our ignorance of the AGN variability with novel extensions to the DRW model, fourth that we use principled Bayesian model comparison tools to estimate the significance of lag measurements instead of ad-hoc measures, fifth that we properly estimate the false positive rate (FPR) of our recoveries by comparing our results to random chance, and finally that we use derivatives of our Bayesian significance measures to put constraints on how many AGN in our sample do or do not contain a clear lag. Our new approach quantifies the limits of what can be done using the traditional \qm{cut and constrain} approach to lag recovery used by \sdss and past \ozdes analyses.

The paper is laid out as follows: in Section~\ref{sec: Data} we discuss the \ozdes and \des data sets that we make use of in this paper. Section~\ref{sec: stats_modellling} outlines our modelling approach for lag measurements, including the novel light curve modelling and significance measures we use in this paper as well as the numerical scheme we use for Bayesian lag fitting. In Section~\ref{sec: signif} we describe how we estimate the FPR of our lags and select our high quality lag sample. This sample is presented, and contrasted with past \ozdes measurements, in Section~\ref{sec: lag_results}.  The limitations of RM with this data set are then examined in Section~\ref{sec: RM_limitations}, along with the implications that these limitations have for our understanding of AGN BLR light curves. In Section~\ref{sec: reverb_frac} we examine an as of yet unexplored question in the field: what fraction of AGN meaningfully reverberate in the simple way that we model in BLR RM. We conclude in Section~\ref{sec: conclusion} by discussing next steps in the application of our pipeline and suggest future applications for population-level BLR RM.

\section{Data \& The \ozdes Sample}
\label{sec: Data}
In this work, we analyse AGN light curves from the $7$-year \ozdes project \citep{OzDES-DR0-Yuan_2015,OZDES-DR1-Childress_2017,OZDES-DR2-Lidman_2020}, a program to acquire spectroscopic measurements for sources with photometric imaging from the Dark Energy Survey \citep[\des;][]{DES_2016_REF}, using the AAOmega spectrograph fed by the Two Degree Field (2dF) fibre positioner \citep{Lewis_2002_2df}. The \des \decam measurements were performed on the CTIO Blanco 4-metre Telescope in Chile at an approximately weekly cadence in in the $g, r, i, z$ filters, while the \ozdes spectra were measured with a roughly monthly cadence with a spectral resolution of $R=1400$ to $1700$ in the wavelength range of $3700\,\angstrom$\ to $8800\,\angstrom$. Both \ozdes and \sdss light curves follow $5-6$ month observational seasons due to observational fields passing near the sun.

Where we refer to the continuum measurements in this work, we follow the convention set by other \ozdes papers and use the $g$-band photometry from \des \citep{Flaugher_2015}. Before using the photometric light curves, we clean them of outlier observations using the method described in Appendix~\ref{app: lightcurve_cleaning}. Emission line response curves are built from spectra using the pipeline of \citet{OzDES-Hoormann_2019} for the \hbeta and \civ sources, and the pipeline of \citet{OzDES-Yu_2021}, which fits for and removes contamination from the Fe\textsc{II} line, for the \mgii sources. As a part of \citet{OzDES-Hoormann_2019}'s pipeline for \hbeta and \civ spectra, observational epochs may be co-added by date (all spectra from the same night) or by run (all measurements from the same observing run, which typically lasted several days). Co-adding by date yields light curves with noisier but more frequent spectra. In this work we examine both data sets independently. This yields a total of $906$ sources with emission line light curves, with $77$ \hbeta sources, $453$ \mgii sources and $376$ \civ sources. A full description of this data set and how it was produced is available in the \ozdes reverberation mapping data release \citep{OzDES-McDougall_2025}.

\section{Statistical Modelling}
\label{sec: stats_modellling}
The core motivation of this paper is a statistically detailed re-analysis of the \ozdes RM sample, and here we describe the three areas of statistical analysis that we advance compared to previous efforts: changes to the modelling of the light curves as a Gaussian Process (i.e. the Bayesian generative model; Sec.~\ref{sec: LC_modelling}), the measures of lag recovery confidence (Bayesian model comparison; Sec~\ref{sec: signif}), and the numerical methods used to properly fit lags without falling victim to the numerical artefacts of codes like \javelin (the Bayesian sampling / integration schemes; Sec~\ref{sec: litmus_and_numerical}).

\subsection{Bayesian Light Curve Modelling}
\label{sec: LC_modelling}
In this section we describe statistical framework for measuring lags with GP based methods in general \citep{Blandford_McKee_1982}, and then in Sections~\ref{sec: mod-DRW_simple},~\ref{sec: mod-tophat}~and~~\ref{sec: mod-jitter} describe the specifics of the DRW model and the variants we employ in this work.

The basis of the GP model is that, while the light curves of AGN vary in a stochastic manner, they exhibit reasonably stationary power-spectral densities (PSDs). This allows us to model them as a GP, in which observations are correlated by some covariance matrix $C$, meaning they obey a multivariate Gaussian likelihood:
\begin{equation}
    \mathcal{L}(D \vert \theta) = \frac{1}{\sqrt{(2\pi)^N\det(C)}} 
    \exp\left(-\frac{1}{2} \vec{y}^TC^{-1}\vec{y} \right)
    ,
    \label{eq: GP_likelihood}
\end{equation}
where $\vec{y}$ is the vector of observations over all light curves after subtracting off their respective means, and $D$ denotes all observational data. The covariance matrix consists of two components added together: the symmetrical positive-definite matrix $S$ which encodes the stochastic signal structure and positive-diagonal matrix $N$ which encodes white-noise measurement uncertainty:
\begin{equation}
    C = S + N, \; 
    \begin{array}l
    N_{ij}=\delta_{ij} E_i E_j\\
    S_{ij}=\braket{y_i,y_j}
    \end{array}
    .
    \label{eq: covar_matrix}
\end{equation}
Here, $E_{i/j}$ are the (assumed Gaussian) measurement uncertainties in observations $y_{i/j}$ and $\delta_{ij}$ is the Kronecker delta such that $N$ is diagonal. Different GPs are distinguished by the structure of $S$, the elements of which are constructed from the signal's covariance function. In the instance of BLR RM, this function takes different forms for covariances within and between light curves. We assume some GP for the continuum light curve (typically a damped random walk, see `The Simple DRW Model' section below), along with a corresponding functional form for its autocovariance function:
\begin{equation}
    \phi_{cc}(t_i, t_j) = \braket{y_c(t_i),y_c(t_j)}
    .
    \label{eq: cc_covar}
\end{equation}
Where $t_{i/j}$ are the epochs of measurements $y_{i/j}$. This covariance function is assumed to be stationary such that it depends only on the time difference between two measurements, $\phi_{cc}(t_i,t_j)=\phi_{cc}(\delta t)$, $\delta t = \abs{t_j-t_i}$. The elements of $S$ corresponding to continuum-continuum measurements are drawn from this equation.

\begin{figure*}
    \centering
    \includegraphics[width=0.95\linewidth]{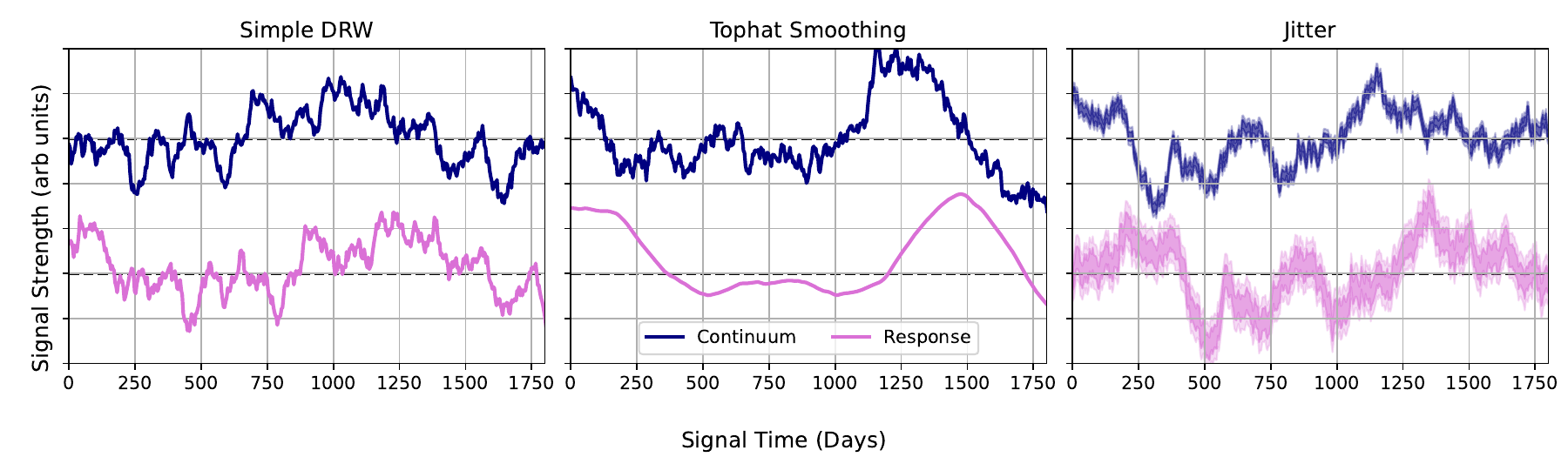}
    \caption{Examples of three light-curve models for mock signals with a timescale $400 \, \dayu$ and a lag of $200 \,\dayu$. From left to right: the simple DRW model with no smoothing, the tophat-smoothing model in which the response is smoothed by a width of $300 \, \dayu$, and the jitter model in which the continuum has a $10\%$ white noise contribution while the line response has $25\%$.}
    \label{fig: Lightcurve_Examples}
\end{figure*}

The BLR response light curve is modelled as a shifted and distorted echo of the continuum, represented by a convolution with some transfer function $\psi(t)$\footnote{More complex descriptions of the transfer function are used in velocity resolved RM \citep[e.g.][]{Bentz_2009_VRRM, Shu_2025_VRRM}, where the observed emission line response is modelled as a convolution of different BLR regions with varying levels of Doppler broadening responding at different lag times.}. In this model, the covariance between the continuum and response, $\phi_{rc}(\delta t)$, is a convolution of the continuum covariance function and the transfer function:
\begin{equation}
    \phi_{rc}(\delta t)
    = \int_{-\infty}^{\infty}{\phi_{cc}(\delta t)\psi(\delta t-t')}dt'
    ,
    \label{eq: conv_transfer}
\end{equation}
and the auto-covariance of the response signal is a double-convolution:
\begin{equation}
    \phi_{rr}(\delta t)
    = \int_{-\infty}^{\infty}\int_{-\infty}^{\infty}{\phi_{cc}(\delta t)\psi(\delta t-t')\psi(\delta t-t'')}dt'dt''
    .
    \label{eq: conv_response}
\end{equation}
Different descriptions of the AGN light curve are characterised by the choice of $\phi_{cc}(t)$ and the functional form of the transfer function. Simple descriptions of the transfer function (e.g. the tophat model we use here) result in a response light curve that is a \qm{shifted, scaled and smoothed} copy of the continuum. In this work we use models that extend slightly beyond this definition, and so suggest the more general term of \qm{call and response} models for instances of single characteristic lag linking stationary GPs. In this work we employ three such models as shown in Figure~\ref{fig: Lightcurve_Examples} and described below.

\paragraph{The Simple Damped Random Walk Model}
\label{sec: mod-DRW_simple}
AGN continuum variability exhibits red noise at short timescales and is stationary and bounded at long timescales, two characteristics well described by the DRW. In the DRW, the continuum auto-covariance of Equation~\ref{eq: cc_covar} obeys the two-sided exponential shape of the Laplace distribution:
\begin{equation}
    \phi_{cc}(t_i,t_j)=\sigma_c^2 \exp \left( -\abs{\frac{t_i-t_j}{\tau}} \right)
    ,
    \label{eq: DRW_covar}
\end{equation}
where $\sigma_c$ is the amplitude of variability of the continuum light curve over infinite time and $\tau$ is the timescale of the fluctuations. In the simplest model for AGN RM, the response is a simple delayed and scaled copy of the continuum, i.e. with the transfer function being a delta function $\psi(t)=\frac{\sigma_r}{\sigma_c}\delta(t-\Delta t)$, where $\sigma_r$ is the amplitude of the response light curve and $\Delta t$ is the time lag. This \qm{Simple DRW} model is the minium complexity model for modelling response lags if the continuum is a DRW, and acts as the baseline for comparing any other variations (e.g. noise processes in or smoothing of the response).

In this simple model, the elements of the signal covariance matrix are, for observations $i$ and $j$ being drawn from the continuum and response functions respectively, built from the covariance function:
\begin{equation}
\phi_{rc}(t_i, t_j) = \sigma_r\sigma_c \exp \left( \frac{-\lvert t_i-t_j-\Delta t \rvert}{\tau} \right)
,
\label{eq: phi_rc_DRW}
\end{equation}
while the response autocovariance is:
\begin{equation}
\phi_{rr}(t_i, t_j) = \sigma_r^2\exp \left( \frac{-\lvert t_i-t_j\rvert}{\tau} \right)
.
\label{eq: phi_rr_DRW}
\end{equation}

\paragraph{The Tophat-Smoothing Model}
\label{sec: mod-tophat}
In most RM lag recovery codes, the BLR response is modelled to have some degree of smoothing compared to the red-noise of the continuum. Though the true transfer function is believed be to complex and possibly multimodal \citep{Li_2016_MICA}, a common approximation (and the one made by \javelin) is to use a tophat smoothing kernel of width $w$ for the transfer function, i.e.:
\begin{equation}
    \psi(t) = \frac{1}{h}\left\{\begin{array}{lr}
        1 , \; \Delta t<t<\Delta t+w \\
        0 , \; \mathrm{otherwise}.
    \end{array}\right.
    \label{eq: tophat}
\end{equation}
When convolved with the exponential covariance function in Equation~\ref{eq: phi_rr_DRW} per Equation~\ref{eq: conv_transfer} of the simple DRW model, this leads to covariance functions between the continuum and response:
\begin{multline}
    \phi_{rc}(t_i,t_j)
    = \sigma_r \sigma_c\frac{1-\exp{-\frac{b}{2}}}{\sqrt{\frac{1}{2}(b+\exp{-b}-1)}}
    \times \\
    \left\{\begin{array}{lr}
        1 - {(
            \cosh\left(\frac{\delta u^{ij}}{\tau}\right) - 1
        )}
        {(
            \exp(\frac{w}{2\tau})-1
        )^{-1}} \
        , \;
        \delta u^{ij} < \frac{w}{2} \\
        \exp\left(\frac{\delta u^{ij}}{\tau}\right)
        {(
            \sinh(\frac{w}{2\tau})
        )}{(
            \exp(-\frac{w}{2\tau})
        )^{-1}}
        , \;
        \delta u^{ij} > \frac{w}{2}
    \end{array}\right.
    \label{eq: rc_covar},
\end{multline}
where, for continuum and response measurements at time $t_i$ and $t_j$, $\delta u^{ij} = \lvert t_i-t_j-\Delta t - \frac{w}{2} \rvert $. The response's self covariance meanwhile, after a double-convolution per Equation~\ref{eq: conv_response}, is:
\begin{multline}
    \phi_{rr}(t_i,t_j) =\sigma_r^2 \times \\
    \left\{\begin{array}{lr}
        \frac{w-\delta t^{ij}}{\tau} + \exp(-\frac{w}{\tau}) \times \cosh(\frac{\delta t^{ij}}{\tau}) - \exp(\frac{-\delta t}{\tau})
        , \; \delta t^{ij} < w\\
        \exp(\frac{-x}{\tau})  \cosh(\frac{w}{\tau})-1
        , \; \delta t^{ij} > w
        \end{array}\right.
        ,
    \label{eq: rr_covar}
\end{multline}
where $\delta t^{ij}=\abs{t_i-t_j}$. Note that in formulating this equation, we have chosen $h$ in Equation~\ref{eq: tophat} such that the amplitude of Equation~\ref{eq: rr_covar} is equal to $\sigma_r$, the variance (i.e.\ signal amplitude) of the response signal. In the limit of $w\rightarrow0$, these covariance equations approach the case of $\psi(t)=\delta t$, in which no smoothing is applied.

Note that there are two ways to measure the characteristic lag for the tophat model: from the middle of the tophat transfer function or from the leading edge, i.e. the time until the earliest response from the BLR or until the average (highest correlation) response. These differ (aside from slight edge effects in the prior boundaries) only by a simple reparameterization, $\Delta t_\text{cent} = \Delta t + \frac{w}{2}$. Where we examine this tophat smoothing model, we present results for both. Because reparameterization does not changed the Bayes factor, these will always be the same for the tophat and tophat-centroid models, but they will differ in the recovered lag and maximum marginal likelihood (See Section~\ref{sec: signif}).

\paragraph{The Jitter Model}
\label{sec: mod-jitter}
There is strong evidence that AGN light curves depart from the idealised DRW model at short timescales \citep[e.g.][]{Mushotzky_2011_nonDRW, Kasliwal_2015_nonDRW, Smith_2018_nonDRW, Stone_2020_nonDRW}. Such deviations may be due to the true underlying continuum emission itself not being a perfect DRW \citep[for example some sources model the light curve as a damped harmonic oscillator, a more flexible higher order GP with a degree of inertia][]{Kelly_2014_nonDRW_DHO, Kasliwal_2017_nonDRW_DHO, Moreno_2019_nonDRW_DHO, Yu_2025_nonDRW_DHO}, from short-lived transient events that disrupt the appearance of this emission \citep[e.g. microlensing][]{Vernardos_2024_microlensing, Best_2025_microlensing}, or from non-stationarity of the signal mean \citep{DallaBonta_2025_detrending, Kroupa_2026a_stationarity}. As a catch-all first order correction to the simple DRW model to account for any such deviations, we also include a model in which both the continuum and response have a \qm{jitter} component, i.e. some portion of their variability being pure white noise so that the DRW is, in effect, fit in a \qm{least squares} type way with any remaining variability being absorbed by this jitter. In this model, the signal covariance matrix is constructed to still have the same signal amplitude / diagonal elements irrespective of the amount of jitter for the continuum and response:
\begin{equation}
    S_{ij} = S_{ij,\text{DRW}}\sqrt{1-\delta_{ij}J_iJ_j}+\delta_{ji}J_{i}J_{j}\sigma_i\sigma_j,
\end{equation}
where $S_\text{DRW}$ is the covariance matrix in Section~\ref{sec: mod-DRW_simple} and $J_i$ and $J_j$ are the jitter amounts ($J_{r/c}\in[0,1]$) for whatever light curves observations $i$ and $j$ are drawn from. In this way, the jitters allow each light curve model to smoothly transition from a pure DRW at $J_{c/r}=0$ to pure unstructured white noise at $J_{r/c}=1$. This is not intended to be an actual model of the AGN variation; a true white noise component is physically impossible for a system with any physical size, and we do not propagate this white noise through the transfer function convolution. Rather, this is meant to represent a crude accounting for any and all ways in which the continuum differs from the simple DRW model, and as a means of quantifying our ignorance of AGN variability. We note that this is very similar to the error calibration method of \cream, in which the measurement uncertainties of the $N$ matrix are scaled by some factor, but here we add some additional variance rather than scaling the existing error budget.

\begin{figure*}
    \centering
    \includegraphics[scale=0.8, trim={0cm 2.0cm 0cm 0cm}, clip]{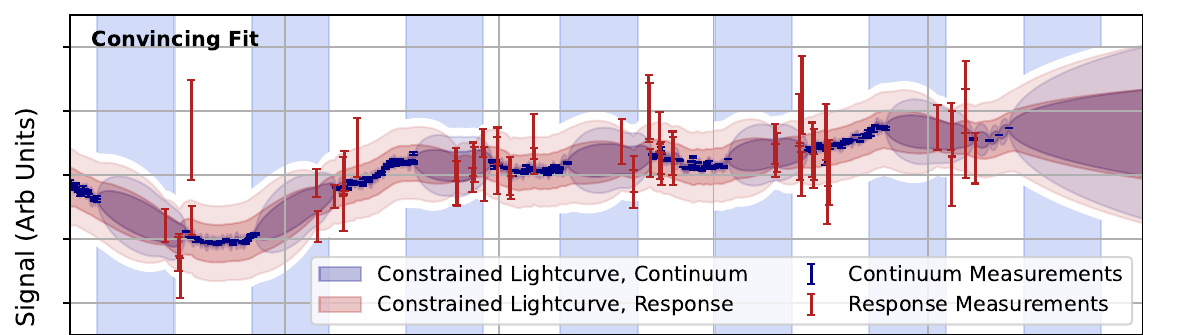}\\
    \includegraphics[scale=0.8, trim={0cm 0cm 0.1cm 0.2cm}, clip]{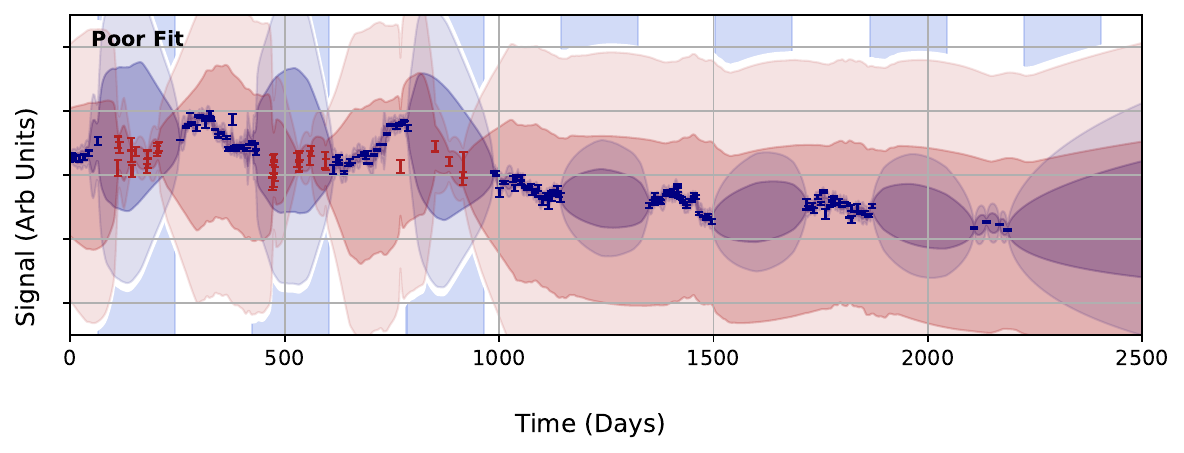}
    \caption{Examples of light curve constraints for two \ozdes sources overlaid for their continuum (blue) and \mgii (dark red) light curves when fitting for a lag with the coupled jitter model, offsetting by the mean recovered lag. The shaded regions represent the uncertainties due to GP stochasticity and after marginalising over uncertainty in the timescale, signal means and signal amplitudes. The top panel shows constraints for \ozdes source $2939782606$. This lag ($\Delta t={77}^{+19}_{-24}$) is fit very convincingly ($\BF{Lag}=32.15$, vanishingly low FPR), and the constrained light curves are tightly fit, agree with one another and the observations. By contrast, the bottom panel shows the same model attempting to fit a lag to \ozdes source $2971128594$, which performs poorly ($\BF{Lag}=0.32$, FPR$\ge50\%$) and is discarded. The light blue background shading highlights the observation-free seasons.}
    \label{fig: constrained_lightcurves}
\end{figure*}

\subsection{Null Hypothesis Models \& Measures of Significance}\label{sec: signif}
The difference of greatest impact between this analysis compared to previous \ozdes RM works is our use of significance criteria based on Bayesian model comparison. For each source (after selecting a light curve model from Section~\ref{sec: LC_modelling}) we have four model hypotheses:
\begin{enumerate} 
    \item The \qm{coupled} model in which the light curves are coupled with some encoded lag; 
    \item An \qm{uncoupled} model, the same light curve model as the lag-bearing one but with the light-curves being different instances of the GP. This is equivalent to setting the lag to infinity;
    \item A \qm{GP-Noise} model in which the continuum light curve is structured as a GP but the response is pure Gaussian white noise;
    \item A \qm{Noise-Noise} model in which both the continuum and noise are unstructured.
\end{enumerate}
By comparing the performance of these hypotheses we can learn what model best describes our observations of the AGN light curves. The most common way to compare two models in a Bayesian framework is the Bayes factor, the ratio of their model evidences (also known as their marginal likelihoods). A model's evidence, representing an overall \qm{goodness of fit} measure for a model, describes how easily a particular generative model and prior distribution can reproduce our observations. It is defined as an integral of the prior-weighted likelihood over all model parameters, $\theta$:
\begin{equation}
    \ev_\text{Model}(D) = \int \pi_\text{Model}(\theta)\mathcal{L}_\text{Model}(D\vert\theta) d\theta
    ,
    \label{eq: evidence_definition}
\end{equation}
where $\pi(\theta)$ represents the prior distribution on the model parameters. In BLR-RM, convention is to use wide uniform priors on all parameters (see Section~\ref{sec: litmus_and_numerical} for specifics). The Bayes factor between two models is $\BF{a-b}=\frac{\ev_a}{\ev_b}$, with a Bayes factor of $1$ indicating ambiguity between the two models, and $\gg1$ or $\ll1$ indicating preference for models $a$ or $b$ respectively.

Using these priors we can compare the four models with three Bayes factors:
\begin{itemize}
    \item $\BF{Lag}$ comparing the coupled ($1$) and uncoupled ($2$) models and telling us about the strength of the lag detection, i.e. whether the apparent alignment of the continuum and response at some lag could be produced by random chance with two stochastic signals,
    \item $\BF{Struc,Resp}$ comparing the uncoupled ($2$) and GP-noise ($3$) models and telling us about how much structured variability is observed in the response light curve, and
    \item $\BF{Struc,Cont}$ comparing the GP-noise ($3$) and noise-noise ($4$) models and telling us how much structured variability is seen in the continuum light curve.
\end{itemize}

The most important of these is $\BF{Lag}$, as it directly tests how well a particular source demonstrates a lag inclusive of all uncertainty in the observations and light curve parameters. In Figure~\ref{fig: constrained_lightcurves} we show examples of two light curves, one with a high $\BF{Lag}$ (strong recovery) and another with a low $\BF{Lag}$ (a poor recovery), demonstrating how these measures quantify our overall confidence in a lag measurement. An important feature here is that the Bayes factor can distinguish between vague and negative results: if $\BF{Lag}\approx1$ it means we cannot confidently say we have seen a lag, but if $\BF{Lag}\ll1$ it means we can confidently say that we haven't.

Ideally we would use only the Bayes factor to compare the models, however proper use of a Bayes factor is contingent on detailed knowledge of a physically justified prior, which we do not have. The traditional uniform prior on lag is suspect in two ways: firstly it is somewhat arbitrary (there is for example no reason for it to be uniform and not log-uniform, and it relies on arbitrary upper and lower bounds), and secondly that it does not describe the true population-informed prior on AGN lags. To remedy this, we adopt a pseudo-Bayesian approach with significance measures derived from Bayesian likelihoods, but significance thresholds built up in a frequentist way (see Section~\ref{sec: selection_criteria} for details).

\begin{figure*}
    \centering
    \includegraphics[width=0.9\linewidth, trim={0.6cm 4.2cm 1cm 0cm}, clip]{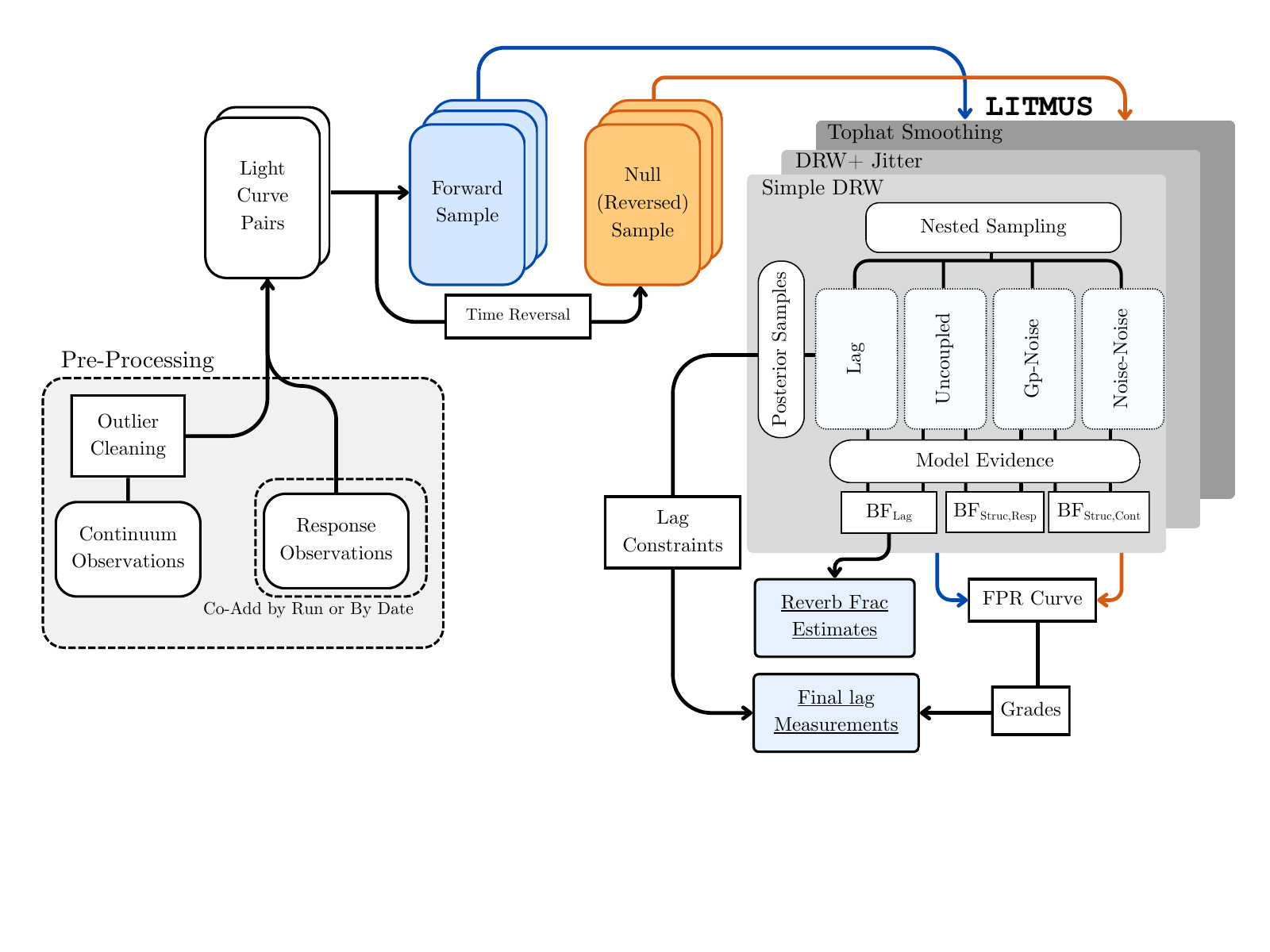}
    \caption{A flowchart showing our full pipeline for lag measurement in this paper. The flow is, in short, to clean the light curves, reverse them to produce forward and reversed samples, run both through \litmus for all light curve models for all model hypotheses, then use the resulting significance measures to tune an FPR curve to identify the most reliable lags. Not shown is the spectral and photometric calibration used to produce the pre-cleaning light curves or the way that the posterior samples and Bayes Factors are used to calculate $\Lrat{Lag}$.}
    \label{fig: flowchart}
\end{figure*}

\paragraph{Alternatives to Uniform Prior Bayes Factor}
Our principle significance measure is the Bayes factor from the uniform lag prior, but it is reasonable to have concerns about results from such a non-physical prior. To mollify those concerns,  we additionally present two other measures of model comparison that do not rely on a uniform prior. One such measure of significance, one that is \qm{prior-free}, is to use the relative likelihood of the best fit lag (highest marginal likelihood) against the evidence of the uncoupled model. 

\begin{figure*}
    \centering
    \includegraphics[width=0.9\linewidth]{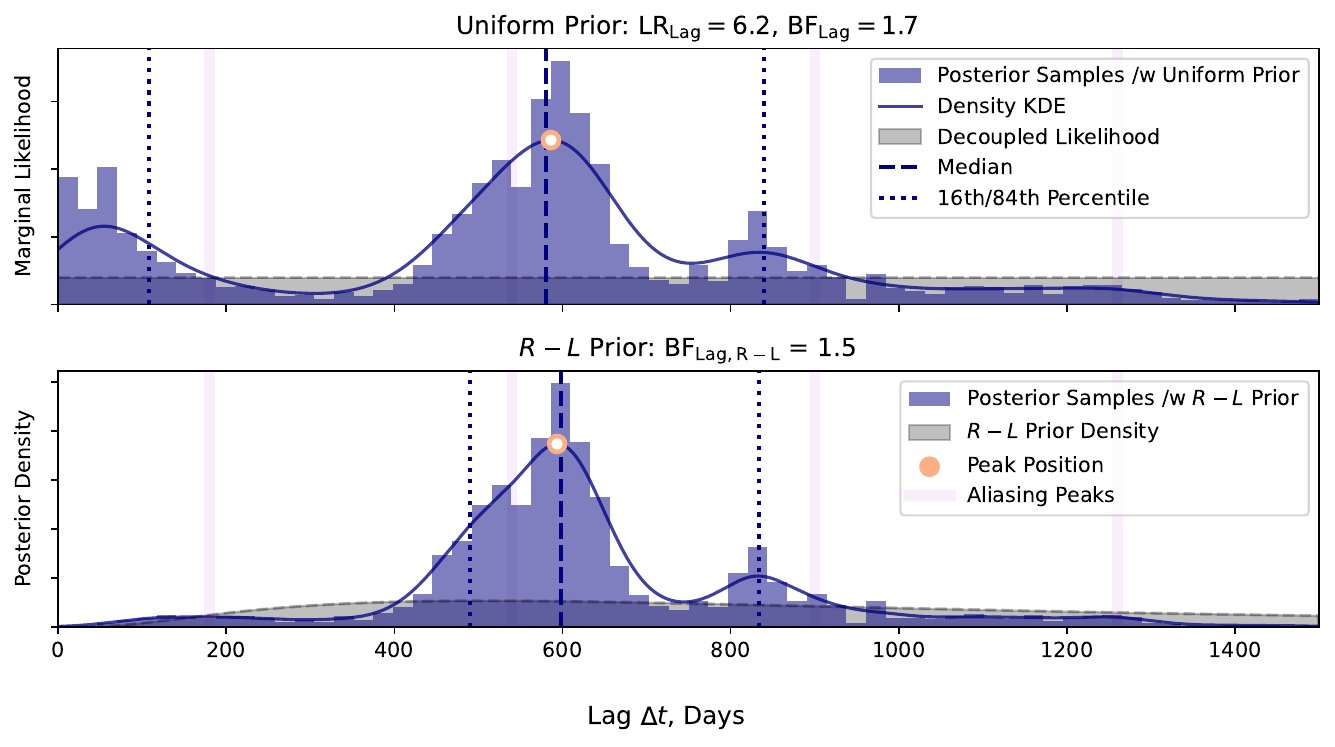}
    \caption{An example of the lag-posterior processing for a potential lag as fit with the simple DRW model, namely that of \ozdes source $2970402464$'s \civ lag. The top panel shows the samples for the uniform lag prior with the height of the shaded region along the bottom of the panel representing the marginal likelihood of the uncoupled null model. The solid line is the KDE-smoothed distribution overlaid on the $128$ bin histogram of the posterior samples, with the peak, used to calculate $\Lrat{Lag}$, shown as a white and orange circle. The vertical dashed and dotted lines show the median along with the \pone and \ptwo percentiles. The lower panel shows the same information after re-weighting by the $R-L$ informed prior from \citet{OzDES-McDougall_2025}, with the log-normal prior itself shown in shaded grey. In both panels, the relative normalisations of the posterior and priors indicate the relative model evidences for the coupled and uncoupled models. The vertical light-purple shaded lines represent the $n+\frac{1}{2}\mathrm{year}$ positions of potential aliasing peaks. Applying the $R-L$ prior concentrates the lag distribution at $\approx 600 \dayu$, but leads to a mild decrease in Bayes factor due to the suppression of the posterior mode at $\Delta t \approx0$. This would not be a particularly convincing source with either prior, as the Bayes factors are not particularly strong ($\BF{Lag}=1.7$, $\BF{Lag,R-L}=1.5$) and the marginal likelihood ratio is only moderate ($\Lrat{Lag}=6.2$).}
    \label{fig: posterior_processing}
\end{figure*}

This marginal likelihood is the integral over all non-lag parameters $\theta^\prime$ at some lag $\Delta t$:
\begin{equation}
    \mathcal{L}_\text{model}(D\vert\Delta t) = \int \pi(\theta^\prime) \mathcal{L}_\text{model}(D\vert\Delta t,\theta^\prime)d\theta^\prime
    .
    \label{eq: marginal_likelihood}
\end{equation}
We can compare the marginal likelihood of this best fit (highest marginal likelihood, occurring at $ \hat{\Delta t}$) lag to that of the uncoupled model, $\Lrat{Lag}= \mathcal{L}_\text{model}(D\vert \hat{\Delta t}) / \ev_\text{uncoupled}$. Compared to $\BF{Lag}$, this can be thought of as measuring the peak of the lag likelihood distribution instead of its average height, meaning that unlike the Bayes factor it has no contributions from secondary modes in the lag distribution. See Figure~\ref{fig: posterior_processing} for a demonstration.

Though we perform our main fits with a uniform prior on lag, we are equipped with a more informative prior in form of the $R-L$ relationship. In the lag axis, for an AGN of redshift $z$, Equation~\ref{eq: R-L_powerlaw} becomes a log-normal prior truncated to match the upper bound of our lag search range / uniform prior:
\begin{multline}
    \pi(\Delta t)=
    \begin{cases}    
    \frac{1}{
        \sqrt{2\pi} \cdot
        k(\Delta t_\text{Max})\cdot
        \ln(10)\Delta t
    }\\\times
    \exp\left(\frac{1}{2} \left(\frac{\log_{10}(\Delta t / ((1+z)\Delta t_{R-L})) }{\sigma}\right)^2\right), \Delta t < \Delta t_\text{Max}\\
    0, \text{otherwise}
    \end{cases},
    \label{eq: R-L_prior}
\end{multline}
where $\Delta t_{R-L}$ is the rest-frame lag predicted by Equation~\ref{eq: R-L_powerlaw}, $z$ is the source redshift, $\sigma$ is the scatter in this $R-L$ relation and $k=\frac{1}{2} \left(1 + \frac{\text{erf}(\log_{10}(\Delta t_{\text{Max}})-\mu)/\sigma)}{\sqrt{2}} \right)$ is a normalisation factor from the truncation of the prior at $\Delta t_\text{Max}$. If we use this physically motivated prior for $\Delta t$ in place of a vague uniform prior, we get $\BF{Lag,R-L}$, a second Bayes factor for the lag. We also accordingly get a new constraint on the lag subject to this prior. Note that we need to make use of some parameters (slope, offset, scatter) for the $R-L$ relations of each line, and we here use the median values presented in \citet{OzDES-McDougall_2025}. We urge some caution in interpreting results derived from these priors, as these are in part tuned using the \ozdes sources that we are re-fitting here and so are subject to a degree of confirmation bias. We discuss a possible future framework for modelling the $R-L$ relationship without bias in Section~\ref{sec: conclusion}.

\subsection{Numerical Modelling \& Analysis of Results}
\label{sec: litmus_and_numerical}

In \citet{McDougall_2025_LITMUS} we demonstrate that RM results are highly sensitive to the numerical fitting scheme, even when adopting a consistent statistical model, due to artefacts arising from the multi-modal nature of the posterior distribution (i.e. the aliasing problem). In this section we outline the numerical details of the lag fitting procedure we use in this paper, as well as other methodological details such as our method of cleaning the continuum light curves of outliers and the prior ranges of our Bayesian models.

\paragraph{Light Curve Pre-Processing \& Priors}
Before fitting, the continuum light curves are cleaned of clear outliers by the method described in Appendix~\ref{app: lightcurve_cleaning}, fitting a rough trend and scatter to each season and trimming observations that are inconsistent with this trend. We then normalise both the continuum and response light curves to approximately zero mean and unit variance per the normalisation method described in \citet{McDougall_2025_LITMUS}. Note that this does not affect any lag fitting, but ensures our priors on mean and amplitude properly capture the signal width. These nuisance parameters are fit with the same arbitrarily wide priors and parameterization described in \citet{McDougall_2025_LITMUS}.

For the observer frame lag, we fit with a uniform prior $\Delta t \in [0,1500] \dayu$, wide enough to capture the longest lag predicted for our sample with the fiducial $R-L$ relations we present in \citet{OzDES-McDougall_2025}.\footnote{As a part of the posterior processing we also re-weight to a prior with a maximum lag of $1000 \dayu$, but find that it has no notable impact on our overall findings.} For the timescale of the DRW in the GP-based models (i.e. all models except noise-noise), we adopt a wide log-uniform prior $\pi(\ln(\tau / \dayu))=U(0.0,\ 10.0)$, following the convention of \javelin. For the tophat-smoothing we similarly use a log-uniform prior $\pi(\ln(b / \dayu))=U(\ln(-2.30,\ 6.91)$, corresponding to upper and lower bounds of $0.1 \; \dayu$ and $1000.0  \; \dayu$, spanning from significantly smaller than the observation cadence to $\approx$ half the observation window. For the jitter models, we fit these with uniform priors $\pi(J_{r/c})\sim U(0,1)$, the widest range possible.

\paragraph{Numerical Scheme}
All of our lag fitting and modelling is performed using \litmus's nested sampling \citep{Skilling_2006_NS} fitting method, which makes use of the \jax-based nested sampling routine \jaxns \citep{Albert_2020_JAXNS} to provide posterior samples and evidence integrals.\footnote{Note that we use Nested sampling instead of the the Laplace Quadrature method presented in \citet{McDougall_2025_LITMUS}. Nested sampling, though computationally slower than quadrature, does not assume Gaussanity along any axis, and so we use it here to avoid any question of the numerical accuracy of our results.}

To ensure convergence we use a number of live samples $N_\mathrm{live} = 50\times\mathrm{Max \; Num \;Modes} \times (\mathrm{Max \; Num \; Parameters}+1)$, where $\mathrm{Max \; Num \;Modes}$ is the maximum number of aliasing modes (for our maximum lag of $1500 \dayu$ this is $4$), and $\mathrm{Max \; Num \; Parameters}$ is the number of parameters in the coupled model ($6$ for the Simple DRW, $7$ for the tophat model and $8$ for the jitter model). We run the nested sampling routine to terminate at an integral uncertainty of $\Delta Z/Z \le 10^{-3}$, and for all runs we output typical diagnostic plots and visually inspect these to ensure convergence.

This fitting is performed for all three light-curve models (DRW, Tophat, \& Jitter) and all four model hypotheses (Coupled, Uncoupled, GP-Noise, \& Noise-Noise) for the entire \ozdes sample, for both the response light curves with spectra co-added by run and by date. When a run is complete, its model evidence and uncertainty are saved along with $10,000$ posterior samples from the re-weighted nested samples.

\paragraph{Posterior Analysis}
From each fit to each model, we draw $10,000$ posterior samples from the nested sampling run through the typical way of re-weighting the nested samples by their contribution to the evidence estimate. Where we report a lag or any other model parameter, we do so using the posterior median value with negative and positive uncertainties found from the \pone\ and \ptwo\ percentiles.\footnote{Because this is a simple re-parameterization of the tophat-smoothing model, rather than re-running the entire fitting pipeline for the centroid model, we create a new chain using $\Delta t \rightarrow \Delta t + \frac{1}{2}\exp(\ln(w))$ to sidestep wasteful numerical cost.} We treat the numerical uncertainty in the evidence integrals as negligible, taking the simple ratio of the average evidences acquired from the nested sampling runs when calculating the Bayes factors for the uniform prior models. 

To estimate marginal likelihood of the lags (Equation~\ref{eq: marginal_likelihood}), we smooth the posterior samples with a Gaussian Kernel Density Estimator (KDE) to get an estimate of $P(\Delta t\vert d)$ with the uniform lag prior, then re-weight the normalisation with the model evidence: $\mathcal{L}(D\vert\ \Delta t) = \mathrm{KDE}(\Delta t)/\pi_\text{Uniform}({\Delta t})\times \BF{Lag}$. We use this to find the lag of peak likelihood, $\hat{\Delta t}$. For comparison with quality cut methods used in other RM works \citet[e.g.]{OzDES-Penton_2025, OzDES-McDougall_2025}, we also estimate the full-width half maximum (FWHM) of this peak and the amount of model evidence contained within this region as estimated from the number of samples that fall within the FWHM. Though we do not examine these metrics here, they are available for all sources and models in electronic tables published online with this paper (see Section~\ref{sec: data_availability} for details).

We do not run a new fit on every source for the $R-L$ informed Bayes Factor, and instead estimate the evidence and lag constraints via importance sampling. The original lag posterior samples from the uniform prior fit are resampled with draws weighted by their importance ratio (the ratio of the old and new posterior densities) $w_i^n=\frac{\pi_{R-L}^n(\Delta t^i_n)}{\pi_\mathrm{ Uniform}(\Delta t^i_n)}$, where $\Delta t^i_n$ is posterior sample $i$ from source $n$ and $\pi_{R-L}^n(\Delta t)$ is the $R-L$ informed prior (i.e. Equation~\ref{eq: R-L_prior}) for that same source. The Bayes factors for the $R_L$ priors are found from importance sampling, adjusting the original model evidence by the average importance ratio across the original posterior samples: $\ev_{R-L} = \ev_\mathrm{ Uniform} \times \mathrm{E}_\text{Post. Samp., n}\left[\frac{\pi_{R-L, n}(\Delta t)}{\pi_\mathrm{ Uniform}(\Delta t)}\right]$.

\section{Estimating the False Positive Rate with a Negative Lag Test}
\label{sec: selection_criteria}
\begin{figure}
    \centering
    \includegraphics[width=1.00\linewidth]{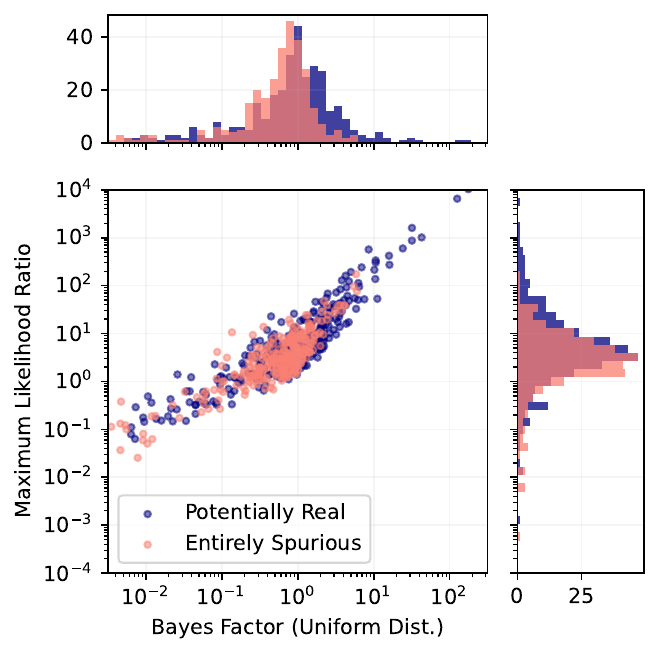}
    \caption{A log-log scatter plot of our two lag significance criteria, $\BF{Lag}$ and $\Lrat{lag}$ (here using the jitter light curve model), for the entire \ozdes sample (co-added by date). The dark blue dots / distributions are for the normal light curves, while the light orange dots / distributions are from the inverted light curves, for which any recovery is spurious. The distributions almost entirely overlap, but the strong recoveries (top right) are hard to produce by coincidence, as evidenced by the paucity of orange dots in this region.}
    \label{fig: noise_dist_example}
\end{figure}

In \citet{McDougall_2025_LITMUS} we demonstrated that measures like the Bayes factor let us estimate the false positive rate. To do so we choose a particular significance cutoff, then compare the lag recoveries to a distribution from a known-to-be-uncoupled null sample of light curves. Thus we essentially use the Bayes factor as a test statistic in a frequentist p-value sense. In that paper, the null sample was generated from mocks with a known population lag-distribution that matched the uniform prior. We avoid this mock-based approach in this work as it relies on our mocks accurately reflecting the reality of our data, and, as shown in Sections~\ref{sec: RM_limitations}~and~\ref{sec: reverb_frac}, there are open questions about our light curve models and the reverberating AGN population. We instead adopt a variation on the \qm{negative lag test} framework used by \citet{SDSS-Shen_2023}, generating our null-sample by reversing both the continuum and response light curves of real data and re-fitting, in effect allowing \litmus to search for a negative lag. Any lag recovered from this inverted sample must be a physically spurious false positive, as effect cannot precede cause.\footnote{This would not be true if the light curves demonstrate any periodicity, but we find in Section~\ref{sec: reverb_frac} that this is unlikely to be the case.} As the confounding factors of aliasing and coincidence should appear equally in positive and negative lags, we can use the number of lags that pass some threshold of significance from the inverted sample as an estimate of the number of similarly spurious recoveries in the true sample. This approach is confounded by the fact that our potential real signal distribution and entirely spurious noise distributions overlap for the vast majority of sources, with only their tails different (for an illustration, see Figure~\ref{fig: noise_dist_example}, along with a demonstration of how $\BF{Lag}$ and $\Lrat{Lag}$ trace one another).

Our FPR estimation method is as follows. First, select some light curve model and whether the response light curves are from the spectra co-added by date or by run. For these choices, we then fit the lag and model evidences (coupled, uncoupled etc.) for both the forward sample (which potentially contain lags) and reversed null sample (which contains entirely spurious recoveries). These evidences then allow us to calculate our significance measures (Bayes factors, max marginal likelihood) for both samples. 

We then discard any sources in which the response is extremely unstructured ($\BF{Struc,Resp}<0.5$) or in which the lag is consistent with zero at $2\sigma$ (these sources being ambiguous as to whether their lags are positive or negative). For the entire sample, (either for all sources or for all sources of a particular emission line type) we then compare the distributions of these significance measures for the forward and reverse curves, with the FPR at some threshold being estimated by the number of definitely spurious recoveries against the number of potentially real ones. In short, if a particular threshold, for example $\BF{Lag}>1.0$, permits $1$ recovery from the null sample and $5$ from the real sample, then the FPR is $\approx1/5=20\%$.

\begin{figure*}
    \centering
    \includegraphics[width=1.0\linewidth]{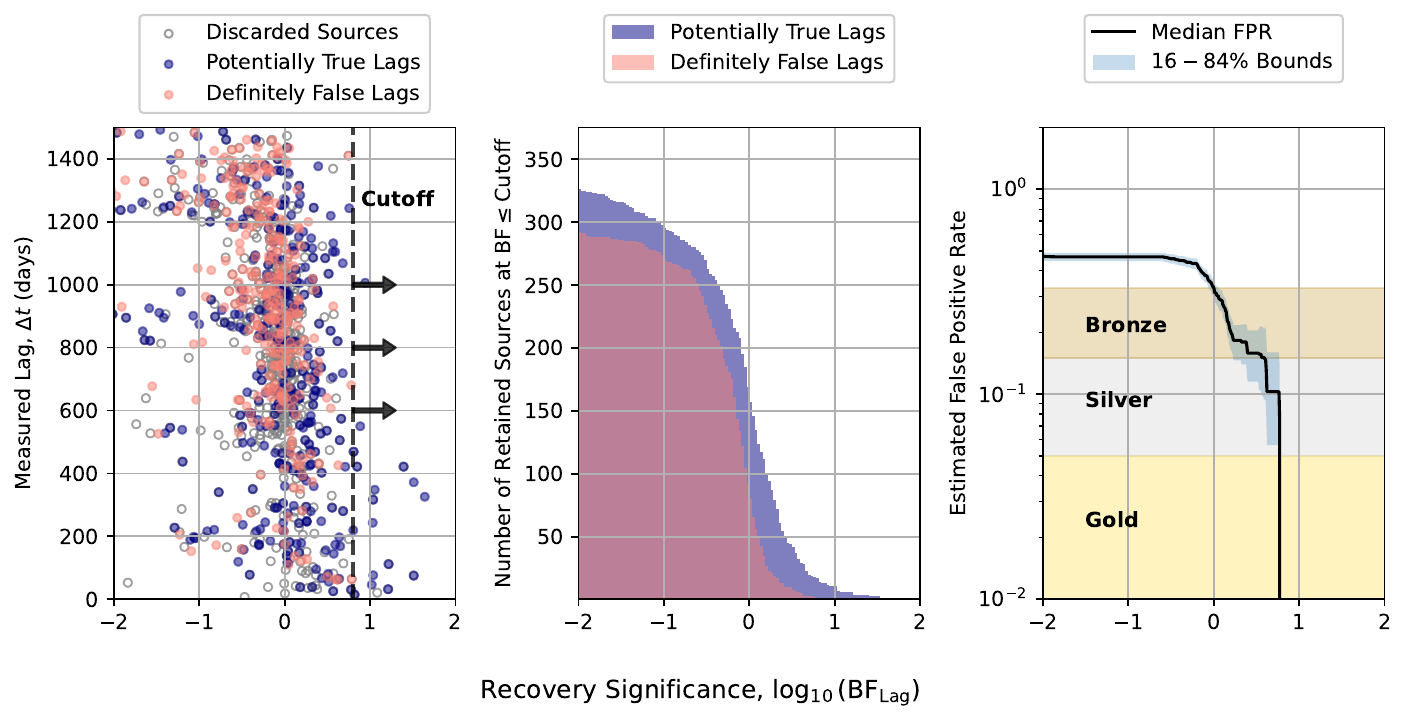}
    \caption{A demonstration of how our FPR estimation procedure works. We first measure lags and significance measures from Section~\ref{sec: signif} (here, the Bayes factor) for both forward and reversed light curves (left panel). This panel shows how low quality recoveries (low Bayes factor) tend to cluster around the aliasing lags ($180 \dayu$, $540 \dayu$ etc) but that this becomes less less true as we go to higher significance. The hollow circles show recoveries that are discarded due to having poor response light curve structure or a lag consistent with zero. We then apply a significance cut (dotted line, left panel) and see how many sources from the forward and reversed samples are above this significance level as we make it increasingly stringent (middle panel). The ratio of these source counts are then the basis for the FPR estimate curve (right panel). Sources at or above the cutoff for $\le33\%$ FPR (by maximum likelihood FPR) are graded as \bronze, $\le15\%$ as \silver and $\le5\%$ as \gold (shaded regions, right panel). This example is for the simple DRW model, spectra co-added by run, and fitting an FPR curve to the entire sample instead of individually per emission line.}
    \label{fig: FPR_plots}
\end{figure*}

We can put uncertainty bounds on this FPR by modelling it as a beta distribution, in which the simple ratio gives the peak likelihood FPR (see Figure~\ref{fig: FPR_plots} for an example illustration of the method). Because the distributions become sparse in the tails where the FPR is lowest, these estimates becomes noisy for our most significant recoveries. To smooth this slightly, we assume the FPR is strictly decreasing for increasing quality of recovery ($\BF{Lag}$, $\Lrat{Lag}$ etc.) and so take the lowest value below some cutoff, e.g. if some cutoff gives an $\text{FPR} < 30\%$ than we enforced an equal or lower FPR for all cutoffs beyond that. For comparison with prior works \citep[e.g.]{OzDES-Penton_2025}, we classify lag recoveries into discrete grades based on their estimated peak-likelihood FPR. A \qm{\bronze} recovery has $\text{FPR}\le33\%$ (twice as likely to be real as spurious), a \qm{\silver} recovery has $\text{FPR}\le15\%$ ($\approx5.5$ times more likely to be real) and a \qm{\gold} recovery has  $\text{FPR}\le5\%$ (less than a $1/20$ chance of being a false positive). 

When choosing which model \& data set to use in our final lag recovery results, we have four decisions: whether to use the response light curves with spectra co-added by date or by run (except for the \mgii sources which area all co-added by run), which light curve model from Section~\ref{sec: stats_modellling} to use, which significance measure from Section~\ref{sec: selection_criteria} to use, and whether to estimate the FPR curve from the entire sample or subdividing by line type. We have a full summary of the number of \gold, \silver and \bronze recoveries for all combinations of these choices in Appendix~\ref{app: recovery_summary}, but we ultimately find that the best method for recovering lags (highest number of \silver recoveries) is to use:
\begin{enumerate}[(i)]
    \item The spectra co-added by run (where available),
    \item Using the simple DRW model, i.e. no tophat smoothing or jitter, 
    \item $\BF{Lag}$, the Bayes factor from a uniform lag prior rather than $\Lrat{Lag}$, and
    \item Subdividing by emission line type when estimating the FPR.
\end{enumerate}
For a more granular analysis of the impact of these choices, see Appendix~\ref{app: RM_reliability}, and in particular Figure~\ref{fig: gradechange_model} and Figure~\ref{fig: gradechange_many} in Appendix~\ref{app: frac_robustness}.

Strictly speaking, the $\BF{Lag,R-L}$ measure (again for co-adding by run) yields the most recoveries for \mgii and \civ. For the concerns of confirmation bias outlined at the end of Section~\ref{sec: signif}, we stress that these $\BF{Lag,R-L}$ results should not be used for any further $R-L$ fitting. The consistency does nevertheless indicate that the $R-L$ relations for \mgii and \civ in \citet{OzDES-McDougall_2025} are reasonably well matched to the true population distributions for these lines, or are at least a closer match than the uniform prior. For non $R-L$ works, lags found in this way may still be useful, though with the caveat that they will inherit bias from any misspecification in the $R-L$ relations we use here. This improvement in lag recovery also indicates that making use of physical priors in lag fitting is a potentially rich topic of future investigation.

\section{Final Reported Lags \& Comparison with Previous \ozdes Results}
\label{sec: lag_results}

Using the preferred criteria outlined in Section~\ref{sec: selection_criteria}, we publish a total of \newlags new AGN lags, comprised of \newhbeta \hbeta lags, \newmgii \mgii lags and \newciv \civ lags. Of these, $15$ are for sources that \ozdes have previously published lags, and of these $2$ have had a significant change from these previous lags. The two significant changes are for \ozdes source $2970730486$'s \civ recovery, which remains at the \bronze confidence level but finds a lag that was longer than \citet{OzDES-Penton_2025}'s upper prior limit, and \ozdes source $2939653045$'s \hbeta recovery which changes by less than $2\sigma$ from a lag of $21.3^{+19.9}_{-14.8}\dayu$ to a \silver confidence $59.7^{+10.7}_{-7.0} \dayu$. The second of these changes can be attributed to the removal of outliers in the continuum light curve, and not to any change in the numerical or statistical modelling compared to past works.

\begin{figure}
    \centering
    \includegraphics[width=1.0\linewidth]{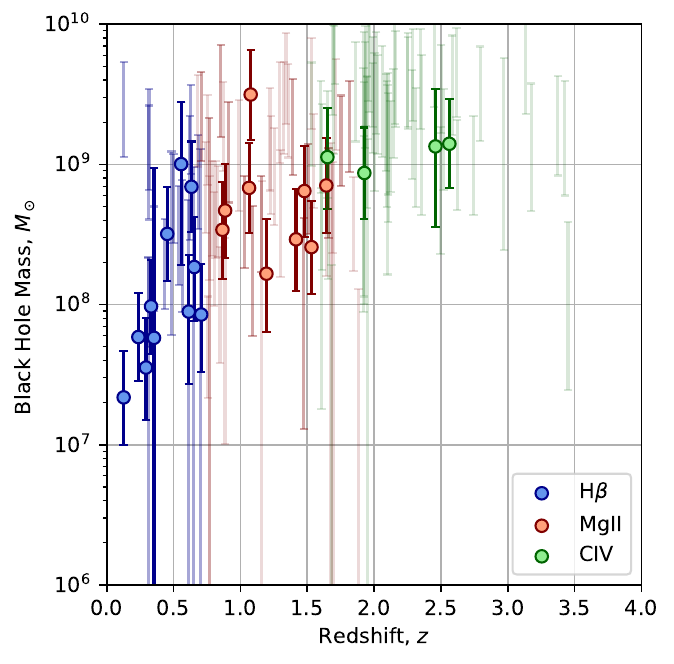}
    \caption{Masses for our final lag sample (presented in Table~\ref{tab: lags_all}) per Equation~\ref{eq: RM_mass} when using the virial factor of $\log_{10}(f)=0.62 \pm 0.07 \pm 0.31$ from \citet{SDSS-Shen_2023} (the first and second uncertainties here being statistical error and population scatter) and the line widths presented in \citet{McDougall_2025_LITMUS}.}
    \label{fig: mass_vs_redshift}
\end{figure}

The results we present here are for the simple DRW model (i.e.\ no tophat smoothing, no jitter). We list the full set of lag measurements for this set (including those below \bronze level and those discarded due to unstructured response curves or lags consistent with zero) in Appendix~\ref{app: lags_all}, and show our estimates on mass for these sources in Figure~\ref{fig: mass_vs_redshift}.

In \citet{OzDES-McDougall_2025}, we demonstrate how population-level trends are highly sensitive to the multi-layered processes of statistical models, numerical schemes and post-fit quality cuts. In this paper we present what we hope is a foundation of a principled start-to-end approach to lag recovery in the low SNR regime that is typical of the industrial scale efforts of \ozdes and \sdss. It is useful then to compare to previous efforts in \ozdes, and to identify what differences arise and why.

\begin{figure*}
    \centering
    \includegraphics[width=1.0\linewidth]{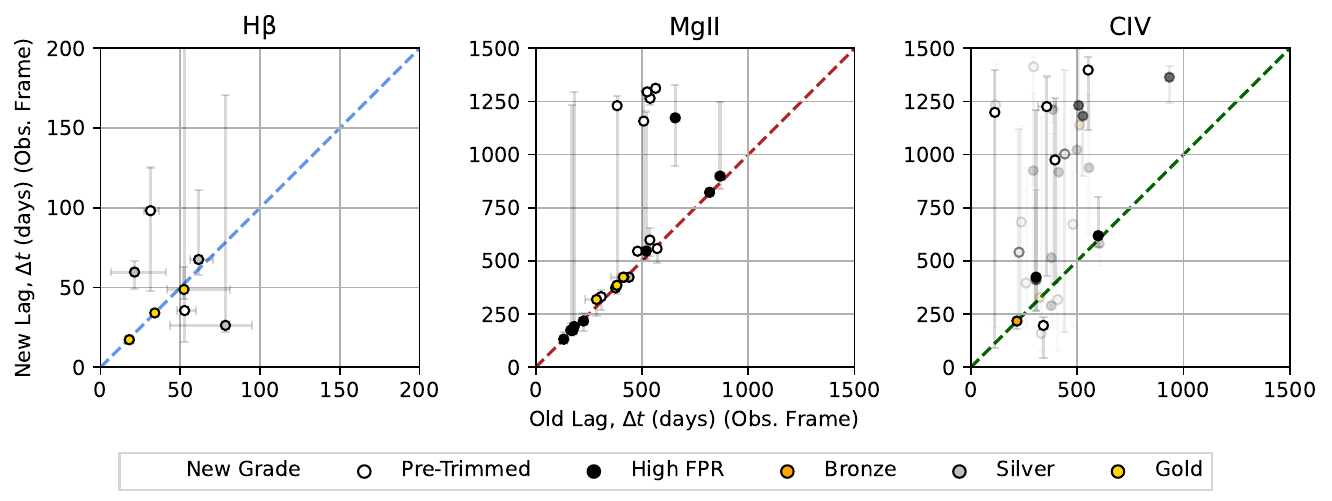}
    \caption{A plot of all $64$ \ozdes lags measured in past works compared to the lags for those sources measured in this work (using the simple DRW model and spectra co-added by run). The panels, from left to right, are for the $8$ \hbeta lags of \citet{OzDES-Malik_2023}, the $29$ \mgii lags from \citet{OzDES-Yu_2023}, and the $31$ \civ lags from \citet{OzDES-Penton_2025} and \citet{OzDES-Hoormann_2019}, with the $1:1$ dotted lines showing the case of the new / old lag measurements being equal. The point colours (\bronze, \silver, \gold) denote the grading of the new recoveries using the Bayes Factor FPR, and the opacity of the \civ lags is lowered for the measurements with old grades of \silver (to $50\%$) and \bronze (to $15\%$). Black points are those at a $>33\%$ estimated FPR, and white points are those trimmed before the negative lag test due to having poor response SNR or a lag consistent with $\Delta t = 0$. Note that the left hand panel has a different axis extent to the others to better show the span of \hbeta lags. 
    }
    \label{fig: OzDES_Vs_New_goldsilverbronze}
\end{figure*}

Table~\ref{tab: OzDES_vs_LITMUS} in Appendix~\ref{app: OzDES_vs_Litmus} gives a full comparative list the $64$ \ozdes lags presented in prior works (\citet{OzDES-Malik_2023} for \hbeta, \citet{OzDES-Yu_2023} for \mgii, and a combination of \citet{OzDES-Penton_2025} and \citet{OzDES-Hoormann_2019} \civ, as summarised in the final \ozdes data release by \citet{OzDES-McDougall_2025}) and compares them to our new measured lags and false positive rates using \litmus. Figure~\ref{fig: OzDES_Vs_New_goldsilverbronze} demonstrates the changes in the sample visually. The \hbeta sample is least affected in terms of reliability, with most lags being retained at the \gold or \silver level. The \mgii sample broadly recovers the same lags except for those in which the newly recovered lag is higher due to the wider lag search range. Most of the previous \mgii sample is discarded due to a high FPR / low significance or the measured lag being consistent with zero. We note that our uncertainties are markedly higher than those from \javelin, in keeping with past examinations of its limitations \citep{Gaskell_2024_javelinerrors}. The \civ sample sees the most marked change, with most of the sample being completely discarded due to unreliability, and most lags significantly changing after correcting the numerical issues of \javelin.

\begin{figure*}
    \centering
    \includegraphics[width=0.9\linewidth]{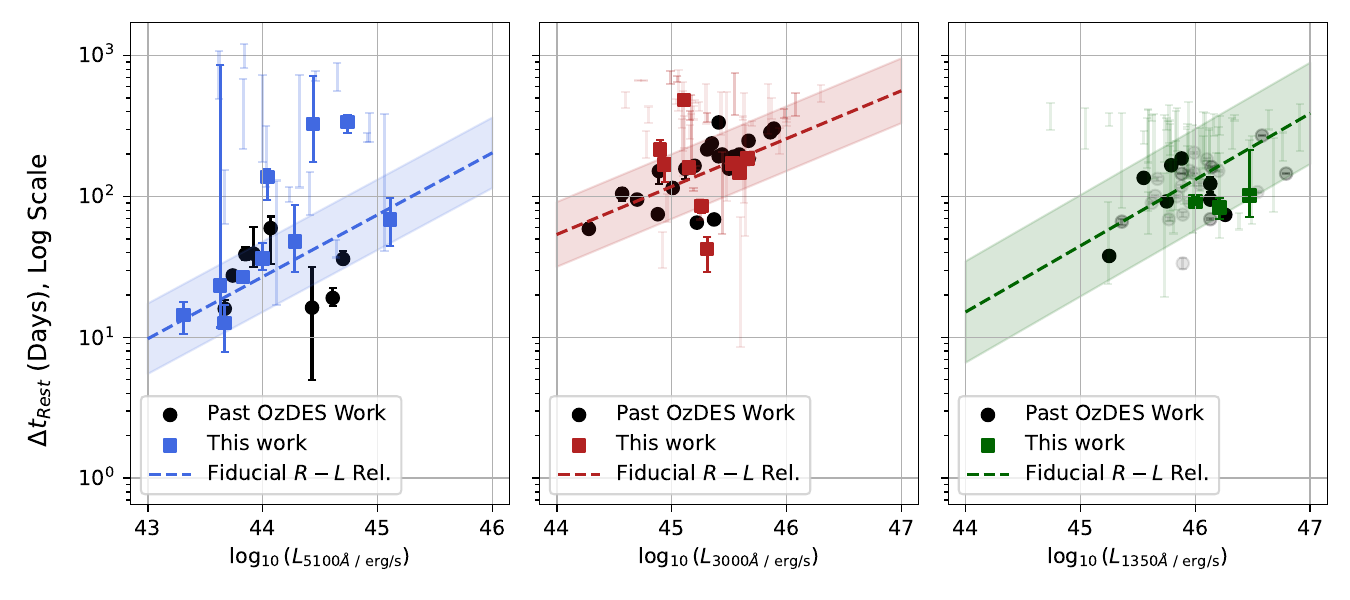}
    \caption{A comparison of the existing \ozdes results \citep{OzDES-Malik_2023, OzDES-Yu_2023, OzDES-Hoormann_2019, OzDES-Penton_2025} compared to our new recoveries illustrated in the $R-L$ plane for \hbeta (left), \mgii (middle) and \civ (right). The opacity of the points denotes the recovery grade, with no marker and lower opacity being \silver and \bronze recoveries, while a points with a square marker and full opacity are \gold standard. Past \ozdes works presented only their \gold standard \hbeta and \mgii recoveries. Underlaid are the fiducial $R-L$ relations we presented in \citet{McDougall_2025_LITMUS} and their $1\sigma$ uncertainty bounds, not accounting for statistical uncertainty in slope or offset. The populations generally agree with or sit higher than past recoveries, a combination of our method discarding many lags close to zero while also being able to recover longer lags.}
    \label{fig: new_vs_ozdes}
\end{figure*}

Figure~\ref{fig: new_vs_ozdes} shows the rest-frame lags of previous \javelin-derived OzDES lags and the new \litmus-derived OzDES lags in the radius-luminosity plane to gauge the impact of the updated lags on the $R-L$ relation. The \mgii sources do not demonstrate as clear a slope as the results of \citet{OzDES-Yu_2023}, though this may be due to the narrow luminosity range of our new recoveries or shorter lags being trimmed as being consistent with zero lag. Many of our suspect recoveries (i.e. \bronze grade) run up against the upper limit of the prior range. Longer lags are associated with poorer signal to noise due to less overlap between the light curves, and so are more vulnerable to false positives. 

\section{Placing Constraints on the Number of Reverberating AGN}
\label{sec: reverb_frac}

The low number of recoveries in Sections~\ref{sec: selection_criteria}~and~\ref{sec: lag_results} motivates us to examine why so few sources present clear lags. In this section we examine a question that has gone under-examined in past RM works: how many of our AGN sources actually have a lag for us to recover? This is of particular interest in the \mgii sample, which has the lowest fractional recovery rate of the three lines. 

In terms of our negative lag FPR test, a fraction of \qm{true negative} sources will cause the distribution of significance measures to overlap between our real sample and our time-reversed null sample. This overlap has the same effect as low signal to noise, which causes both samples to give only ambiguous results (i.e. dragging their distributions in Figure~\ref{fig: FPR_plots}'s left panel towards $\BF{Lag}=0$). Through the frame of false positive rates and lag significance, ambiguous positives and true negatives have a degenerate appearance. Owing to our Bayesian model comparison framework, however, we are able to break this degeneracy and examine the overall population of AGN in a Bayesian way.

\litmus's provision of marginal likelihoods for the coupled (lag bearing) and uncoupled (lag free) models means that we can distinguish between sources in which we do not see a lag and sources in which we see that there is not a lag. This allows us to put meaningful constraints on what fraction of AGN, $f$ are reverberating in our sample. We can estimate this in a simple Bayesian way with a marginal likelihood:
\begin{equation}
    \mathcal{L}(D \vert f, \mathcal{M}) \propto \prod_i \left( f\times \frac{\BF{Lag}^i}{1+\BF{Lag}^i}  + (1-f)\times \frac{1}{1+\BF{Lag}^i} \right)
    .
    \label{eq: frac_likelihood}
\end{equation}

\begin{figure*}
    \centering
    \includegraphics[width=1.0\linewidth]{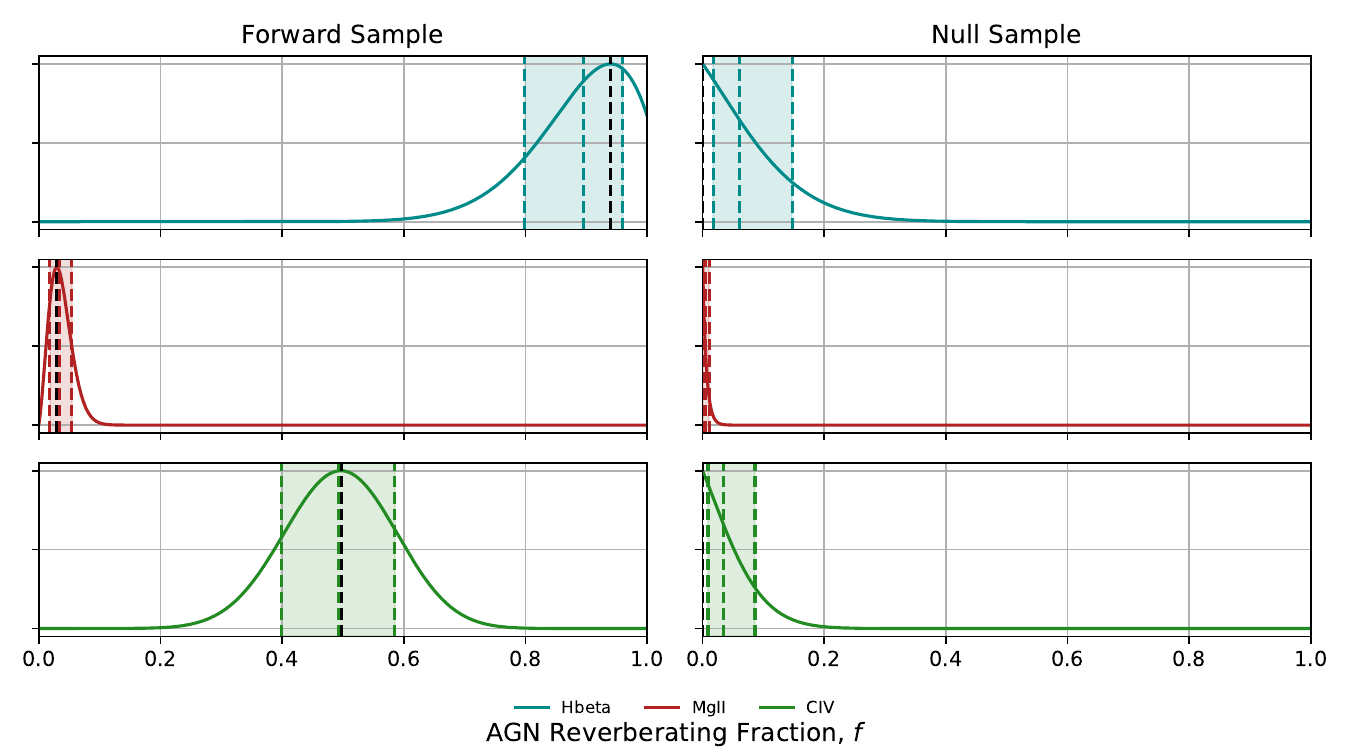}
    \caption{Constraints on the fraction of AGN with simple reverberations ($f$ in Equation~\ref{eq: frac_likelihood}) when using the simple DRW model and light curves from spectra co-added by run, divided by line or for the entire sample. The left panels show the results for the real light curves, i.e. those we believe may contain lags, while the right panel is the results after time-reversing those light curves to create the null sample. In the real sample, we identify some non-zero fraction of AGN that reverberate for all lines (consistent with $100\%$ for \hbeta), while also successfully identifying the negative sample as having no lags (consistent with $f=0$ for all lines). The coloured dashed lines show the median and \pone-\ptwo percentiles bounds, while the black dashed lines show the peak likelihood. The constraints here are when using no jitter or smoothing, and using the entire sample. See Appendix~\ref{app: frac_robustness} for how different assumptions impact these constraints.}
    \label{fig: reverb_fraction}
\end{figure*}

If we apply Equation~\ref{eq: frac_likelihood} to our uniform Bayes factors and the simple DRW model, we acquire the constraints on $f$ for the different lines shown in the left panels of Figure~\ref{fig: reverb_fraction}. It inspires confidence that the \hbeta sample is so readily coupled, with $f\approx1$ easily achievable. The \civ sources prefer a coupled fraction of the mid $10$'s of percent, though \civ is known to have complexities to its RM light curves \citep[e.g. outflows, see][]{Denney_2012} and so a lower fraction is not surprising (\civ but can reasonably support $f=1$ under more generous / vague conditions, see Appendix~\ref{app: frac_robustness} for details). Most interestingly, the \mgii sample has an extremely low reverberating fraction. While the \civ fraction may be pulled higher by a factor of a few with a more realistic prior (e.g. from an $R-L$ scaling relationship), such a consistently low value (a maximum of $f_{\mgiimath}\approx20\%$ is the highest achieved in any combination of models and datasets) may indicate that many, perhaps even most, \mgii sources do not manifest their reverberations in the same simple way as \hbeta. This analysis performed on the inverted (negative lag) sample successfully and strongly constrains $f\approx0$, i.e. it confirms that these inverted curves are in fact all uncoupled (right panels of Figure~\ref{fig: reverb_fraction}). This validates both our approach here and our negative lag tests in lag recovery.

Most lag recovery regimes \citep[e.g. \pyroa, \PyCCF, the \sdss analysis in]{SDSS-Shen_2023} all rely on the assumption that a simple lag does in fact exist to be recovered from every source. If this is not true, as appears to be the case for the bulk of the \mgii sample, then we should proceed with caution when analysing such sources. Our self-tuning and model free FPR estimate is inured against the complication of lag-free sources when reporting our lags, but this gives rise to an area of interest for future study. 

We note that, because the uniform prior we use to estimate $\BF{Lag}$ is arbitrarily broad, it will systematically under-estimate this Bayes Factor and so bias $f$ downwards.\footnote{We examine the impact of this, along with those of our choice of model and data selection, in Appendix~\ref{app: frac_robustness}.} In this way, our constraints here should be seen as lower bounds on $f$ for each line in a physical sense, but a measure of how many lags can be recovered in our call and response models in this sample. To properly estimate $f$, it should be calculated with Bayes factors acquired from a physically meaningful prior, i.e. from an appropriate $R-L$ relation. To avoid confirmation bias, it may be of interest to fit $f$ simultaneously with the $R-L$ parameters ($\alpha, \beta, \sigma$ in Equation~\ref{eq: R-L_powerlaw}) with the entire sample in a hierarchical way. The authors intend to investigate this in a future work.

\paragraph{What Does it Mean for a Source to be Uncoupled?}

\begin{figure*}
    \centering
    \includegraphics[scale=0.8, trim={0cm 0cm 0.1cm 0.2cm}, clip]{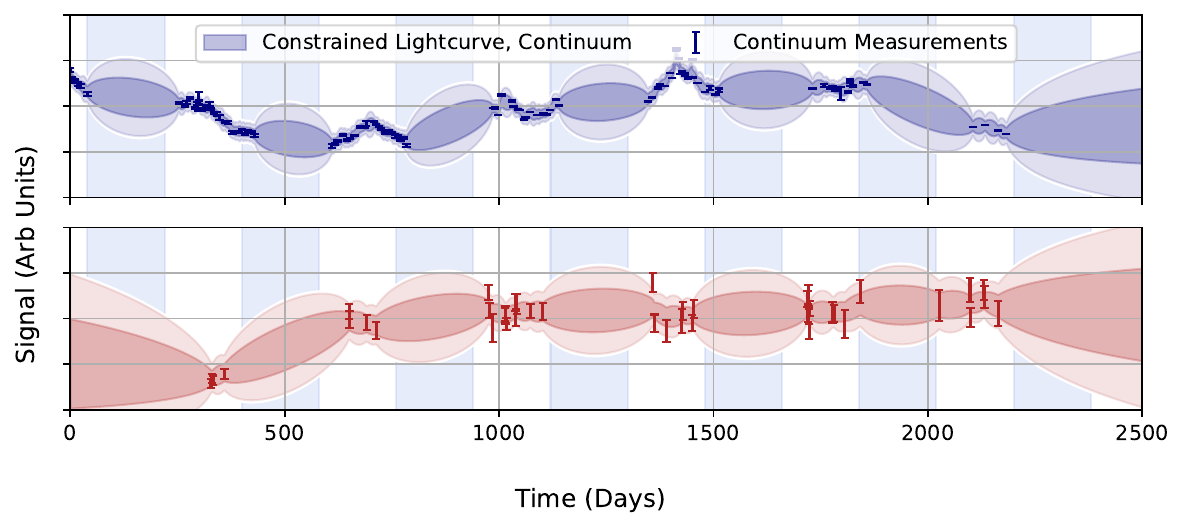}
    \caption{An example of a strongly uncoupled source, showing the posterior predictive light curves for \ozdes source $2970932715$'s \mgii light curve as fit with the uncoupled jitter model. Both the continuum and response demonstrate clear structure ($\BF{Struc,Cont}=2.9\times10^{57}$ and $\BF{Struc,Resp}=7.4\times10^{8}$), but have distinctly different patterns of variation such that no lag can convincingly bring them into alignment. The continuum has a down-up-down pattern, while the response rises over the first two years of observation and then remains relatively flat. This tension leads to the coupled model being strongly disfavoured ($\BF{Lag}=5.1\times10^{-4}$), identifying this as a clearly uncoupled source.}
    
    \label{fig: lc_constraint_jitternull}
\end{figure*}

The observation of $f\ne1$ arises from the fact that there are sources for which coupling is not just disfavoured, but strongly rejected by the Bayes factor, i.e. cases in which we have not only no lag but a clear and confirmed absence of any coupling between the continuum and response. In Figure~\ref{fig: lc_constraint_jitternull}, we an example of a source that demonstrates strong decoupling: both continuum and response can be constrained to well defined light curves, but there is no reasonable shifting and scaling that can bring them into agreement with one another to reflect  lag. There has been past investigation of temporary decoupling \citep[the \qm{BLR holiday}][]{Goad_2016_BLRHoliday, Dehghanian_2019} or transfer functions that go beyond the simple use of a simple smoothing timescale \citep[e.g. see the Gaussian transfer functions of \mica]{Li_2016_MICA}, and so there is good reason for us to interpret this decoupling as a real physical phenomenon in our sources. For \mgii in particular, there is good evidence that its reverberation is qualitatively dissimilar to that of \hbeta, particularly when AGN are in a state of high activity \citep{SDSS_Fries_2024}.

There are a few important caveats in our estimates of $f$ and our definition of a lag-free source. An important detail is that $f$ here represents not the fraction of AGN that do or do not experience reverberations, but specifically the number that reverberate in way described by our light curve models up to our maximum observer frame lag of $1500 \dayu$. Any deviation from \qm{simple} reverberation (i.e. a call and response model) will violate this assumption, even in systems that do physically reverberate. We note however that the \ozdes sample's sparse cadence in its response light curves renders it relatively insensitive to the particulars of the transfer function (see Section~\ref{sec: RM_limitations} and Appendix~\ref{app: RM_reliability}, as well as work by \citet{OzDES-Yu_2019}), and so it is unlikely that the low fraction is due to our simple smoothing model.

\section{Reliability of Light Curve Models \& Limitations of Reverberation Mapping}
\label{sec: RM_limitations}
Our use of multiple light curve models (varying the amount of jitter or the response smoothness see Section~\ref{sec: LC_modelling}) is mainly intended to find the best means of BLR-RM lag recovery, but also offers a means to probing how well the commonly adopted light curve models actually serve to describe the observed variability of AGN. In this section, we use the tophat smoothing and jitter models as a means of examining how well the commonly used \qm{shifted, scaled and smoothed DRW} model matches the data. In this section we examine the performance of the jitter and tophat-smoothing models to see how necessary they are, or are not, in the fitting BLR lags.

\subsection{Smoothness \& Structure of the Response Light Curves}
\label{sec: limitations_smoothing_and_jitter}

Though the jitter and tophat smoothing models from Section~\ref{sec: LC_modelling} are primarily used to refine our lag measurements in this work, they also offer a means by which we may interrogate the nature of the line response function. We give a precise count of how many \ozdes AGN exhibit strong, weak, or intermediate \qm{degrees of} smoothing or jitter (along with the regime by which we quantify this) in Appendix~\ref{app: RM_reliability}. Here, we summarise the most important findings.

The tophat lag model lets us estimate the smoothing timescale of the response function, or at least put limits on our ability to resolve this timescale. The cadence of our spectroscopic measurements limits how short of a timescale we can constrain, and the length of our survey sets an upper limit on this same value. Nevertheless, we can still use our (log-scale) constraints on the smoothing timescale to provide some meaningful insight into how prevalent the smoothing of the BLR response is compared to the continuum. In more than half of the sources we cannot meaningfully constrain the smoothing scale at all (see Table~\ref{tab: smoothing_constraint_table}), and in almost all of the remaining cases we are only able to say which end of our window it rests beyond, i.e. that a source has very little smoothing or that it has very much. Intermediate smoothing is most prevalent in the \hbeta sample while the \civ line has many sources that prefer or require strong smoothing, which may be informative of the \civ BLR transfer function or simply a product of these more distant sources being subject to stronger cosmological redshift that shifts their smoothing timescale upwards. The \mgii sample strongly prefers no smoothing, though this might be related to the many \mgii sources with white-noise-like unstructured light curves. Speaking broadly, the inclusion of the tophat smoothing does not significantly improve our lag recovery, and in fact makes it somewhat worse due to broadening of the parameter space (see Appendix~\ref{app: RM_reliability} and in particular that appendix's Figure~\ref{fig: gradechange_many} for details).

The strength of the response jitter allows us to examine how well structured the continuum response light curves are. Where many sources have well constrained strong jitter, i.e. $J_r\approx1$, this tells us that the light curve has little DRW-like structure. This is not a question of low SNR (which gives \qm{unconstrained} jitter), but that the BLR light curve varies in a white-noise like way beyond what can be explained by measurement uncertainty. In Table~\ref{tab: jitter_constraint_table}, we find that most \hbeta sources either have very strong structure or can at least be reasonably described by a pure DRW model. The \civ sample has a few (about one in ten) sources that are unstructured, but roughly a quarter have SNR too poor to make any statement. Giving rise to some concern is the fact that the \mgii sample has nearly half of its sources that are clearly unstructured or that have a strong white noise component. It is very likely that this is in some way related to the poor recovery rate for the \mgii sources; many \mgii light curves simply do not show good structure.

\subsection{Non-DRW Short-Timescale Variations in the Continuum Signal}
\label{sec: limitations_cont_jitter}

Examining the jitter component of the response light curves allows us to examine whether they have meaningful structure. The continuum light curves, by contrast, have clear and consistent structure over nearly all sources (all but five have $\BF{Struc,Const}\ge10^2$ across all models). Nevertheless, the strength of jitter here lets us probe the amount that our observations of the continuum variations differ from a pure DRW on short timescales. the white-noise component describes the amount of signal variation, in excess of the measurement uncertainty, that differs from the PSD of the DRW. Constraining the continuum jitter for all AGN, this comes out to a subdominant but significant component, with the median amount of jitter being $\approx12.2\%^{+6.7\%}_{-13.3\%}$. This noise is present in $755-813$ of the $826$ continuum light curves (i.e. $91.4-98.4 \%$ of the sample), with only $87$ ($10.5 \%$) being confidently jitter-noise-free (see Figure~\ref{fig: jitterfrac} and Table~\ref{tab: jitter_constraint_table} in Appendix~\ref{app: RM_reliability}).

We suggest four potential sources for the jitter component: 
\begin{itemize}
    \item Outlier epochs not cleaned by our trimming in Appendix~\ref{app: lightcurve_cleaning} or some similar confounding signal with a small amplitude,
    \item Miscalibration of the continuum error bars (though this is unlikely given the rigour of the \des pipeline),
    \item A short-timescale noise process in the AGN light curve super-imposed onto our observations of the prevailing DRW signal of the disk, or
    \item A deviation of the true PSD of the accretion disk's prevailing variability from that of the pure DRW.
\end{itemize}
We consider the last two these to be the most likely, as there is already good evidence of a break from the DRW in AGN light curves over small timescales \citep{Kasliwal_2015_nonDRW, Smith_2018_nonDRW, Stone_2020_nonDRW}. We note that miscalibration on the order of $10\%$ of the entire signal variation would be difficult to explain, as the individual uncertainties can reach percent-level error and the contribution seems to be quite consistent over the sample. We are unable to confirm a variation in the strength or timescale of the jitter-noise with source redshift, which would confirm a source-based astrophysical origin. 

If this noise reflects transient outlier epochs in the signal, either from a data reduction artefact or a physical origin like microlensing, we would expect it to be well described by modelling the light curve as a mixture model of a GP and some stationary noise process. If it is a secondary but stationary signal super-imposed onto variations in the disk, we might model this by extending our noise model to be a GP with a flexible timescale instead of the conservative white-noise of our jitter model in this work. If these short time-scale variations are a true feature of the accretion disk variability, we should expect to see them reflected in the BLR response, i.e. whether they are convolved with the transfer function / exhibit a lag in high SNR sources. As a simple first step, future work may investigate whether the coupled light curves are better explained by jitter being applied only in the continuum ($J_r=0$) or by an equal amount of white noise in the response signal ($J_r=J_c$). If these rapid variations are a true feature of the driving variability of the AGN. We suggest suggest modelling the continuum with more expressive GP kernel \citep[e.g. higher order GP's as done in][]{Kelly_2014_nonDRW_DHO, Kasliwal_2017_nonDRW_DHO, Moreno_2019_nonDRW_DHO, Yu_2025_nonDRW_DHO, Kroupa_2026a_stationarity}, or a sum of simple kernels at different timescale.

\begin{figure}
    \centering
    \includegraphics[width=1.0\linewidth]{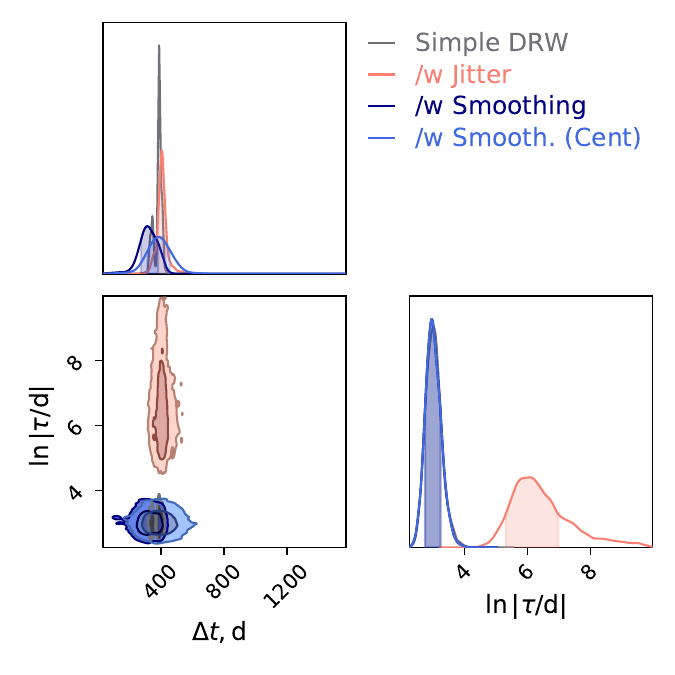}
    \caption{An example of the parameter constraints, specifically for the lag $\Delta t$ and DRW log-timescale $\ln\vert\tau\vert$, for the three light curve models (and the centroid variation of the tophat smoothing model) for source $2943200932$-\mgii (a \gold standard recovery in all models).  We see that the recovered lag is only weakly sensitive to the model choice, but the DRW timescale $\tau$ is strongly affected by adding jitter.}
    \label{fig: contour_example}
\end{figure}

\paragraph{Impact on Lag Recovery}
The presence of jitter-noise in the signal has two impacts: obscuring short lags and distorting our estimates of the timescale of variability. The ambiguity suggested by the jitter model dominates at short timescales, making it difficult to constrain near-zero lags. We can estimate the scale at which this noise begins to have an impact from the structure function \citep{Hughes_1992_strucfunc, Kozlowski_2016, JiaJia_2024_strucfunc}, a counter-part to the covariance function that measures the average difference between two points on the light curve (i.e. how much structure exists at that time-scale). For a DRW, the structure function is:
\begin{equation}
    \mathrm{SF}_c(\Delta t) = \sqrt{1-\phi_{cc}(\Delta t)}=\sqrt{1-\exp\left(-\abs{ \frac{\Delta t}{\tau}}\right)}
    \label{eq: struc_func}
\end{equation}
For a jitter component $J$, the two are of equal power at a timescale: 
\begin{equation}
    \Delta t = - \tau \ln(1-J^2)    
    .
    \label{eq: jitter_timescale}
\end{equation}
This provides a rough timescale at which the white noise uncertainty begins to dominate and obscure lag recoveries. For the continuum curves in our sample, the median rest-frame variability timescale is $\tau \approx 530\dayu$. At a median white noise strength of $J_c=12.2\%$, Equation~\ref{eq: jitter_timescale} tells us that it becomes difficult to observe lags at or below a characteristic scale of $\approx 7.7 \dayu$. This will move higher / lower for sources with more or less noise and longer / shorter variability timescales, but we note that all of our reported lags sit comfortably above this limit. The similarity of this scale to the continuum observational cadence is interesting and may lead one to be suspicious that it is our model is simply failing to exclude jitter at timescales below this value. This is however not supported by our posterior fits, in which most sources strongly exclude the possibility of $J=0$ at several sigma, which would not be the case in a noise-floor scenario.

\begin{figure}
    \centering
    \includegraphics[width=0.9\linewidth]{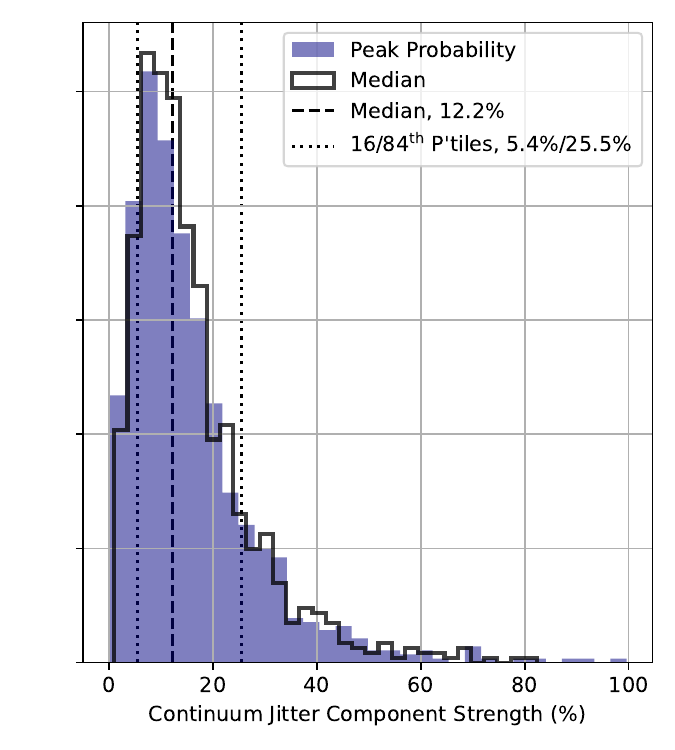}
    \caption{Histogram showing the number of continuum light curves at varying strengths of their non-DRW white noise component, as measured from either the most probable (navy, shaded) or posterior median (black, unshaded) value, with vertical dotted and dashed lines showing the sample median and \pone / \ptwo percentiles.} 
    \label{fig: jitterfrac}
\end{figure}

\begin{figure*}
    \centering
    \includegraphics[scale=0.8, trim={0cm 2.0cm 0cm 0cm}, clip]{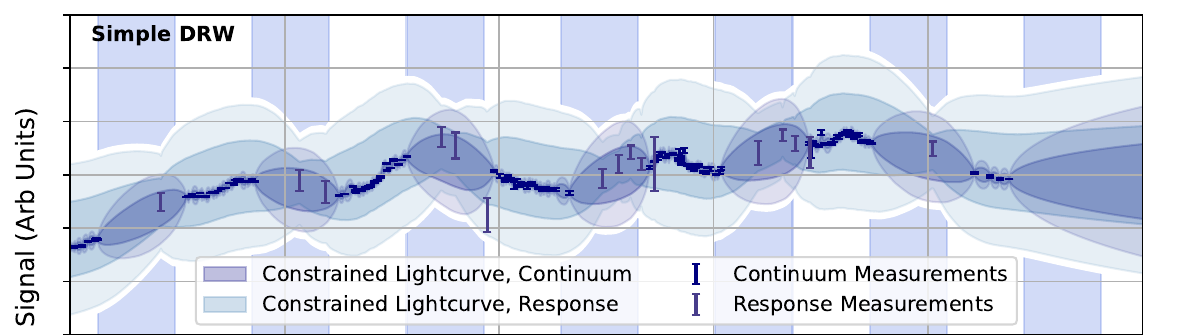}\\
    \includegraphics[scale=0.8, trim={0cm 0cm 0.1cm 0.2cm}, clip]{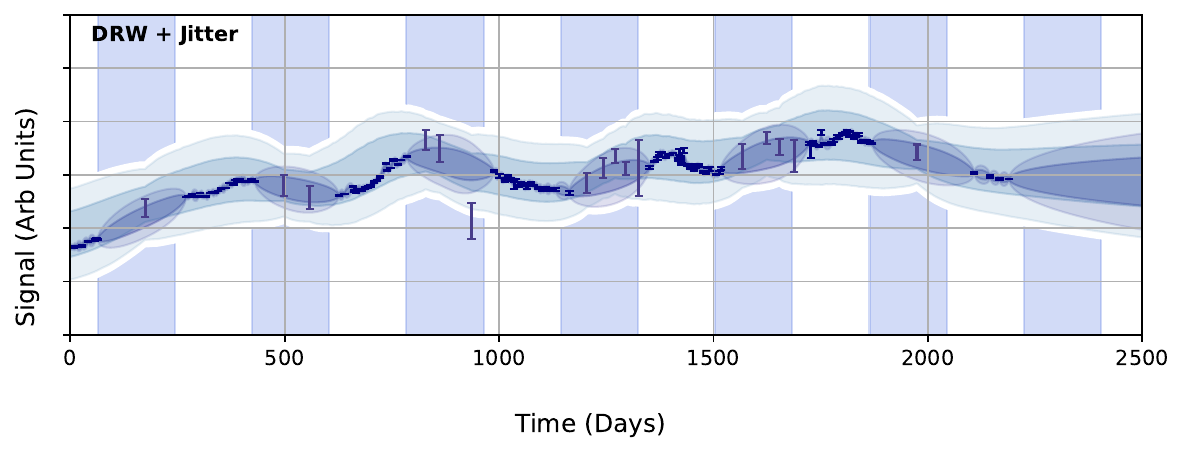}
    \caption{A demonstration of how the addition of jitter / non-DRW white noise to the light curve model changes the lag constraints for a single source. Both panels show the coupled light curve fits for \ozdes source $2925552152$'s \hbeta response, with the top panel showing constraints with the simple DRW model and the bottom for the jitter model. Both models recover similar lags ($\Delta t = {114}_{-40}^{+48}$) for the simple model and $\Delta t = {121}_{-44}^{+55}$ for the jitter model. Without jitter, the variability timescale is forced to be short to fit for rapid variations in the continuum. With jitter, these rapid variations are absorbed by the white noise component and the fitter is able to track the smooth long-timescale variations in both light curves. The vagueness of the jitter model means the Bayes factor decreases from $\BF{Lag}=16.1$ to $7.03$ when adding jitter, but the self-tuning nature of our FPR estimation method means that the quality of the recovery changes from being discarded to a \gold standard.}
    \label{fig: lc_constraint_jitternojitter}
\end{figure*}

One area the short-timescale fluctuations captured by the continuum jitter do have a significant impact is in our estimates of the variability timescale of our sources. When we do not model rapid fluctuations explicitly (i.e. when we use a simple DRW model) the timescale is forced to short values to capture these rapid changes (e.g. the timescale constraints in Figure~\ref{fig: contour_example}). Of the $352$ sources with well constrained timescales $237$ of them see a significant increase in timescale when modelling the continuum with the jitter light curve model.\footnote{\qm{Well constrained} here meaning the \pone and \ptwo posterior percentiles on $\tau$ lying between $30$ and $1000$ days, and a \qm{significant increase} meaning $\tau_\mathrm{DRW+jitter}/\tau_\mathrm{DRW}>1$ for $>84\%$ of the posterior samples.} This has a secondary effect in lag recovery: a shorter timescale means we cannot interpolate as smoothly between our observational seasons (e.g. the light curves in Figure~\ref{fig: lc_constraint_jitternojitter}), which may make it hard to identify or reject lags that fall within these gaps. For this reason, we suggest that investigating the modelling of this noise (e.g. by the methods we describe above) may be an important next step in combating aliasing and the seasonal windowing function.

\subsection{Impact of Light Curve Model \& Other Choices on Lag Recovery Rate}
\label{sec: limitations_model}

In Section~\ref{sec: signif} we chose our fiducial choices of data, significance measure and light curve model to give the highest number of decent (i.e. \silver) recoveries, which was accomplished with the simple DRW, co-added by run, using $\BF{Lag}$ as a significance measure. It is worth examining the performance of difference choices, both as a check on the robustness of our pipeline and to investigate the nature of our sample. We provide additional detail in Appendix~\ref{app: RM_reliability}, but here focus on the impact of different light curve models.

\begin{figure}
    \centering
    \includegraphics[width=0.45\linewidth]{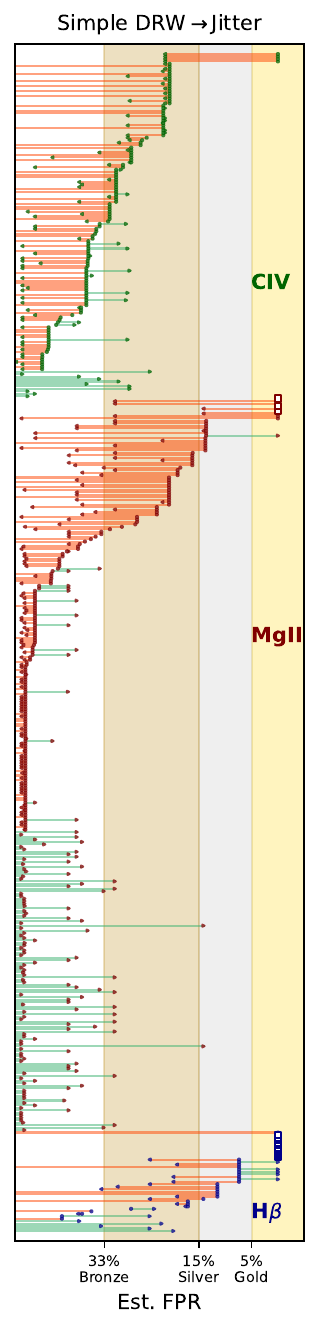}
    \includegraphics[width=0.45\linewidth]{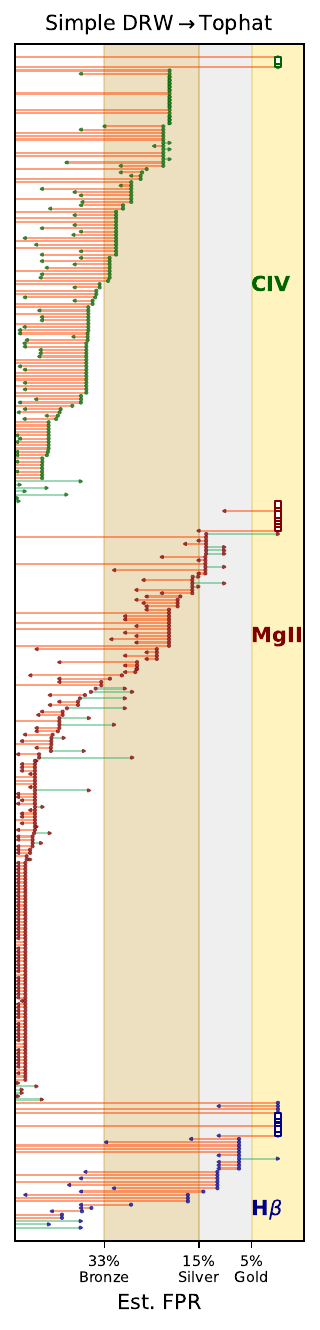}\\
    \includegraphics[width=0.8\linewidth]{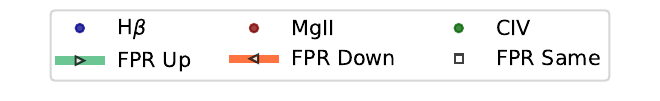}
    \caption{A demonstration of how our false positive rate and lag recovery grades change if we re-run our pipeline after switching from the simple DRW model to include jitter (left panel) or tophat smoothing in the response (right panel). Sources are organised by line type and then sorted by the reliability of their fiducial recovery.}
    \label{fig: gradechange_model}
\end{figure}

Firstly, we note that jitter model yields markedly more \hbeta \gold recoveries than the simple DRW. This is a result of a few moderate recoveries improving in resolution (see Figure~\ref{fig: gradechange_model}), but comes at the cost of many otherwise significant results vanishing. By contrast, the \mgii sample sees no improvement, counter to the findings of Section~\ref{sec: limitations_smoothing_and_jitter} in which the \mgii response curves are found to have many sources with strong non-DRW noise compared to \hbeta or \civ. We note that allowing for white-noise in the \mgii sample markedly improves some of its weakest recoveries, but leads to significant losses elsewhere. 

\mgii has the most \gold recoveries from using $\Lrat{Lag}$ as a significance measure in concert with the tophat centroid model. Again this is interesting as in Section~\ref{sec: limitations_smoothing_and_jitter} in which \mgii was found to have many unstructured light curves but no real preference for smoothing. This suggests that the smoothing is prevalent in some subset of the \mgii sample with well structured lag-bearing light curves. Similarly, while the \civ sample seems to prefer strong smoothing in the response light curve, it sees no major improvement from any change in model choice, though this may be a product of its poor SNR. 

\section{Conclusion}
\label{sec: conclusion}
From the analysis in this work we have made a number of novel findings pertaining to the state of industrial scale reverberation mapping, significant both physically and methodologically. We present the framework of what we believe to be a principled pipeline for reliable lag constraints and false positive screening, and from it are able to state some conclusions about the limits of current RM techniques on \ozdes-like surveys, and in what areas we should focus our attention to improve these techniques. Our most interesting physical finding is that significantly fewer AGN present discernible lags in the \mgii sample than previously believed.

Regarding the current state of RM techniques, we can draw a few conclusions from our findings here. Firstly, we find that the sub-$10\%$ retention rate for traditional \qm{cut and constrain} RM arrived at in past \ozdes works are quite reasonable for modern industrial scale surveys. Secondly, the numerical issues discussed in \citet{McDougall_2025_LITMUS} are very much an effect of leading importance, particularly in the low SNR case of the \civ sample. Thirdly, the use of stringent significance criteria that are consistent with our lag recovery methods is not only an abstractly \qm{more principled} approach in contrast to the historical patchwork mix of fitting methods, but is actually a necessary step in screening for inherent true negatives of decoupled AGN, which we find occur at a much higher than anticipated rate in our data. Finally, we find that, absent a proper understanding of how many AGN reverberate in the simple \qm{call and response} manner and a reliable and physically principled lag prior, a negative lag test in the style of \sdss is a necessary component of lag recovery. Estimates of merit fail to properly account for the look-elsewhere effect or the incidence of true negatives.

We find that, for SNR as low / cadence as sparse as \ozdes's spectroscopic light curves, exact choices of noise model or transfer function are not particularly important for the lag recovery rate. Nevertheless, there is overwhelming evidence that the DRW alone is not the best model of AGN continuum variability, with some form of short timescale variability being present, either as an overlaid process (of possible physical interest), a source of outlier contamination (e.g. lensing) or a non-DRW like Gaussian process. While there is already ongoing work to characterise the short timescale behaviour of AGN light curves, we suggest that this should be investigated in the context of reverberation mapping to identify which elements of these rapid fluctuations are echoed in the broad emission line response. This is of physical interest by itself, but accurate modelling of this relevant to RM in particular to prevent the short-timescale fitting issue we discuss in Section~\ref{sec: limitations_cont_jitter}. 

On a physical basis, our most interesting finding is that, while \hbeta sources reverberate consistently to $100\%$ of the time and \civ (consistent with prior vagueness of constraints at these redshifts) reverberates \qm{some-to-all of the time}, the \mgii sample is has only $\approx5\%$ of its sources demonstrate simple reverberations in a simple DRW model, and even our most generous modelling can allow for clear reverberation in only $\approx28\%$. Some part, though not a majority, of the lag free \mgii sources can be attributed to some short-scale noise like process in the continuum light curves. Two potential sources are of first and highest interest: either this is an issue of the \ozdes pipeline (\mgii is known to be complicated by the existence of the flanking iron lines), or this is a real physical decoupling of the \mgii region from the simple RM model. This has significant implications for reverberation mapping efforts going forward: the more sources that are uncoupled in our samples, the more chances we have for false positives and so the more convincing our evidence for a lag recovery must be for us to trust it. A proper understanding of this high rate of decoupling and its source will be an important component in large scale RM programs like \sdss and \ozdes, and going forward DESI \citep{Alfarsy_2026_stacking} and 4MOST/TiDES \citep{Frohmaier_2025_4mostRM}.

\paragraph{Suggestions for Future Work}
In this work we have examined nearly every part of the RM pipeline except the initial data reduction that produces our light curves. The most obvious next step is to apply this pipeline, or a variant of it, to the data set of \ozdes's sister survey \sdss. With a similar physical footprint and science goals but different internal methods, seeing if our Bayesian modelling here yields a higher \mgii reverberation fraction when the accordant light curves are produced by different means will help determine how much of this result can be explained by methodology instead of physics.

In this work, we tune our false positive rate curve by comparing to time-reversed light curves due in a \qm{negative lag test}. This requires us to discard measured lags that are consistent with zero, i.e. those that are neither positive nor negative. A more complete approach would be instead to tune the FPR curve as was done in \citet{McDougall_2025_LITMUS}, where we simulated two large samples of mock light curves. However, this cannot be done reliably unless we have reliable and realistically detailed models from which to generate mocks. A more detailed understanding of the light curves would be an important step in generating mocks that represent the true population of AGN and so we suggest that this is a valuable area to direct future efforts in RM methods. We suggest, in particular, investigation of the short-timescale continuum noise processes we measure in this work, as these also offer benefits to our source-by-source lag constraints.

This work, along with the findings of \citet{McDougall_2025_LITMUS}, suggest that industrial scale RM is at the limits of what can be constrained with the traditional \qm{cut and constrain} approach. Rather, given $\ge90\%$ of our sources are of marginal significance, the frontier now is in leveraging these marginal sources in aggregate to constrain the population. In this work we also found that the constraints of the $R-L$ relationship as a prior on lag were a powerful tool for sorting signal from noise (see Figure~\ref{fig: gradechange_many}'s right most panel for a demonstration). This is not surprising: improving lag recoveries in weak SNR sources by using population-based constraints is the basis of stacked RM. The next step is then somewhat obvious: abandon the \qm{cut and constrain} regime in favour of a hierarchical model that fits the hyper-parameters of the $R-L$ relationship prior, along with the reverberating fraction $f$, for each line simultaneously with the lags for each source. The authors intend to pursue this hierarchical approach in the future.

\section*{Contribution Statement}
Analysis, programming, calculations: HM; Writing: HM; Original figures: HM; Editing: TMD, CL, PM, ZY, GL, BP; \ozdes Project conception and coordination: TMD, CL, PM; Data generation and/or curation: all authors.

\section*{Acknowledgements}
We acknowledge and pay respect to the traditional owners of the land on which the University of Queensland, University of Sydney, and Macquarie University are situated, upon whose unceded, sovereign, ancestral lands we work. We pay respects to their Ancestors and descendants, who continue cultural and spiritual connections to Country.

HGM, TMD, AP acknowledge support for early stages of this project from an Australian Research Council (ARC) Laureate Fellowship (project number FL180100168) and for later stages from the ARC Centre of Excellence for Gravitational Wave Discovery, OzGrav (CE230100016). HGM has been supported by the Australian Government Research Training Program (RTP) award. We are grateful to the Australian public for enabling this science. 

Based in part on data acquired at the Anglo-Australian Telescope, under program Ab/2013B/012]. We acknowledge the traditional owners of the land on which the AAT stands, the Gamilaroi people, and pay our respects to elders past and present.

Calculations were made using \python \citep{VanRossum_2009_python} and with the aid of \texttt{numpy} \citep{harris_2020_numpy}. Plots and figures were generated with the aid of \texttt{matplotlib} \citep{Hunter_2007_matplotlib} and \texttt{chainconsumer} \citep{Hinton_2016_chainconsumer}. All initial lag fitting was done with \litmus, which is built using the probabilistic programming language \numpyro \citep{numpyro} and the \python framework \jax \citep{jax}. Fits are performed with the nested sampling package \jaxns \citep{Albert_2020_JAXNS}.

\section*{Data Availability}
\label{sec: data_availability}
The full set of light curves, along with their posterior samples and evidences for the models described in Section~\ref{sec: LC_modelling}, are available at \red{[Zenodo link to be provided upon acceptance]}


\bibliographystyle{apj}
\bibliography{bib_main,  bib_Hbetasources, bib_MgIIsources, bib_CIVsources, bib_OzDES, bib_LITMUS, bib_ozdeslitmus}



\appendix
\section{Light Curve Cleaning}
\counterwithin{figure}{section}
\label{app: lightcurve_cleaning}
In this section we describe the method by which we clean the outlier epochs from the continuum observations prior to lag fitting / model comparison. Firstly, we divide each continuum light curve into annual seasons. We then describe each season with a generative model in which observations are drawn from a mixture model with two components: a time-polynomial foreground model with Gaussian scatter, and a stationary background Gaussian distribution, i.e. a model with a likelihood given by, for a set of observations of strength $y_i$ at times $t^i$:
\begin{equation}
    \mathcal{L}(\{t^i,y^i\}\vert \theta) = (1-q)\times\mathcal{N}\left( \mu_\text{fg}(t^i), \; \sigma_\text{fg}\right) + q \times \mathcal{N}\left( \mu_\text{bg}, \; \sigma_\text{bg}\right)
    ,
    \label{eq: cleaning_likelihood}
\end{equation}
where $\mu_\text{fg}(t^i)=\sum_{j=0}^N{c_j t^i_jk}$ is an $N$\textsuperscript{th} order polynomial fit of the foreground distribution. Here $\mathcal{N}$ represents a normal distribution, $\sigma_\text{fg}$ and $\sigma_\text{bg}$ are the scatters of the foreground and background distributions, $\mu_\text{bg}$ is the mean of the background noise and $q$ is the relative strength of the outlier noise. Here, $\theta$ represents the set of all foreground and background parameters.

Starting from a maximum likelihood estimation with $q=0$ and $\sigma_\text{fg}$=0, we fit the coefficients in a Bayesian way with arbitrarily large priors, marginalising over the parameters using \numpyro's implementation of the No U-Turn Sampler (NUTS). At each posterior sample we calculate the distance of each epoch from the foreground model mean relative to the scatter, (i.e. the \qm{$z$ test statistic}, $z=E[\frac{y^i-\mu(t^i)}{\sigma_\text{fg}}]$). If a point has $\braket{Z}_\mathrm{samples} > 3$, classify it as an outlier. This is repeated for every season in a light curve and for every light curve in the sample.

\begin{figure*}
    \centering
    \includegraphics[width=0.9\linewidth]{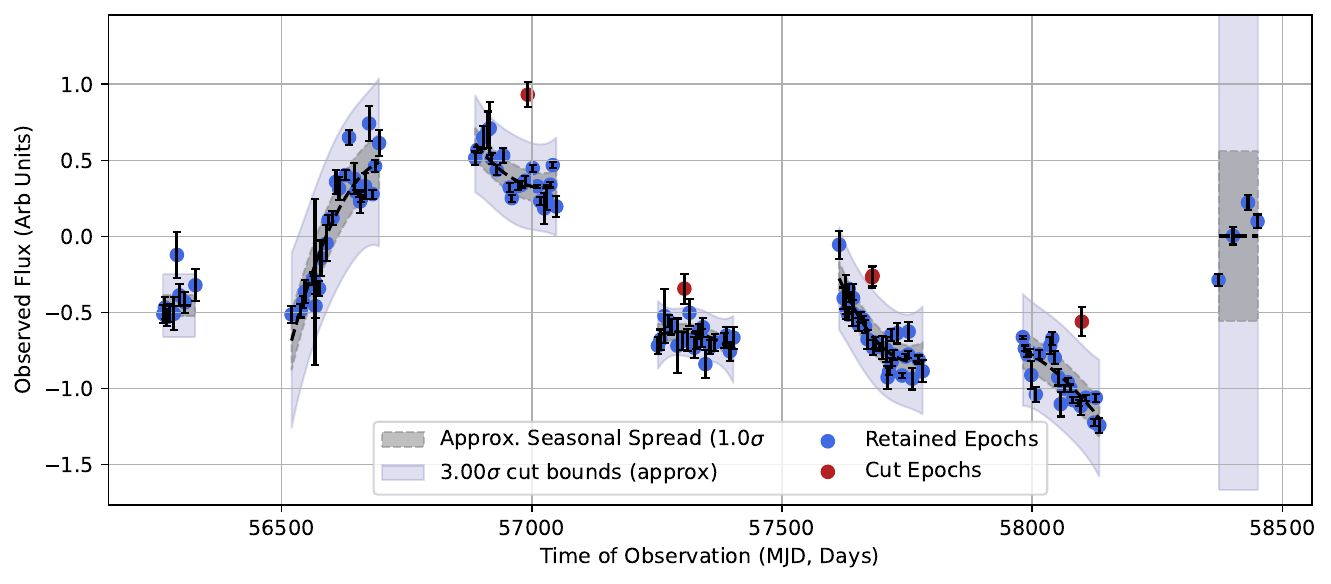}
    \caption{An example of the light curve cleaning for the g-band continuum of \ozdes source $2970518535$. The navy and red points show the retained and removed epochs of observation. The grey bands show the polynomial fit and $1\sigma$ scatter bounds for each season, marginalised over parameter uncertainty, while the blue bounds show the $3\sigma$ bounds. Epochs are removed if their deviation from the season trend, averaged over the posterior samples, is $>3\sigma$.}
    \label{fig: lightcurve_cleaning_example}
\end{figure*}

All $727$ continuum curves are inspected using diagnostic plots similar to \ref{fig: lightcurve_cleaning_example}. If a season is appears to over-fit the data, its polynomial order is reduced and the fit re-run. The first and last seasons have fewer epochs to fit from, and so we use only a zeroth order polynomial, i.e. a stationary foreground distribution for these. The limited data of the first season means the foreground and background distributions become difficult to distinguish in some cases, and some clear and obvious outliers must be trimmed by hand.

At worst case, this method pushes our continuum light curves to be smoother. If this cutting method is too aggressive, it would lead to an upwards shift of $\tau$ (removing short term variability) and a downwards shift of the continuum jitter (for jitter models) in the model fitting, but should not bias any of our lag measurements.

\section{Details on Lag Reliability \& Modelling Choices}
\label{app: RM_reliability}
In Section~\ref{sec: RM_limitations} we briefly summarise a few areas in which we investigated the robustness of our lag recovery / appropriateness of our light curve models. In this appendix we provide more expansive and detailed data on this summary. 

\subsection{Impact from Modelling Choices On Lag Recovery}
\label{app: recovery_summary}

\begin{figure*}
    \centering
    \includegraphics[width=0.24\linewidth]{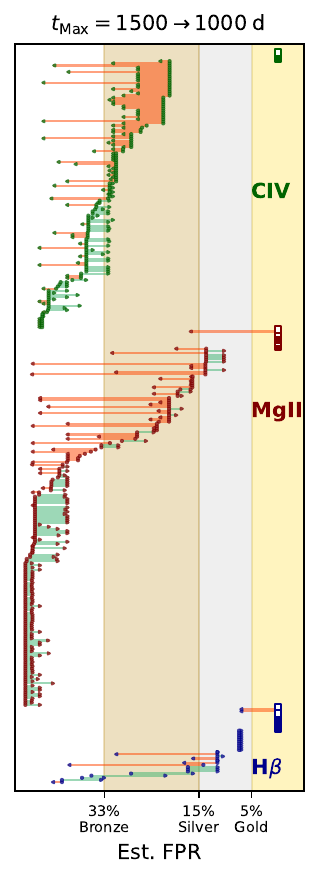}
    \includegraphics[width=0.24\linewidth]{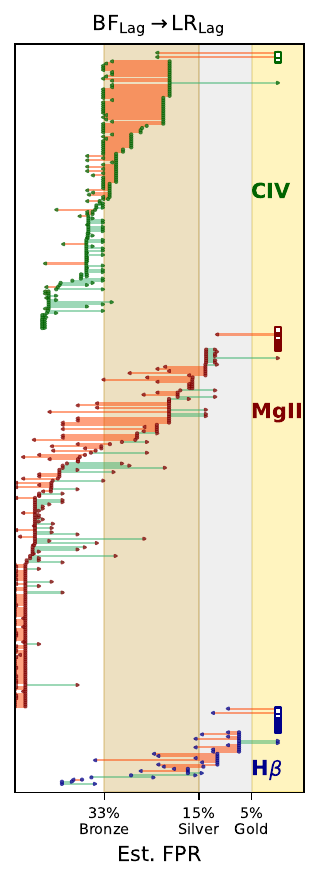}
    \includegraphics[width=0.243\linewidth]{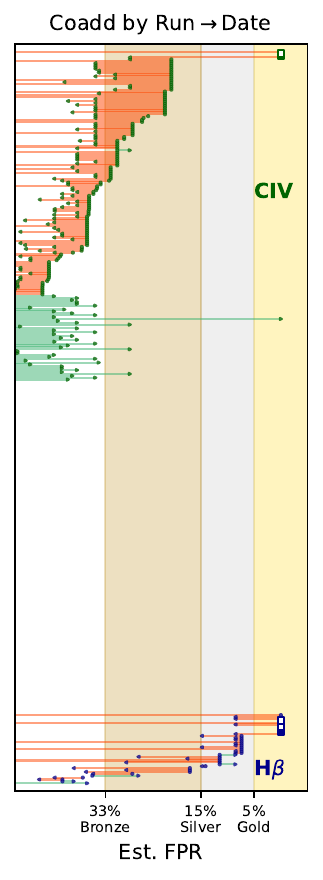}
    \includegraphics[width=0.24\linewidth]{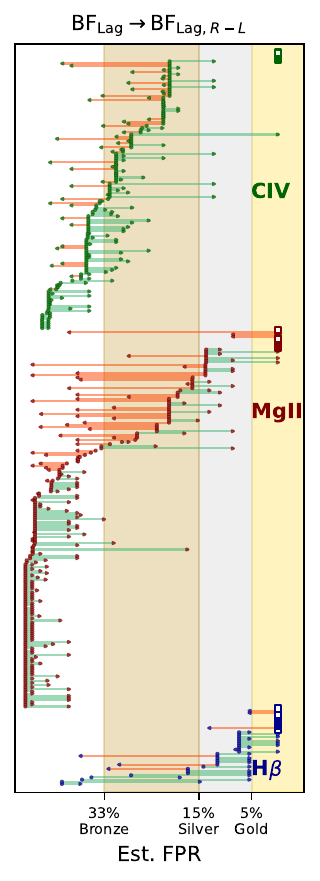}\\
    \includegraphics[width=0.5\linewidth]{media/FPR_dist_legend.pdf}
    \caption{A demonstration of how our false positive rate and lag recovery grades change if we re-run our pipeline after switching away from our fiducial choices in Section~\ref{sec: selection_criteria}. Sources are organised by line type and then sorted by the reliability of their fiducial recovery. Sources by The left panel shows the impact of using a maximum bound in our lag prior of $1000 \dayu$ instead of $1500 \dayu$, the middle shows the impact of using $\Lrat{Lag}$ instead of $\BF{Lag}$ as our measure of significance, the third panel shows the impact of using spectra co-added by date instead of by run to construct our response light curves and the right panel shows the impact of using an $R-L$ informed prior on lag from our fiducial relations presented in \citet{OzDES-McDougall_2025}. The general trend in all of these is that while there is a marginal increase in our trust for a few weak recoveries, and many of our most reliable fits (gold recoveries) are immune to different decisions, we overall lose more resolution in our recoveries  (more good sources moving to the left) than we gain in for our marginal recoveries. When using an $R-L$ prior, \hbeta and \civ see significant increases in the number and reliability of their recoveries, while \mgii sees an interesting split of an overall increase in quality but with many sources being made less reliable. For the third panel, no \mgii sources are shown as only have spectra coadded by run.}
    \label{fig: gradechange_many}
\end{figure*}

In Section~\ref{sec: limitations_model} we discuss that the choice of light curve model or lag significance measure does not strongly affect our lag recovery numbers, and that the simple DRW model with the uniform prior Bayes factor was a decent choice for all lines. In Table~\ref{tab: grade_summaries} we list a breakdown of the \gold, \silver and \bronze recoveries for all variations on these choices, and in Figure~\ref{fig: gradechange_many} we show this impacts the source-by-source false positive rate.

If prioritising the number of \gold recoveries instead of \silver, the \hbeta sample is fit much better by the jitter model and the \mgii sample by the tophat smoothing model. Replacing the uniform lag prior with the $R-L$ relationships from \citet{OzDES-McDougall_2025} gives the greatest gains to the \hbeta sample, in particular when using the jitter model. The \mgii sample sees marked gains when using the tophat smoothing model with the prior, and \civ sees a marginal improvement when using the simple DRW model. This is interesting, as the \mgii sample does not show strong preference for smoothing. The fact that \hbeta sees such a ready improvement here is not unexpected, as the $R-L$ relationship for \hbeta is the most broadly studied and so least likely to be misaligned with the true population. That \civ sees so little improvement may indicate that more work is needed to refine the $R-L$ relationship for this line.

The choice of model also slightly shifts the recovered lag, particularly for the tophat smoothing models. If the lag is measured from the start of the tophat or the centroid, the lag differs by $\frac{w}{2} \dayu$. Because the jitter model loosens the light curve constraints, it also generates vaguer constraints ($\BF{Lag}\rightarrow1.0$, wider error bars on $\Delta t$).

\subsection{Detailed Constraints on Smoothing Timescale \& Short Timescale Noise}

The tophat lag model lets us estimate the smoothing timescale of the response function, as we discuss in Section~\ref{sec: RM_limitations}, and we discuss in Section~\ref{sec: limitations_cont_jitter} how the jitter model of light curve noise allows us to examine how much the continuum light curves differ from a perfect DRW. In those sections, we refer to the response light curves preferring long or short timescales, and the continuum light curves having week or strong jitter. In this appendix we elaborate on and quantify what we mean by these terms. 

\begin{figure*}
    \centering
    \includegraphics[width=0.9\linewidth]{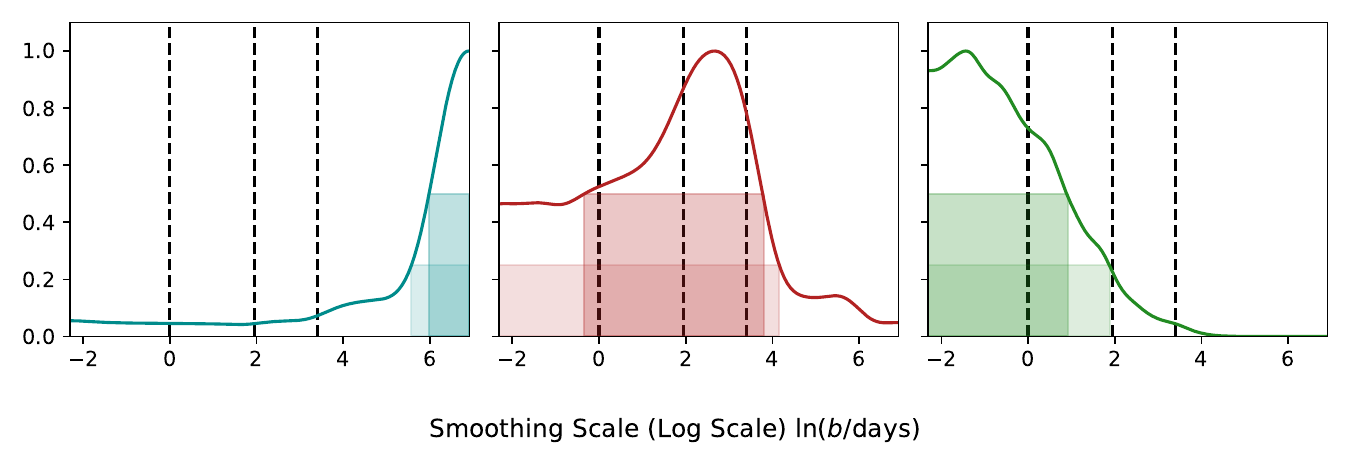}
    \caption{Examples of the posterior distributions of the smoothing timescale for three AGN, demonstrating what we consider to be a strong or weak constraint. The top and bottom panels show distributions for an \hbeta (\ozdes ID $2925344542$) and \civ (\ozdes ID $2940140085$) source with clear constraints at high and low degrees of smoothing, while the middle panel shows the weak constraints for an \mgii (\ozdes ID $2970914647$) source, the only AGN with a moderate degree of smoothing (neither above nor below our constraint range). The dotted lines are, from left to right, placed at $1$, $7$ and $30$ days. All of these sources have weak coupling ($\BF{Lag}<10$) and so these constraints are for the uncoupled null model.}
    \label{fig: smoothing_constraint_examples}
\end{figure*}

Constraints on smoothing timescale are typically quite broad in our sample, meaning that measuring the peak or median of the distribution is often uninformative. This is coupled with the survey window setting upper and lower bounds for what sort of smoothing scales we can meaningfully measure: if the smoothing timescale is shorter than the $\approx 30 \dayu$ cadence of the spectroscopic observations in \ozdes or of a similar scale to the $6$-year baseline then it is difficult to constrain. We quantify the constraints on smoothing timescale for each source here by taking the posterior distributions for $\ln{w}$ for each source and, via use of a Gaussian KDE, estimate the FWHM and (Full-Width Quarter Maximum ) FWQM of the distribution. If the FWHM spans the entire prior we say that the smoothing is unconstrained. If only the FWQM does or if the FWHM does not touch the prior edges we say that it is weakly constrained (or that it \sqm{prefers} short or long smoothing. If neither the FWHM nor FWQM touch the prior bounds, the smoothing is \qm{strongly} constrained (see examples in Figure~\ref{fig: smoothing_constraint_examples}). 

In this fitting, we use the posterior distribution for $\ln{b}$ for the uncoupled model except in cases where coupling is very strongly observed ($\BF{Lag}\ge10$, to avoid any biasing of the smoothing timescale that would arise from trying to resolve a tension while fitting a spurious lag. We summarise the number of sources constrained at different smoothing lengths for each line in Table~\ref{tab: smoothing_constraint_table}. Using this classification, the majority of sources exhibit low smoothing. The \hbeta sample has roughly equal numbers of long and short smoothing timescale while \mgii strongly prefers short smoothing and \civ has generally poor constraints but nevertheless shows a stronger affinity for long smoothing timescales than \hbeta or \mgii.

We can quantify the strength of the jitter in much the same way that we constrain the smoothing timescale in the previous section, i.e. using the wether the FHWM and FWQM of the posterior make contact with the prior boundaries. We show the summarise these constraints in Table~\ref{tab: jitter_constraint_table}.

\begin{table}[]
    \centering
    \begin{tabular}{c|c|c|c|c}
    \hline
        Constraint & \hbeta &  \mgii & \civ & All \\ \hline
        Unconstrained & 18 & 41 & 215 & 274 \\ \hline
        Constrained Short & 16 & 341 & 36 & 393  \\ \hline
        Prefers Short & 6 & 27 & 28 & 61  \\ \hline
        Strongly Constrained & 6 & 8 & 1 & 15  \\ \hline
        Weakly Constrained & 2 & 3 & 0 & 5  \\ \hline
        Prefers Long & 15 & 18 & 74 & 107  \\ \hline
        Constrained Long & 14 & 15 & 22 & 51  \\ \hline
    \end{tabular}
    \caption{The number of sources for each of the three reverberating lines that can be constrained to short or long smoothing timescales at various levels of confidence. The \hbeta sample has roughly equal numbers of long and short smoothing timescale while \mgii strongly prefers short smoothing and \civ has generally poor constraints but a slight preference for long timescales.}
    \label{tab: smoothing_constraint_table}
\end{table}

\begin{table}[]
    \centering
    \begin{tabular}{c||c||c|c|c|c}
    \hline
        Constraint & Cont. & \hbeta & \mgii & \civ & All \\ \hline
        \makecell{Unconstrained} & 2 & 13 & 25 & 102 & 140 \\ \hline
        \makecell{No Noise} & 87 & 43 & 101 & 133 & 277 \\ \hline
        \makecell{Prefers No Noise}& 0 & 12 & 25 & 97 & 134 \\ \hline
        \makecell{Vague Noise} & 58 & 4 & 56 & 8 & 68 \\ \hline
        \makecell{Clear Noise} & 755 & 0 & 57 & 3 & 60 \\ \hline
        \makecell{Weak Structure} & 2 & 2 & 17 & 14 & 33 \\ \hline
        \makecell{Entirely Noise} & 2 & 3 & 172 & 19 & 194 \\ \hline
    \end{tabular}
    \caption{The number of sources for each of the continuum and each of the three reverberating lines that can be constrained to have weak or strong non-DRW jitter components. The \hbeta sample is well structured (very few sources with strong noise), and while the \civ sample has more unconstrained sources it still broadly shows strong structure. The \mgii sample, meanwhile, has roughly a third of its sources demonstrating extremely poor structure.}
    \label{tab: jitter_constraint_table}
\end{table}

\begin{table*}
\centering
\begin{tabular}{c|c||c|c|c||c|c|c||c|c|c||c|c|c}
\hline
\multicolumn{14}{c}{Response LCs Co-Added By Date}\\ \hline
& & \multicolumn{3}{|c||}{\hbeta} & \multicolumn{3}{|c||}{\mgii} & \multicolumn{3}{|c||}{\civ} & \multicolumn{3}{|c}{All}\\ \hline
\Gape[7mm][0pt]{LC Model} & \makecell{Significance \\ Measure} &   \rotatebox[origin=c]{90}{\gold} & \rotatebox[origin=c]{90}{\silver} & \rotatebox[origin=c]{90}{\bronze}  & \rotatebox[origin=c]{90}{\gold} & \rotatebox[origin=c]{90}{\silver} & \rotatebox[origin=c]{90}{\bronze}  & \rotatebox[origin=c]{90}{\gold} & \rotatebox[origin=c]{90}{\silver} & \rotatebox[origin=c]{90}{\bronze}  & \rotatebox[origin=c]{90}{\gold} & \rotatebox[origin=c]{90}{\silver} & \rotatebox[origin=c]{90}{\bronze} \\[8pt] \hline
\multirow{3}{*}{Simple DRW}  & $\BF{Lag}$ 
 & 5 & 20 & 21 & 9 & 22 & 54 & 3 & 3 & 14 & 17 & 45 & 89 \\
& $\Lrat{Lag}$   & 6 & 19 & 29 & 9 & 22 & 47 & 3 & 3 & 7 & 18 & 44 & 83 \\
& $\BF{Lag,R-L}$  & 12 & 20 & 20 & 9 & 24 & 38 & 3 & 10 & 39 & 24 & 54 & 97 \\
\hline \multirow{3}{*}{\textbackslash w Jitter} 
& $\BF{Lag}$  & 12 & 17 & 25 & 4 & 9 & 32 & 4 & 4 & 19 & 20 & 30 & 76 \\
& $\Lrat{Lag}$   & 6 & 25 & 31 & 5 & 5 & 13 & 3 & 3 & 12 & 14 & 33 & 56 \\
& $\BF{Lag,R-L}$  & 21 & 26 & 26 & 6 & 6 & 68 & 4 & 4 & 40 & 31 & 36 & 134 \\
\hline \multirow{3}{*}{\textbackslash w Smoothing} 
& $\BF{Lag}$   & 4 & 13 & 19 & 8 & 20 & 50 & 2 & 2 & 10 & 14 & 35 & 79 \\
& $\Lrat{Lag}$   & 5 & 15 & 28 & 9 & 14 & 41 & 4 & 4 & 4 & 18 & 33 & 73 \\
& $\BF{Lag,R-L}$   & 14 & 22 & 22 & 17 & 21 & 32 & 3 & 3 & 30 & 34 & 46 & 84 \\
\hline \multirow{3}{*}{\textbackslash w Smooth. (Cent)} 
& $\BF{Lag}$   & 4 & 15 & 19 & 8 & 20 & 49 & 2 & 2 & 10 & 14 & 37 & 78 \\
& $\Lrat{Lag}$  & 6 & 9 & 19 & 13 & 22 & 37 & 4 & 4 & 39 & 23 & 35 & 95 \\
& $\BF{Lag,R-L}$  & 6 & 10 & 18 & 17 & 21 & 33 & 2 & 2 & 26 & 25 & 33 & 77 \\
\hline
\multicolumn{14}{c}{Response LCs Co-Added By Run}\\ \hline
\multirow{3}{*}{Simple DRW}  
& $\BF{Lag}$  & 11 & 28 & 28 & - & - & - & 4 & 4 & 69 & 24 & 54 & 151 \\
& $\Lrat{Lag}$   & 11 & 21 & 32 & - & - & - & 3 & 3 & 21 & 23 & 46 & 100 \\
& $\BF{Lag,R-L}$  & 14 & 24 & 24 & - & - & - & 5 & 11 & 53 & 28 & 59 & 115 \\
\hline \multirow{3}{*}{\textbackslash w Jitter} 
& $\BF{Lag}$  & 15 & 15 & 24 & - & - & - & 0 & 0 & 45 & 19 & 24 & 101 \\
& $\Lrat{Lag}$   & 10 & 27 & 32 & - & - & - & 0 & 0 & 21 & 15 & 32 & 66 \\
& $\BF{Lag,R-L}$  & 19 & 22 & 22 & - & - & - & 3 & 9 & 31 & 28 & 37 & 121 \\
\hline \multirow{3}{*}{\textbackslash w Smoothing} 
& $\BF{Lag}$  & 6 & 12 & 15 & - & - & - & 2 & 2 & 23 & 16 & 28 & 88 \\
& $\Lrat{Lag}$   & 8 & 8 & 15 & - & - & - & 2 & 2 & 10 & 19 & 24 & 68 \\
& $\BF{Lag,R-L}$  & 10 & 18 & 18 & - & - & - & 3 & 3 & 16 & 30 & 42 & 66 \\
\hline \multirow{3}{*}{\textbackslash w Smooth. (Cent)} 
& $\BF{Lag}$  & 8 & 17 & 19 & - & - & - & 3 & 3 & 26 & 19 & 34 & 94 \\
& $\Lrat{Lag}$  & 10 & 10 & 18 & - & - & - & 4 & 4 & 10 & 27 & 36 & 65 \\
& $\BF{Lag,R-L}$  & 11 & 21 & 21 & - & - & - & 4 & 4 & 21 & 32 & 46 & 75 \\
\hline

\end{tabular}
\caption{A summary of how many \bronze, \silver \& \gold recoveries are made for each light curve model and significance measurement if grouping by line or using the entire sample collectively, with results for co-adding spectra by date and by run when generating the response curves in the top and bottom panels. The \mgii column for the by-date section are left blank as all \mgii sources are co-added by run in the \mgii pipeline of \citet{OzDES-Yu_2021}.}
\label{tab: grade_summaries}
\end{table*}

\clearpage
\onecolumn
\section{Robustness of Reverberating Fraction Estimates}
\label{app: frac_robustness}
\begin{figure*}
    \centering
    \includegraphics[width=\linewidth]{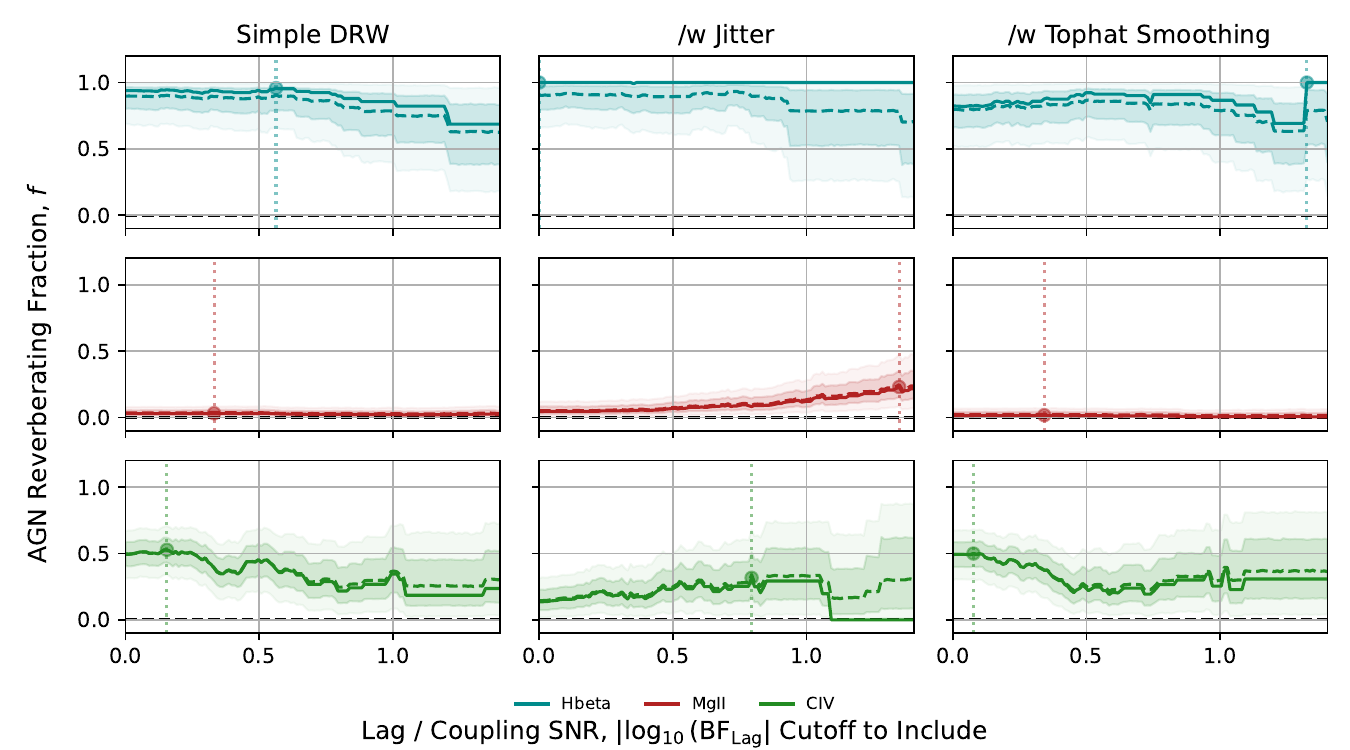}
    \caption{A demonstration of how the constraints on the reverberation fractions shown in Figure~\ref{fig: reverb_fraction} change as we include only sources with increasingly convincing Bayes factors to accept or reject lags ($\lvert \log_{10}(\BF{Lag})\rvert \ge \text{Some Cutoff}$). As the cutoff increases (going to the right) the constraints become looser, as we use less sources, but also less subject to the biased under-estimation from the uninformative uniform lag prior used in calculating $\BF{Lag}$. The dots and dashed vertical lines indicate the position of the peak fraction shown in Table~\ref{tab: reverb_frac_summary}.}
    \label{fig: reverb_frac_slidingscale_BF}
\end{figure*}

Our estimates of the reverberation fraction for each line in Section~\ref{sec: reverb_frac} assume a single homogeneous population of AGN signals and unbiased measurements of $\BF{Lag}$. Because the uniform prior we use to estimate $\BF{Lag}$ is arbitrarily broad, it will in fact systematically under-estimate this Bayes Factor and so bias $f$ downwards. We can make a coarse correction for this bias by only using samples with extreme values for $\BF{Lag}$, moving towards likelihood dominated sources that can very clearly be seen to be coupled or uncoupled. As we increase this threshold, the constraints on $f$ evolve as shown in the Figure~\ref{fig: reverb_frac_slidingscale_BF}.

\begin{figure*}
    \centering
    \includegraphics[width=\linewidth]{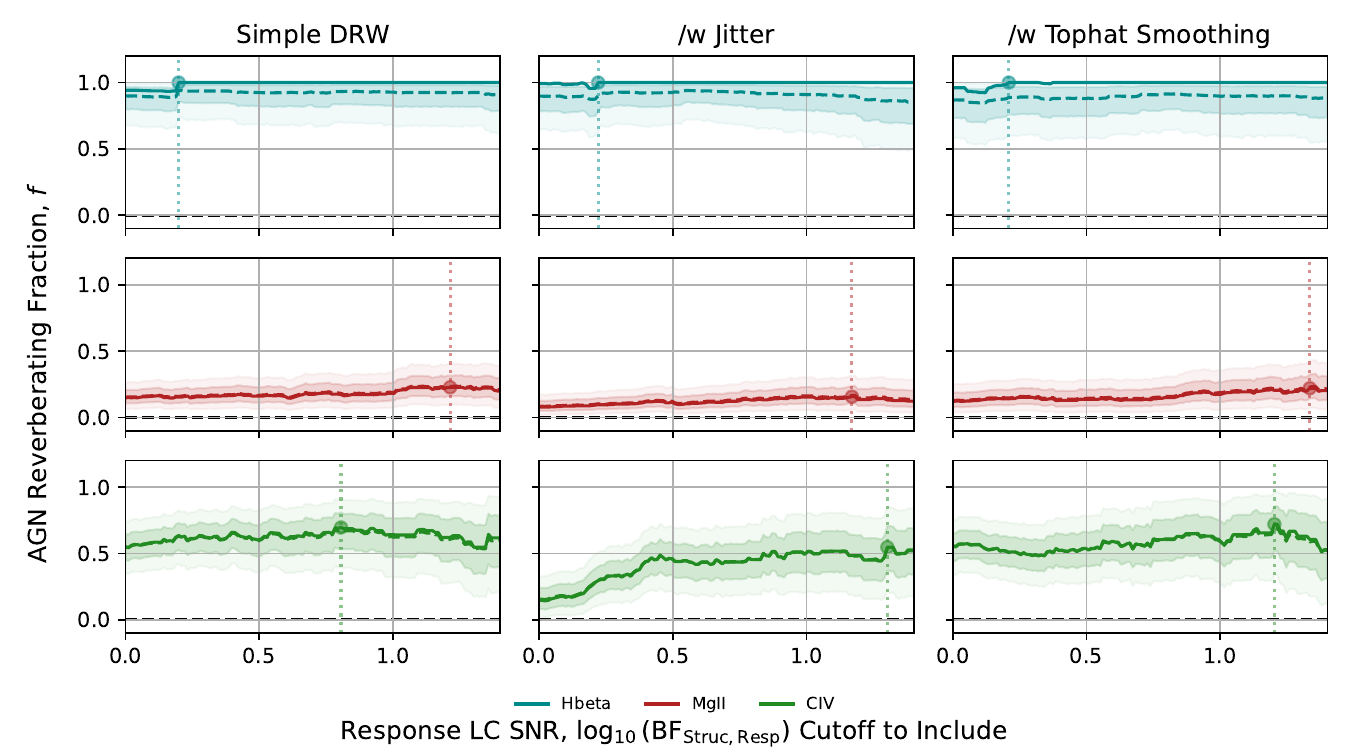}
    \caption{A demonstration of how the constraints on the reverberation fractions shown in Figure~\ref{fig: reverb_fraction} change as we include only sources with increasingly well structured light curves ($\BF{Struc,Resp} \ge \text{Some Cutoff}$). Using only well structured light curves improves the fraction of \mgii sources that reverberate, but still fails to get much higher than about one in four. The dots and dashed vertical lines indicate the position of the peak fraction shown in Table~\ref{tab: reverb_frac_summary}.}
    \label{fig: reverb_frac_slidingscale_BFstruc}
\end{figure*}

We note also that in Section~\ref{sec: limitations_smoothing_and_jitter} we found that the \mgii sample has an unusually high number of \qm{strong jitter} response curves, i.e. there is a large subset of the \mgii sample in which the response has a distinct lack of structure, more than can be explained by low SNR alone. This motivates us to see if the low reverberation fraction for \mgii may be explained instead by retaining sources above some cutoff for $\BF{Struc,Resp}$. Figure~\ref{fig: reverb_frac_slidingscale_BFstruc} we show how the constraints on $f$ vary as we apply increasingly stringent cutoffs on $\BF{Struc,Resp}$ and the magnitude of $\BF{Lag}$

In Table~\ref{tab: reverb_frac_summary} we summarise the constraints on $f$ for all lines with either all sources or for the highest value that can be achieved when using these cutoffs. Generally, we see high $f$ (i.e. more clearly reverberating AGN) when co-adding spectra by run rather than date. It is easy to get constraints for \hbeta in which $f\approx1$ with any model or cutoff, with the lowest estimate being $f_{\hbetamath} \approx 0.79$ and the highest being $f_{\hbetamath} \approx0.94$ and both of these being consistent with $100\%$ reverberation to within $2\times$ the error bounds. \civ prefers to have roughly half the sources presenting simple lags. When using the jitter model to account for its strong non-DRW component (see Section~\ref{sec: limitations_smoothing_and_jitter}~and~\ref{app: RM_reliability}) and permitting only well structured or light curves or convincing lag measurements, the reverberating fraction for \mgii can reach as high as $\approx$ one in four. Nevertheless, there is no combination of model and selection criteria in which we can say that all, or even most, of the \mgii sample reverberates (subject to the caveats that we discuss in Section~\ref{sec: reverb_frac}).

\begin{table*}
    \centering
    \begin{tabular}{|c|c|c|c|c|c|c|c|c|c|c|}
    \hline
    \multicolumn{10}{|c|}{Response LCs Co-Adde d By Run}\\ \hline
Line & \multicolumn{3}{|c|}{Simple DRW}& \multicolumn{3}{|c|}{Jitter}& \multicolumn{3}{|c|}{Tophat Smoothing}\\ \hline
& \makecell{All\\Sources}& \makecell{Highest\\for\\Strong\\$\BF{Lag}$} &\makecell{Highest\\for\\Strong\\Struc.} & \makecell{All\\Sources}& \makecell{Highest\\for\\Strong\\$\BF{Lag}$} &\makecell{ Highest\\for\\Strong\\Struc.} & \makecell{All\\Sources}& \makecell{Highest\\for\\Strong\\$\BF{Lag}$} &\makecell{Highest\\for\\Strong\\Struc.} \\ \hline
\hbeta & $0.90^{+0.07}_{-0.10}$ & $0.90^{+0.07}_{-0.11}$ & $0.94^{+0.04}_{-0.09}$
& $0.91^{+0.07}_{-0.11}$ & $0.90^{+0.07}_{-0.11}$  & $0.92^{+0.06}_{-0.10}$
& $0.80^{+0.11}_{-0.14}$ & $0.79^{0.15}_{0.26}$  & $0.89^{+0.08}_{-0.13}$
\\
\mgii &$0.03^{+0.02}_{-0.02}$ & $0.04^{+0.02}_{-0.02}$  & $0.28^{+0.13}_{-0.11}$
& $0.05^{+0.03}_{-0.03}$ & $0.25^{+0.10}_{-0.09}$  & $0.25^{+0.11}_{-0.09}$
& $0.02^{+0.02}_{-0.01}$ & $0.02^{+0.02}_{-0.02}$  & $0.26^{+0.10}_{-0.10}$
\\
\civ & $0.49^{+0.09}_{-0.09}$ & $0.52^{+0.09}_{-0.09}$  & $0.70^{+0.17}_{-0.21}$
 & $0.15^{+0.07}_{-0.07}$ & $0.34^{+0.17}_{-0.16}$  & 	$0.58^{+0.22}_{-0.24}$
& $0.49^{+0.09}_{-0.09}$ &$0.50^{+0.09}_{-0.09}$ & $0.69^{+0.18}_{-0.23}$
\\ \hline
    \multicolumn{10}{|c|}{Response LCs Co-Added By Date}\\ \hline
Line & \multicolumn{3}{|c|}{Simple DRW}& \multicolumn{3}{|c|}{Jitter}& \multicolumn{3}{|c|}{Tophat Smoothing}\\ \hline
& \makecell{All\\Sources}& \makecell{Highest\\for\\Strong\\$\BF{Lag}$} &\makecell{Highest\\for\\Strong\\Struc.} & \makecell{All\\Sources}& \makecell{Highest\\for\\Strong\\$\BF{Lag}$} &\makecell{Highest\\for\\Strong\\Struc.}& \makecell{All\\Sources}& \makecell{\makecell{Highest\\for\\Strong\\$\BF{Lag}$}} &\makecell{Highest\\for\\Strong\\Struc.}\\ \hline
\hbeta & $0.86^{+0.09}_{-0.12}$ & $0.84^{+0.12}_{-0.22}$  & $0.88^{+0.08}_{-0.11}$
& $0.83^{+0.10}_{-0.13}$ & $0.83^{+0.12}_{-0.22}$ & $0.87^{+0.08}_{-0.12}$
& $0.79^{+0.12}_{-0.14}$ & $0.84^{+0.12}_{-0.21}$  & $0.87^{+0.08}_{-0.12}$
\\

\mgii & $-$ & $-$ & $-$ & $-$ & $-$ & $-$  & $-$ & $-$ & $-$ \\

\civ & $0.44^{+0.08}_{-0.08}$ & $0.43^{+0.08}_{-0.08}$  & $0.59^{+0.10}_{-0.11}$
& $0.26^{+0.08}_{-0.08}$ & 	$0.27^{+0.08}_{-0.08}$  & $0.44^{+0.10}_{-0.10}$
& $0.43^{+0.08}_{-0.08}$ & 	$0.43^{+0.08}_{-0.08}$  & $0.55^{+0.11}_{-0.11}$
\\
\hline
    \end{tabular}
    \caption{Constraints on the reverberating fraction for all lines across all light curve types and models. The \qm{All Sources} values are when using all sources in the sample, while the \qm{Highest for strong $\BF{Lag}$} values are the maximum that can be achieved when removing the marginal $\BF{Lag}\approx1$ sources, i.e. the highest $f$ that can be achieved when varying the cutoff as in Figure~\ref{fig: reverb_frac_slidingscale_BF}. The values under \qm{Highest for Strong Struc.} are the highest possible $f$ when accepting only curves with high $\BF{struc,resp}$, i.e. only the sources with the best structured light response curves, as shown in Figure~\ref{fig: reverb_frac_slidingscale_BFstruc}.}
    \label{tab: reverb_frac_summary}
\end{table*}

\clearpage
\twocolumn

\newpage
\onecolumn
\section{\ozdes vs. \litmus}
\label{app: OzDES_vs_Litmus}
In Section~\ref{sec: lag_results} we summarise the ways in which the use of \litmus and our new significance criteria pipeline changes our measured lags and the footprint of the \ozdes lag sample in the $R-L$ plane. Table~\ref{tab: OzDES_vs_LITMUS} gives a list of all $64$ lags as listed in \citet{OzDES-McDougall_2025}, along with the new lag measurements for these same sources and accordant estimated false positive rate from this work.

\begin{longtable}{c|c|c|c|c|c}
    \centering
\ozdes ID & Line Type & \makecell{Previously\\Published Lag\\ (Obs. Frame) } & \makecell{New Lag \\ (Obs. Frame)} & FPR ($\BF{Lag}$) & FPR ($\Lrat{Lag}$) \\ \hline
2971056445 & \hbeta & $18.0^{+2.7}_{-2.7}$ & $17.3^{+2.6}_{-2.7}$ & $0.0\%$ & $0.0\%$ \\
2925826924 & \hbeta & $34.0^{+1.6}_{-0.8}$ & $34.1^{+0.8}_{-0.7}$ & $0.0\%$ & $0.0\%$ \\
2939653045 & \hbeta & $21.3^{+14.8}_{-19.9}$ & $59.7^{+10.7}_{-7.0}$ & $8.4\%$ & $9.3\%$ \\
2938232574 & \hbeta & $78.3^{+34.7}_{-16.4}$ & $26.3^{+4.2}_{-144.4}$ & $8.4\%$ & $9.3\%$ \\
2970848737 & \hbeta & $52.3^{+10.5}_{-28.7}$ & $48.8^{+6.2}_{-14.0}$ & $0.0\%$ & $0.0\%$ \\
2938089267 & \hbeta & $52.7^{+4.6}_{-6.9}$ & $35.6^{+20.0}_{-1117.5}$ & $50.0\%$ & $50.0\%$ \\
2971046252 & \hbeta & $31.3^{+3.8}_{-5.3}$ & $98.2^{+50.6}_{-27.1}$ & $50.0\%$ & $50.0\%$ \\
2939010304 & \hbeta & $61.5^{+5.3}_{-9.0}$ & $67.6^{+9.5}_{-43.3}$ & $8.4\%$ & $0.0\%$ \\
2940144229 & \mgii & $175.0^{+9.2}_{-8.7}$ & $173.7^{+15.9}_{-16.1}$ & $48.1\%$ & $50.0\%$ \\
2925491917 & \mgii & $286.0^{+54.0}_{-19.0}$ & $319.0^{+79.1}_{-9.3}$ & $0.0\%$ & $0.0\%$ \\
2970909027 & \mgii & $571.0^{+14.5}_{-6.5}$ & $558.6^{+68.8}_{-23.3}$ & $50.0\%$ & $50.0\%$ \\
2943200932 & \mgii & $382.0^{+14.5}_{-16.4}$ & $385.6^{+40.2}_{-18.1}$ & $0.0\%$ & $0.0\%$ \\
2939379534 & \mgii & $225.0^{+23.7}_{-21.4}$ & $217.6^{+48.7}_{-36.0}$ & $46.6\%$ & $37.8\%$ \\
2940669226 & \mgii & $132.0^{+9.4}_{-12.1}$ & $132.3^{+22.4}_{-34.4}$ & $48.1\%$ & $50.0\%$ \\
2970471108 & \mgii & $564.0^{+14.3}_{-22.4}$ & $1312.1^{+5.4}_{-11.6}$ & $50.0\%$ & $50.0\%$ \\
2940304200 & \mgii & $181.0^{+10.7}_{-10.3}$ & $191.7^{+32.1}_{-1103.9}$ & $48.0\%$ & $50.0\%$ \\
2971010612 & \mgii & $383.0^{+5.8}_{-14.4}$ & $1230.3^{+851.6}_{-45.6}$ & $50.0\%$ & $50.0\%$ \\
2925344837 & \mgii & $818.0^{+2.5}_{-16.0}$ & $822.8^{+5.6}_{-19.5}$ & $48.0\%$ & $49.8\%$ \\
2925568292 & \mgii & $412.0^{+27.0}_{-12.5}$ & $423.8^{+9.1}_{-21.7}$ & $0.0\%$ & $0.0\%$ \\
2940687692 & \mgii & $171.0^{+4.4}_{-5.2}$ & $176.8^{+12.1}_{-1057.0}$ & $50.0\%$ & $50.0\%$ \\
2971138367 & \mgii & $537.0^{+5.6}_{-4.4}$ & $598.5^{+76.6}_{-58.2}$ & $50.0\%$ & $50.0\%$ \\
2925731614 & \mgii & $165.0^{+8.3}_{-11.1}$ & $174.3^{+28.1}_{-32.1}$ & $46.2\%$ & $34.7\%$ \\
2970946076 & \mgii & $376.0^{+2.0}_{-4.0}$ & $373.4^{+3.2}_{-12.3}$ & $0.0\%$ & $0.0\%$ \\
2925756177 & \mgii & $520.0^{+9.9}_{-11.8}$ & $547.4^{+40.2}_{-655.4}$ & $48.0\%$ & $50.0\%$ \\
2925403880 & \mgii & $415.0^{+62.9}_{-10.2}$ & $423.6^{+21.7}_{-12.1}$ & $0.0\%$ & $0.0\%$ \\
2938718310 & \mgii & $524.0^{+5.3}_{-5.3}$ & $1295.1^{+34.4}_{-18.0}$ & $50.0\%$ & $50.0\%$ \\
2971163280 & \mgii & $656.0^{+11.7}_{-23.9}$ & $1173.0^{+228.0}_{-155.3}$ & $34.8\%$ & $43.0\%$ \\
2971008992 & \mgii & $508.0^{+9.4}_{-11.3}$ & $1157.7^{+646.4}_{-142.8}$ & $50.0\%$ & $50.0\%$ \\
2939193043 & \mgii & $439.0^{+11.7}_{-9.4}$ & $423.7^{+16.8}_{-15.2}$ & $13.5\%$ & $0.0\%$ \\
2939644652 & \mgii & $308.0^{+24.6}_{-14.9}$ & $331.9^{+63.8}_{-32.5}$ & $50.0\%$ & $50.0\%$ \\
2971214955 & \mgii & $479.0^{+9.5}_{-14.5}$ & $546.1^{+7.6}_{-6.2}$ & $50.0\%$ & $50.0\%$ \\
2938756405 & \mgii & $538.0^{+4.7}_{-10.0}$ & $1265.0^{+31.2}_{-51.6}$ & $50.0\%$ & $50.0\%$ \\
2925515125 & \mgii & $867.0^{+23.1}_{-22.4}$ & $899.1^{+61.2}_{-346.5}$ & $48.0\%$ & $49.9\%$ \\
2938498296 & \civ & $396.0^{+11.3}_{-11.3}$ & $975.2^{+579.1}_{-289.2}$ & $50.0\%$ & $50.0\%$ \\
2938970755 & \civ & $217.0^{+17.1}_{-17.1}$ & $217.5^{+37.4}_{-20.0}$ & $28.0\%$ & $0.0\%$ \\
2938937393 & \civ & $294.0^{+5.8}_{-5.8}$ & $924.7^{+69.3}_{-367.3}$ & $34.9\%$ & $36.0\%$ \\
2939629447 & \civ & $604.0^{+9.8}_{-9.8}$ & $581.1^{+104.8}_{-49.1}$ & $38.4\%$ & $36.0\%$ \\
2925783603 & \civ & $113.0^{+9.0}_{-9.0}$ & $1199.0^{+1109.1}_{-197.8}$ & $50.0\%$ & $50.0\%$ \\
2939433380 & \civ & $556.0^{+11.5}_{-11.5}$ & $938.1^{+313.8}_{-342.5}$ & $49.1\%$ & $49.4\%$ \\
2970913529 & \civ & $442.0^{+5.9}_{-5.9}$ & $1003.2^{+838.1}_{-393.7}$ & $50.0\%$ & $50.0\%$ \\
2937538328 & \civ & $379.0^{+9.4}_{-9.4}$ & $515.0^{+149.8}_{-580.4}$ & $34.9\%$ & $36.3\%$ \\
2938664659 & \civ & $413.0^{+11.6}_{-11.6}$ & $917.5^{+452.7}_{-303.0}$ & $34.9\%$ & $40.1\%$ \\
2940923721 & \civ & $480.0^{+8.6}_{-8.6}$ & $672.1^{+481.0}_{-541.8}$ & $50.0\%$ & $50.0\%$ \\
2940479063 & \civ & $237.0^{+9.7}_{-9.7}$ & $683.1^{+458.6}_{-452.6}$ & $50.0\%$ & $50.0\%$ \\
2938353110 & \civ & $599.0^{+15.2}_{-15.2}$ & $618.7^{+52.5}_{-184.3}$ & $42.8\%$ & $36.0\%$ \\
2970717582 & \civ & $527.0^{+7.1}_{-7.1}$ & $1182.0^{+283.9}_{-58.8}$ & $34.9\%$ & $36.0\%$ \\
2940271483 & \civ & $499.0^{+7.6}_{-7.6}$ & $1022.7^{+108.0}_{-93.6}$ & $34.9\%$ & $36.0\%$ \\
2938924759 & \civ & $552.0^{+12.7}_{-12.7}$ & $1398.0^{+282.6}_{-62.2}$ & $50.0\%$ & $50.0\%$ \\
2939280338 & \civ & $307.0^{+3.6}_{-3.6}$ & $424.0^{+158.9}_{-785.7}$ & $38.4\%$ & $38.5\%$ \\
2970571450 & \civ & $295.0^{+10.9}_{-10.9}$ & $1412.9^{+1114.4}_{-46.1}$ & $50.0\%$ & $50.0\%$ \\
2925793267 & \civ & $227.0^{+11.7}_{-11.7}$ & $541.1^{+328.2}_{-581.3}$ & $50.0\%$ & $50.0\%$ \\
2938968217 & \civ & $934.0^{+9.0}_{-9.0}$ & $1364.1^{+123.3}_{-52.7}$ & $34.9\%$ & $36.0\%$ \\
2925473757 & \civ & $409.0^{+10.7}_{-10.7}$ & $318.9^{+240.4}_{-126.3}$ & $50.0\%$ & $50.0\%$ \\
2925733402 & \civ & $116.0^{+10.7}_{-10.7}$ & $1234.0^{+1043.6}_{-192.1}$ & $50.0\%$ & $50.0\%$ \\
2940830092 & \civ & $507.0^{+4.3}_{-4.3}$ & $1231.4^{+15.1}_{-265.1}$ & $50.6\%$ & $49.4\%$ \\
2970730486 & \civ & $512.0^{+4.9}_{-4.9}$ & $1139.4^{+6.8}_{-87.5}$ & $31.1\%$ & $31.7\%$ \\
2940361224 & \civ & $325.0^{+6.5}_{-6.5}$ & $328.3^{+14.3}_{-34.8}$ & $0.0\%$ & $0.0\%$ \\
2971212466 & \civ & $386.0^{+3.9}_{-3.9}$ & $1212.0^{+63.7}_{-35.4}$ & $41.1\%$ & $36.0\%$ \\
2938404439 & \civ & $331.0^{+5.4}_{-5.4}$ & $159.3^{+83.2}_{-1018.1}$ & $50.0\%$ & $50.0\%$ \\
2938718841 & \civ & $378.0^{+5.6}_{-5.6}$ & $290.3^{+104.7}_{-249.2}$ & $40.8\%$ & $38.5\%$ \\
2925599608 & \civ & $260.0^{+8.2}_{-8.2}$ & $396.7^{+209.9}_{-540.6}$ & $50.0\%$ & $50.0\%$ \\
2925437937 & \civ & $307.0^{+5.4}_{-5.4}$ & $411.5^{+144.3}_{-424.0}$ & $34.9\%$ & $36.3\%$ \\
2937961955 & \civ & $357.3^{+43.0}_{-42.0}$ & $1225.9^{+796.7}_{-141.5}$ & $50.0\%$ & $50.0\%$ \\
2970556875 & \civ & $341.3^{+16.0}_{-23.0}$ & $196.8^{+153.4}_{-40.1}$ & $50.0\%$ & $50.0\%$ \\
    \caption{A comparison of the previously measured / published lags from the \ozdes sample for \hbeta \citep{OzDES-Malik_2023}, \mgii \citep{OzDES-Yu_2021, OzDES-Yu_2023} and \civ \citep{OzDES-Hoormann_2019, OzDES-Penton_2025} as compared to our new measurements here, along with our estimates of the FPR based on $\BF{Lag}$ and $\Lrat{Lag}$. Measurements here are for our \qm{primary} sample of light curves from spectra co-added by run fit using the simple DRW model.}
    \label{tab: OzDES_vs_LITMUS}
\end{longtable}

\newpage
\section{Table of Final Lags}
\label{app: lags_all}
This table provides the full list of recovered lags at \bronze or higher level with our selected data set, (\hbeta and \civ light curves from spectra co-added by run) model (the simple DRW) and significance criteria ($\BF{Lag}$). The full set of lag recoveries for all models, data sets and selection criteria are available in electronic tables online. See Section~\ref{sec: data_availability} for details.

\begin{longtable}{l|c|c|c|c|c|c|c|c|c|c}
    \centering
\makecell{\ozdes\\Source ID} & Line & Redshift & \makecell{$\log_{10}\left(\frac{L_\lambda}{ \left[\text{erg/s}\right]}\right)$\textsuperscript{$\dagger$}} & \makecell{Observer \\ Frame Lag} & \makecell{Rest \\ Frame Lag} & $\BF{Lag}$ & Grade  & \makecell{FPR From\\$\BF{Lag}$} & \makecell{Vel. Disp.\\$\sigma_v$ (km/s)}&\makecell{Est. Mass\\($M_\odot$)}\\ \hline
2971056445 & \hbeta & $0.127$ & $43.31$ & $16.2^{+3.9}_{-4.3}$ & $14.4^{+3.5}_{-3.8}$ & $2.25$ & \gold & 0.0\% & $1585\pm33$ & ${2.2}^{+2.8}_{-1.3}\times10^8$\\
2925826924 & \hbeta & $0.237$ & $43.83$ & $33.3^{+1.0}_{-0.9}$ & $26.9^{+0.8}_{-0.7}$ & $1.20$ & \gold & 0.0\% & $1623\pm16$ & ${3.1}^{+4.7}_{-2.2}\times10^8$\\
2925798541 & \hbeta & $0.294$ & $43.67$ & $16.4^{+5.7}_{-6.3}$ & $12.7^{+4.4}_{-4.8}$ & $0.81$ & \gold & 0.0\% & $1876\pm3$ & ${1.8}^{+2.3}_{-1.1}\times10^8$\\
2970848737 & \hbeta & $0.332$ & $44.00$ & $48.2^{+13.8}_{-7.9}$ & $36.2^{+10.4}_{-5.9}$ & $1.01$ & \gold & 0.0\% & $1748\pm13$ & ${1.8}^{+1.9}_{-0.9}\times10^9$\\
2938089267 & \hbeta & $0.354$ & $43.63$ & $31.5^{+1120.5}_{-15.6}$ & $23.3^{+827.6}_{-11.5}$ & $0.77$ & \gold & 0.0\% & $1440\pm40$ & ${4.3}^{+8.0}_{-3.7}\times10^8$\\
2970738404 & \hbeta & $0.453$ & $44.04$ & $199.8^{+28.4}_{-62.1}$ & $137.5^{+19.5}_{-42.7}$ & $0.63$ & \gold & 0.0\% & $1532\pm20$ & ${1.1}^{+1.5}_{-0.7}\times10^9$\\
2938313335 & \hbeta & $0.557$ & $44.44$ & $507.6^{+609.2}_{-234.4}$ & $326.0^{+391.3}_{-150.5}$ & $0.64$ & \gold & 0.0\% & $1911\pm19$ & ${3.6}^{+4.5}_{-2.1}\times10^7$\\
2938444601 & \hbeta & $0.614$ & $44.28$ & $77.3^{+63.8}_{-30.5}$ & $47.9^{+39.5}_{-18.9}$ & $1.50$ & \gold & 0.0\% & $1645\pm6$ & ${5.9}^{+6.3}_{-3.0}\times10^7$\\
2940843708 & \hbeta & $0.634$ & $44.74$ & $552.0^{+61.4}_{-93.1}$ & $337.8^{+37.6}_{-57.0}$ & $0.88$ & \gold & 0.0\% & $1681\pm6$ & ${7.5}^{+8.6}_{-4.0}\times10^8$\\
2925552152 & \hbeta & $0.655$ & $45.11$ & $113.7^{+47.8}_{-40.3}$ & $68.7^{+28.9}_{-24.3}$ & $1.21$ & \gold & 0.0\% & $1813\pm2$ & ${1.5}^{+1.9}_{-0.9}\times10^9$\\
2939010304\textsuperscript{*} & \hbeta & $0.707$ & $-$ & $67.6^{+43.3}_{-9.5}$ & $39.6^{+25.4}_{-5.5}$ & $0.63$ & \gold & 0.0\% & $1639\pm4$ & ${5.7}^{+6.2}_{-2.9}\times10^8$\\
2939875865 & \hbeta & $0.129$ & $43.84$ & $1021.3^{+333.4}_{-106.4}$ & $904.6^{+295.3}_{-94.2}$ & $0.04$ & \silver & 11.5\% & $1831\pm9$ & ${5.8}^{+90.6}_{-88.7}\times10^7$\\
2939653045 & \hbeta & $0.310$ & $44.64$ & $57.8^{+6.6}_{-9.5}$ & $44.1^{+5.1}_{-7.3}$ & $0.51$ & \silver & 7.4\% & $1867\pm5$ & ${5.5}^{+16.1}_{-12.6}\times10^7$\\
2938232574 & \hbeta & $0.314$ & $44.12$ & $26.7^{+143.8}_{-4.3}$ & $20.3^{+109.4}_{-3.3}$ & $0.56$ & \silver & 7.4\% & $2075\pm7$ & ${1.0}^{+1.8}_{-0.8}\times10^9$\\
2925657995 & \hbeta & $0.315$ & $43.62$ & $820.2^{+589.9}_{-171.9}$ & $623.7^{+448.6}_{-130.7}$ & $0.09$ & \silver & 11.5\% & $1592\pm11$ & ${8.9}^{+13.5}_{-6.2}\times10^7$\\
2937976373 & \hbeta & $0.317$ & $44.65$ & $788.4^{+378.4}_{-52.2}$ & $598.6^{+287.3}_{-39.6}$ & $0.40$ & \silver & 7.4\% & $1671\pm7$ & ${8.5}^{+11.2}_{-5.2}\times10^7$\\
2925351861 & \hbeta & $0.355$ & $43.67$ & $154.4^{+53.9}_{-67.5}$ & $114.0^{+39.8}_{-49.8}$ & $0.03$ & \silver & 11.5\% & $1809\pm3$ & ${1.3}^{+4.4}_{-3.5}\times10^8$\\
2970575166 & \hbeta & $0.434$ & $44.23$ & $146.5^{+20.7}_{-14.7}$ & $102.2^{+14.4}_{-10.2}$ & $0.57$ & \silver & 7.4\% & $1499\pm49$ & ${2.8}^{+6.9}_{-4.2}\times10^8$\\
2925633662 & \hbeta & $0.486$ & $44.00$ & $425.4^{+655.1}_{-164.7}$ & $286.3^{+440.8}_{-110.9}$ & $0.02$ & \silver & 11.5\% & $1716\pm4$ & ${1.1}^{+1.1}_{-0.6}\times10^8$\\
2938088395 & \hbeta & $0.501$ & $44.91$ & $391.1^{+32.5}_{-25.8}$ & $260.6^{+21.6}_{-17.2}$ & $0.21$ & \silver & 7.4\% & $1863\pm4$ & ${2.5}^{+2.9}_{-1.3}\times10^9$\\
2940102912 & \hbeta & $0.541$ & $44.41$ & $180.5^{+47.8}_{-66.8}$ & $117.1^{+31.0}_{-43.3}$ & $0.18$ & \silver & 7.4\% & $1864\pm14$ & ${3.2}^{+3.8}_{-1.8}\times10^8$\\
2925383757 & \hbeta & $0.564$ & $44.03$ & $243.1^{+253.2}_{-59.6}$ & $155.4^{+161.9}_{-38.1}$ & $0.16$ & \silver & 11.5\% & $1722\pm30$ & ${9.0}^{+13.5}_{-6.1}\times10^8$\\
2939517578 & \hbeta & $0.613$ & $45.06$ & $82.4^{+538.3}_{-16.6}$ & $51.1^{+333.7}_{-10.3}$ & $0.50$ & \silver & 7.4\% & $1599\pm4$ & ${7.0}^{+7.7}_{-3.6}\times10^8$\\
2940771836 & \hbeta & $0.622$ & $43.83$ & $666.5^{+446.9}_{-313.1}$ & $410.9^{+275.5}_{-193.0}$ & $0.35$ & \silver & 7.4\% & $1538\pm5$ & ${1.9}^{+2.1}_{-1.0}\times10^8$\\
2925598080 & \hbeta & $0.629$ & $44.46$ & $1174.6^{+91.1}_{-94.2}$ & $721.1^{+55.9}_{-57.8}$ & $0.01$ & \silver & 14.2\% & $1712\pm7$ & ${3.2}^{+3.7}_{-1.7}\times10^8$\\
2939644306 & \hbeta & $0.651$ & $44.32$ & $276.3^{+929.5}_{-85.1}$ & $167.4^{+563.0}_{-51.5}$ & $0.03$ & \silver & 11.5\% & $1838\pm9$ & ${9.7}^{+11.1}_{-5.2}\times10^7$\\
2925858108 & \hbeta & $0.684$ & $44.93$ & $562.8^{+97.3}_{-151.9}$ & $334.2^{+57.8}_{-90.2}$ & $0.28$ & \silver & 7.4\% & $1385\pm3$ & ${2.2}^{+2.5}_{-1.2}\times10^7$\\
2971077718\textsuperscript{*} & \hbeta & $0.702$ & $-$ & $157.2^{+1202.6}_{-50.4}$ & $92.4^{+706.6}_{-29.6}$ & $0.47$ & \silver & 7.4\% & $1929\pm4$ & ${2.6}^{+10.1}_{-8.6}\times10^8$\\
2939782606 & \mgii & $0.866$ & $45.31$ & $79.1^{+17.3}_{-24.7}$ & $42.4^{+9.3}_{-13.3}$ & $1.03$ & \gold & 0.0\% & $2350\pm321$ & ${7.0}^{+8.3}_{-3.8}\times10^8$\\
2925491917 & \mgii & $0.886$ & $44.94$ & $319.0^{+9.3}_{-79.1}$ & $169.1^{+5.0}_{-41.9}$ & $1.03$ & \gold & 0.0\% & $2583\pm216$ & ${2.3}^{+2.9}_{-1.3}\times10^9$\\
2943200932 & \mgii & $1.065$ & $45.66$ & $385.6^{+18.1}_{-40.2}$ & $186.7^{+8.7}_{-19.5}$ & $3.51$ & \gold & 0.0\% & $1805\pm133$ & ${3.4}^{+9.7}_{-6.1}\times10^7$\\
2939473162 & \mgii & $1.077$ & $45.11$ & $1007.6^{+4.1}_{-13.8}$ & $485.1^{+2.0}_{-6.6}$ & $0.93$ & \gold & 0.0\% & $2440\pm117$ & ${7.6}^{+14.2}_{-7.6}\times10^8$\\
2925810744 & \mgii & $1.195$ & $44.90$ & $470.9^{+78.0}_{-19.2}$ & $214.5^{+35.5}_{-8.8}$ & $0.80$ & \gold & 0.0\% & $1496\pm109$ & ${5.0}^{+7.6}_{-3.4}\times10^8$\\
2970745846 & \mgii & $1.416$ & $45.26$ & $206.7^{+11.4}_{-21.6}$ & $85.6^{+4.7}_{-8.9}$ & $2.10$ & \gold & 0.0\% & $1867\pm163$ & ${4.7}^{+5.4}_{-2.5}\times10^8$\\
2925568292 & \mgii & $1.479$ & $45.53$ & $423.8^{+21.7}_{-9.1}$ & $171.0^{+8.8}_{-3.7}$ & $0.84$ & \gold & 0.0\% & $2157\pm210$ & ${6.4}^{+7.2}_{-3.4}\times10^8$\\
2970946076 & \mgii & $1.532$ & $45.59$ & $373.4^{+12.3}_{-3.2}$ & $147.5^{+4.8}_{-1.3}$ & $1.50$ & \gold & 0.0\% & $1323\pm96$ & ${6.9}^{+7.7}_{-3.6}\times10^8$\\
2925403880 & \mgii & $1.643$ & $45.15$ & $423.6^{+12.1}_{-21.7}$ & $160.3^{+4.6}_{-8.2}$ & $1.38$ & \gold & 0.0\% & $1093\pm67$ & ${3.7}^{+4.3}_{-2.0}\times10^8$\\
2939484586 & \mgii & $0.706$ & $45.05$ & $1106.1^{+110.3}_{-4.8}$ & $648.4^{+64.6}_{-2.8}$ & $0.58$ & \silver & 14.1\% & $1369\pm249$ & ${5.4}^{+6.8}_{-3.0}\times10^8$\\
2925458055 & \mgii & $0.766$ & $45.17$ & $293.3^{+627.4}_{-15.4}$ & $166.1^{+355.2}_{-8.7}$ & $0.55$ & \silver & 14.1\% & $2701\pm259$ & ${4.7}^{+5.6}_{-2.6}\times10^8$\\
2970606689 & \mgii & $0.853$ & $44.99$ & $1220.0^{+226.6}_{-50.2}$ & $658.4^{+122.3}_{-27.1}$ & $0.53$ & \silver & 14.1\% & $1005\pm269$ & ${1.7}^{+2.4}_{-1.0}\times10^8$\\
2970951477 & \mgii & $0.912$ & $45.55$ & $1106.8^{+331.1}_{-383.7}$ & $578.9^{+173.2}_{-200.7}$ & $0.55$ & \silver & 14.1\% & $1255\pm403$ & ${4.6}^{+7.3}_{-3.0}\times10^8$\\
2970411872 & \mgii & $1.031$ & $45.31$ & $708.6^{+87.3}_{-25.1}$ & $348.9^{+43.0}_{-12.3}$ & $0.55$ & \silver & 14.1\% & $3303\pm196$ & ${4.1}^{+4.5}_{-2.1}\times10^9$\\
2970914103 & \mgii & $1.091$ & $45.25$ & $170.9^{+30.7}_{-27.4}$ & $81.8^{+14.7}_{-13.1}$ & $0.64$ & \silver & 13.6\% & $1112\pm159$ & ${3.8}^{+4.5}_{-2.0}\times10^8$\\
2937419923 & \mgii & $1.155$ & $45.16$ & $820.8^{+51.4}_{-67.1}$ & $380.9^{+23.9}_{-31.1}$ & $0.61$ & \silver & 13.6\% & $1395\pm63$ & ${4.6}^{+5.0}_{-2.4}\times10^8$\\
2970752138 & \mgii & $1.392$ & $46.08$ & $1244.3^{+45.0}_{-350.3}$ & $520.2^{+18.8}_{-146.4}$ & $0.75$ & \silver & 13.1\% & $1649\pm78$ & ${4.8}^{+5.6}_{-2.6}\times10^8$\\
2970698702 & \mgii & $1.475$ & $45.44$ & $175.0^{+333.6}_{-40.8}$ & $70.7^{+134.8}_{-16.5}$ & $0.58$ & \silver & 14.1\% & $1479\pm235$ & ${2.1}^{+5.4}_{-2.9}\times10^8$\\
2937778190 & \mgii & $1.507$ & $45.19$ & $282.4^{+6.6}_{-6.4}$ & $112.6^{+2.6}_{-2.6}$ & $0.67$ & \silver & 13.6\% & $3476\pm219$ & ${3.8}^{+4.2}_{-2.0}\times10^9$\\
2939193043 & \mgii & $1.659$ & $45.21$ & $423.7^{+15.2}_{-16.8}$ & $159.3^{+5.7}_{-6.3}$ & $0.70$ & \silver & 13.6\% & $2712\pm166$ & ${6.7}^{+7.3}_{-3.5}\times10^8$\\
2939627477 & \mgii & $1.753$ & $45.98$ & $1070.0^{+86.0}_{-81.6}$ & $388.7^{+31.2}_{-29.6}$ & $0.67$ & \silver & 13.6\% & $996\pm275$ & ${2.6}^{+3.8}_{-1.6}\times10^8$\\
2938398472 & \mgii & $1.816$ & $45.71$ & $575.6^{+15.3}_{-47.9}$ & $204.4^{+5.5}_{-17.0}$ & $0.65$ & \silver & 13.6\% & $592\pm107$ & ${1.0}^{+1.5}_{-0.7}\times10^8$\\
2970604169 & \mgii & $0.682$ & $44.76$ & $1116.9^{+10.1}_{-13.3}$ & $664.0^{+6.0}_{-7.9}$ & $0.31$ & \bronze & 20.6\% & $1967\pm252$ & ${1.1}^{+1.2}_{-0.6}\times10^9$\\
2925623118 & \mgii & $0.741$ & $44.60$ & $853.9^{+109.7}_{-117.1}$ & $490.5^{+63.0}_{-67.3}$ & $0.53$ & \bronze & 16.3\% & $3359\pm234$ & ${1.9}^{+2.1}_{-1.0}\times10^9$\\
2940084258 & \mgii & $0.755$ & $44.71$ & $1170.0^{+7.3}_{-8.2}$ & $666.7^{+4.1}_{-4.7}$ & $0.21$ & \bronze & 20.6\% & $2171\pm162$ & ${3.1}^{+7.2}_{-3.7}\times10^8$\\
2940924073 & \mgii & $0.756$ & $44.92$ & $69.4^{+28.4}_{-15.0}$ & $39.5^{+16.2}_{-8.6}$ & $0.42$ & \bronze & 16.3\% & $1840\pm309$ & ${1.2}^{+1.7}_{-0.7}\times10^9$\\
2970924806 & \mgii & $0.768$ & $45.06$ & $1258.8^{+148.0}_{-542.8}$ & $712.0^{+83.7}_{-307.0}$ & $0.16$ & \bronze & 23.3\% & $2200\pm99$ & ${6.3}^{+6.8}_{-3.2}\times10^8$\\
2937556864 & \mgii & $0.788$ & $44.78$ & $403.8^{+113.9}_{-69.4}$ & $225.8^{+63.7}_{-38.8}$ & $0.42$ & \bronze & 16.3\% & $2833\pm185$ & ${3.1}^{+3.4}_{-1.7}\times10^9$\\
2937880615 & \mgii & $0.805$ & $45.70$ & $606.4^{+12.3}_{-32.7}$ & $336.0^{+6.8}_{-18.1}$ & $0.09$ & \bronze & 29.8\% & $2045\pm66$ & ${2.2}^{+2.4}_{-1.1}\times10^9$\\
2925794565 & \mgii & $0.815$ & $44.96$ & $149.9^{+39.6}_{-30.7}$ & $82.6^{+21.8}_{-16.9}$ & $0.16$ & \bronze & 23.3\% & $2448\pm106$ & ${2.5}^{+2.8}_{-1.3}\times10^9$\\
2937448560 & \mgii & $0.831$ & $45.16$ & $537.8^{+46.9}_{-68.4}$ & $293.7^{+25.6}_{-37.4}$ & $0.49$ & \bronze & 16.3\% & $2172\pm148$ & ${1.5}^{+1.6}_{-0.8}\times10^9$\\
2938233883 & \mgii & $0.848$ & $45.30$ & $739.3^{+424.1}_{-130.1}$ & $400.0^{+229.5}_{-70.4}$ & $0.42$ & \bronze & 16.3\% & $3203\pm179$ & ${3.4}^{+4.1}_{-1.9}\times10^8$\\
2939970002 & \mgii & $0.872$ & $46.00$ & $655.5^{+33.8}_{-59.4}$ & $350.1^{+18.1}_{-31.7}$ & $0.13$ & \bronze & 26.8\% & $2570\pm85$ & ${1.9}^{+2.1}_{-1.0}\times10^9$\\
2970681444 & \mgii & $0.885$ & $45.10$ & $476.8^{+746.4}_{-215.1}$ & $252.9^{+396.0}_{-114.1}$ & $0.29$ & \bronze & 20.6\% & $1811\pm275$ & ${8.2}^{+9.9}_{-4.5}\times10^8$\\
2938749340 & \mgii & $0.989$ & $45.84$ & $936.4^{+313.5}_{-380.2}$ & $470.8^{+157.6}_{-191.2}$ & $0.12$ & \bronze & 26.8\% & $1672\pm119$ & ${1.5}^{+1.7}_{-0.8}\times10^9$\\
2937590486 & \mgii & $1.161$ & $45.48$ & $295.9^{+902.0}_{-95.3}$ & $136.9^{+417.4}_{-44.1}$ & $0.24$ & \bronze & 20.6\% & $2017\pm408$ & ${1.5}^{+2.0}_{-0.9}\times10^9$\\
2940088809 & \mgii & $1.222$ & $45.20$ & $1056.0^{+17.4}_{-39.8}$ & $475.2^{+7.8}_{-17.9}$ & $0.23$ & \bronze & 20.6\% & $3290\pm177$ & ${2.4}^{+2.8}_{-1.3}\times10^9$\\
2940367031 & \mgii & $1.227$ & $45.11$ & $642.0^{+192.1}_{-147.2}$ & $288.3^{+86.2}_{-66.1}$ & $0.17$ & \bronze & 23.3\% & $2094\pm191$ & ${9.9}^{+12.1}_{-5.5}\times10^8$\\
2940028080 & \mgii & $1.254$ & $45.09$ & $709.9^{+84.4}_{-18.8}$ & $315.0^{+37.4}_{-8.3}$ & $0.34$ & \bronze & 20.6\% & $1811\pm275$ & ${8.2}^{+9.9}_{-4.5}\times10^8$\\
2925417838 & \mgii & $1.294$ & $45.43$ & $1049.1^{+224.3}_{-418.1}$ & $457.3^{+97.8}_{-182.3}$ & $0.12$ & \bronze & 27.0\% & $2583\pm216$ & ${2.3}^{+2.9}_{-1.3}\times10^9$\\
2925469559 & \mgii & $1.299$ & $45.10$ & $699.0^{+591.0}_{-216.3}$ & $304.1^{+257.0}_{-94.1}$ & $0.34$ & \bronze & 20.6\% & $1496\pm109$ & ${5.0}^{+7.6}_{-3.4}\times10^8$\\
2939622876 & \mgii & $1.325$ & $46.30$ & $1222.4^{+219.4}_{-136.6}$ & $525.7^{+94.3}_{-58.8}$ & $0.07$ & \bronze & 31.9\% & $2448\pm106$ & ${2.5}^{+2.8}_{-1.3}\times10^9$\\
2925858928 & \mgii & $1.341$ & $45.24$ & $1085.4^{+18.2}_{-135.4}$ & $463.7^{+7.8}_{-57.8}$ & $0.47$ & \bronze & 16.3\% & $3303\pm196$ & ${4.1}^{+4.5}_{-2.1}\times10^9$\\
2940318743 & \mgii & $1.358$ & $45.94$ & $664.6^{+52.6}_{-192.2}$ & $281.8^{+22.3}_{-81.5}$ & $0.35$ & \bronze & 18.8\% & $3290\pm177$ & ${2.4}^{+2.8}_{-1.3}\times10^9$\\
2937774380 & \mgii & $1.532$ & $45.19$ & $983.3^{+129.9}_{-92.5}$ & $388.4^{+51.3}_{-36.5}$ & $0.37$ & \bronze & 18.8\% & $3476\pm219$ & ${3.8}^{+4.2}_{-2.0}\times10^9$\\
2938273176 & \mgii & $1.547$ & $46.00$ & $876.9^{+27.6}_{-19.1}$ & $344.3^{+10.8}_{-7.5}$ & $0.39$ & \bronze & 18.4\% & $1967\pm252$ & ${1.1}^{+1.2}_{-0.6}\times10^9$\\
2940655235 & \mgii & $1.616$ & $45.17$ & $773.2^{+237.7}_{-325.8}$ & $295.6^{+90.9}_{-124.6}$ & $0.26$ & \bronze & 20.6\% & - & -\\
2925836943 & \mgii & $1.644$ & $44.80$ & $1009.4^{+140.1}_{-130.4}$ & $381.8^{+53.0}_{-49.3}$ & $0.13$ & \bronze & 26.8\% & $1255\pm403$ & ${4.6}^{+7.3}_{-3.0}\times10^8$\\
2970784222 & \mgii & $1.678$ & $45.63$ & $281.8^{+1177.5}_{-40.4}$ & $105.2^{+439.7}_{-15.1}$ & $0.29$ & \bronze & 20.6\% & $2216\pm198$ & ${4.0}^{+10.2}_{-7.2}\times10^8$\\
2925750275 & \mgii & $1.688$ & $45.62$ & $974.5^{+18.6}_{-34.6}$ & $362.5^{+6.9}_{-12.9}$ & $0.38$ & \bronze & 18.8\% & $1369\pm249$ & ${5.4}^{+6.8}_{-3.0}\times10^8$\\
2938657376 & \mgii & $1.693$ & $45.64$ & $249.3^{+612.0}_{-108.4}$ & $92.6^{+227.3}_{-40.2}$ & $0.25$ & \bronze & 20.6\% & $2171\pm162$ & ${3.1}^{+7.2}_{-3.7}\times10^8$\\
2971239599 & \mgii & $1.713$ & $45.04$ & $1286.2^{+44.7}_{-64.2}$ & $474.1^{+16.5}_{-23.7}$ & $0.17$ & \bronze & 23.1\% & $2642\pm217$ & ${2.7}^{+3.0}_{-1.4}\times10^9$\\
2925707000 & \mgii & $1.847$ & $45.87$ & $1122.6^{+302.2}_{-119.4}$ & $394.3^{+106.1}_{-41.9}$ & $0.23$ & \bronze & 20.6\% & $1093\pm67$ & ${3.7}^{+4.3}_{-2.0}\times10^8$\\
2925420688 & \mgii & $1.884$ & $45.60$ & $40.6^{+164.7}_{-16.1}$ & $14.1^{+57.1}_{-5.6}$ & $0.31$ & \bronze & 20.6\% & $1805\pm133$ & ${3.4}^{+9.7}_{-6.1}\times10^7$\\
2940013145\textsuperscript{*} & \civ & $1.651$ & $-$ & $221.9^{+86.2}_{-68.5}$ & $83.7^{+32.5}_{-25.9}$ & $0.79$ & \gold & 0.0\% & $4174\pm82$ & ${1.1}^{+1.4}_{-0.6}\times10^9$\\
2938970755 & \civ & $1.927$ & $46.21$ & $244.6^{+40.4}_{-42.4}$ & $83.6^{+13.8}_{-14.5}$ & $1.02$ & \gold & 0.0\% & $3601\pm27$ & ${8.7}^{+9.7}_{-4.6}\times10^8$\\
2938852284 & \civ & $2.459$ & $46.47$ & $350.6^{+386.1}_{-104.3}$ & $101.4^{+111.6}_{-30.2}$ & $1.05$ & \gold & 0.0\% & $4220\pm59$ & ${1.3}^{+2.1}_{-1.0}\times10^9$\\
2940361224 & \civ & $2.563$ & $46.00$ & $327.4^{+26.7}_{-16.4}$ & $91.9^{+7.5}_{-4.6}$ & $1.63$ & \gold & 0.0\% & $4340\pm79$ & ${1.4}^{+1.5}_{-0.7}\times10^9$\\
2937774380\textsuperscript{*} & \civ & $1.532$ & $-$ & $470.8^{+136.1}_{-206.1}$ & $185.9^{+53.8}_{-81.4}$ & $0.35$ & \bronze & 20.6\% & $4052\pm142$ & ${2.3}^{+3.0}_{-1.4}\times10^9$\\
2925608973\textsuperscript{*} & \civ & $1.601$ & $-$ & $651.9^{+293.9}_{-309.4}$ & $250.6^{+113.0}_{-119.0}$ & $0.08$ & \bronze & 30.8\% & $2974\pm142$ & ${1.6}^{+2.3}_{-1.0}\times10^9$\\
2940231193\textsuperscript{*} & \civ & $1.606$ & $-$ & $591.5^{+235.8}_{-106.5}$ & $227.0^{+90.5}_{-40.9}$ & $0.09$ & \bronze & 30.8\% & $1004\pm750$ & ${1.6}^{+5.2}_{-1.4}\times10^8$\\
2939647824\textsuperscript{*} & \civ & $1.610$ & $-$ & $178.2^{+264.9}_{-53.8}$ & $68.3^{+101.5}_{-20.6}$ & $0.13$ & \bronze & 30.8\% & $4398\pm100$ & ${9.8}^{+17.2}_{-8.0}\times10^8$\\
2938238274\textsuperscript{*} & \civ & $1.652$ & $-$ & $555.1^{+128.0}_{-84.3}$ & $209.3^{+48.3}_{-31.8}$ & $0.17$ & \bronze & 27.9\% & $2969\pm647$ & ${1.4}^{+1.9}_{-0.8}\times10^9$\\
2925733889\textsuperscript{*} & \civ & $1.653$ & $-$ & $969.1^{+94.9}_{-417.0}$ & $365.3^{+35.8}_{-157.2}$ & $0.22$ & \bronze & 25.0\% & - & -\\
2940401883\textsuperscript{*} & \civ & $1.657$ & $-$ & $640.3^{+626.5}_{-198.6}$ & $241.0^{+235.8}_{-74.8}$ & $0.16$ & \bronze & 29.5\% & - & -\\
2971040058\textsuperscript{*} & \civ & $1.657$ & $-$ & $116.4^{+93.9}_{-49.4}$ & $43.8^{+35.3}_{-18.6}$ & $0.03$ & \bronze & 32.0\% & $3989\pm65$ & ${5.1}^{+7.9}_{-3.6}\times10^8$\\
2925424419\textsuperscript{*} & \civ & $1.682$ & $-$ & $40.4^{+100.1}_{-18.9}$ & $15.1^{+37.3}_{-7.1}$ & $0.17$ & \bronze & 27.9\% & $2327\pm123$ & ${5.8}^{+13.6}_{-6.9}\times10^7$\\
2970695909\textsuperscript{*} & \civ & $1.685$ & $-$ & $585.4^{+171.7}_{-71.1}$ & $218.0^{+63.9}_{-26.5}$ & $0.29$ & \bronze & 21.8\% & - & -\\
2925424201\textsuperscript{*} & \civ & $1.693$ & $-$ & $1245.4^{+143.3}_{-139.6}$ & $462.4^{+53.2}_{-51.8}$ & $0.33$ & \bronze & 20.6\% & $3542\pm123$ & ${4.7}^{+5.1}_{-2.4}\times10^9$\\
2940332845\textsuperscript{*} & \civ & $1.698$ & $-$ & $672.2^{+630.9}_{-194.9}$ & $249.1^{+233.9}_{-72.2}$ & $0.07$ & \bronze & 30.8\% & $4577\pm126$ & ${3.9}^{+5.9}_{-2.7}\times10^9$\\
2938000350\textsuperscript{*} & \civ & $1.707$ & $-$ & $1025.3^{+30.0}_{-29.9}$ & $378.8^{+11.1}_{-11.0}$ & $0.07$ & \bronze & 30.8\% & $5012\pm126$ & ${7.7}^{+8.4}_{-4.0}\times10^9$\\
2970386160 & \civ & $1.861$ & $45.84$ & $916.5^{+291.6}_{-316.3}$ & $320.4^{+101.9}_{-110.5}$ & $0.07$ & \bronze & 30.8\% & $3669\pm74$ & ${3.3}^{+4.1}_{-1.9}\times10^9$\\
2937895789 & \civ & $1.865$ & $46.91$ & $969.5^{+320.8}_{-368.2}$ & $338.4^{+112.0}_{-128.5}$ & $0.04$ & \bronze & 32.0\% & - & -\\
2939622630 & \civ & $1.920$ & $45.78$ & $378.9^{+631.3}_{-183.5}$ & $129.8^{+216.2}_{-62.8}$ & $0.24$ & \bronze & 21.8\% & $4172\pm63$ & ${1.6}^{+3.2}_{-1.5}\times10^9$\\
2940510474 & \civ & $1.923$ & $45.24$ & $118.5^{+149.7}_{-48.7}$ & $40.5^{+51.2}_{-16.7}$ & $0.02$ & \bronze & 32.0\% & $4467\pm103$ & ${5.8}^{+10.1}_{-4.7}\times10^8$\\
2938070615 & \civ & $1.925$ & $-$ & $727.9^{+692.8}_{-331.2}$ & $248.8^{+236.9}_{-113.2}$ & $0.20$ & \bronze & 26.1\% & $4325\pm110$ & ${3.4}^{+5.5}_{-2.5}\times10^9$\\
2940398921 & \civ & $1.926$ & $45.59$ & $458.9^{+741.3}_{-228.6}$ & $156.8^{+253.4}_{-78.1}$ & $0.17$ & \bronze & 27.9\% & $3390\pm240$ & ${1.3}^{+2.6}_{-1.2}\times10^9$\\
2939679498 & \civ & $1.933$ & $45.58$ & $710.4^{+340.8}_{-244.3}$ & $242.2^{+116.2}_{-83.3}$ & $0.25$ & \bronze & 21.8\% & $4217\pm89$ & ${3.3}^{+4.3}_{-2.0}\times10^9$\\
2938937393 & \civ & $1.936$ & $45.77$ & $934.2^{+373.2}_{-229.9}$ & $318.2^{+127.1}_{-78.3}$ & $0.17$ & \bronze & 27.9\% & $4190\pm40$ & ${4.3}^{+5.3}_{-2.5}\times10^9$\\
2939722065 & \civ & $1.945$ & $45.81$ & $299.8^{+235.7}_{-139.0}$ & $101.8^{+80.0}_{-47.2}$ & $0.10$ & \bronze & 30.8\% & $4745\pm74$ & ${1.7}^{+2.6}_{-1.2}\times10^9$\\
2939629447 & \civ & $1.949$ & $45.93$ & $565.5^{+112.0}_{-163.8}$ & $191.7^{+38.0}_{-55.6}$ & $0.25$ & \bronze & 21.8\% & $3604\pm67$ & ${2.0}^{+2.3}_{-1.1}\times10^9$\\   
2940128834 & \civ & $1.951$ & $45.73$ & $97.0^{+335.9}_{-39.9}$ & $32.9^{+113.8}_{-13.5}$ & $0.36$ & \bronze & 20.6\% & $4182\pm51$ & ${4.2}^{+10.7}_{-6.6}\times10^8$\\
2925508629 & \civ & $1.957$ & $45.04$ & $1035.2^{+213.1}_{-393.0}$ & $350.1^{+72.1}_{-132.9}$ & $0.50$ & \bronze & 20.6\% & $4822\pm69$ & ${6.4}^{+7.6}_{-3.6}\times10^9$\\
2925382434 & \civ & $1.963$ & $45.59$ & $942.2^{+289.6}_{-426.2}$ & $318.0^{+97.7}_{-143.8}$ & $0.06$ & \bronze & 32.0\% & $4821\pm51$ & ${5.6}^{+7.2}_{-3.3}\times10^9$\\
2970501798 & \civ & $1.990$ & $45.78$ & $660.0^{+23.7}_{-73.3}$ & $220.7^{+7.9}_{-24.5}$ & $0.42$ & \bronze & 20.6\% & $3764\pm83$ & ${2.5}^{+2.7}_{-1.3}\times10^9$\\
2939006752 & \civ & $1.991$ & $45.25$ & $1016.3^{+149.0}_{-73.5}$ & $339.8^{+49.8}_{-24.6}$ & $0.32$ & \bronze & 20.6\% & $3767\pm127$ & ${3.9}^{+4.3}_{-2.0}\times10^9$\\
2940545993 & \civ & $2.004$ & $45.94$ & $723.0^{+684.3}_{-297.0}$ & $240.7^{+227.8}_{-98.9}$ & $0.35$ & \bronze & 20.6\% & $3706\pm51$ & ${2.4}^{+3.8}_{-1.8}\times10^9$\\
2939082045 & \civ & $2.015$ & $46.12$ & $720.8^{+176.2}_{-315.7}$ & $239.1^{+58.4}_{-104.7}$ & $0.67$ & \bronze & 20.6\% & $4591\pm30$ & ${3.9}^{+4.9}_{-2.3}\times10^9$\\
2938053850 & \civ & $2.026$ & $46.07$ & $648.0^{+159.1}_{-249.9}$ & $214.2^{+52.6}_{-82.6}$ & $0.33$ & \bronze & 20.6\% & $4038\pm97$ & ${2.7}^{+3.3}_{-1.5}\times10^9$\\
2939967017 & \civ & $2.061$ & $46.07$ & $799.7^{+419.4}_{-316.2}$ & $261.2^{+137.0}_{-103.3}$ & $0.26$ & \bronze & 21.8\% & $3883\pm45$ & ${3.0}^{+4.1}_{-1.8}\times10^9$\\
2938627200 & \civ & $2.089$ & $46.20$ & $811.5^{+153.9}_{-295.1}$ & $262.7^{+49.8}_{-95.5}$ & $0.05$ & \bronze & 32.0\% & $3131\pm56$ & ${2.0}^{+2.4}_{-1.1}\times10^9$\\
2937538328 & \civ & $2.096$ & $45.98$ & $529.2^{+459.7}_{-159.6}$ & $170.9^{+148.5}_{-51.6}$ & $0.41$ & \bronze & 20.6\% & $3511\pm112$ & ${1.6}^{+2.3}_{-1.1}\times10^9$\\
2939003048 & \civ & $2.100$ & $45.76$ & $426.5^{+632.8}_{-198.4}$ & $137.6^{+204.1}_{-64.0}$ & $0.28$ & \bronze & 21.8\% & $3582\pm92$ & ${1.3}^{+2.4}_{-1.1}\times10^9$\\
2925825614 & \civ & $2.104$ & $45.64$ & $723.0^{+473.8}_{-308.4}$ & $232.9^{+152.6}_{-99.4}$ & $0.07$ & \bronze & 30.8\% & $3250\pm171$ & ${1.8}^{+2.7}_{-1.2}\times10^9$\\
2925462019 & \civ & $2.106$ & $45.83$ & $430.7^{+341.1}_{-181.2}$ & $138.7^{+109.8}_{-58.3}$ & $0.30$ & \bronze & 21.8\% & $3542\pm56$ & ${1.3}^{+1.9}_{-0.9}\times10^9$\\
2940335936 & \civ & $2.124$ & $46.18$ & $567.1^{+69.2}_{-95.1}$ & $181.5^{+22.2}_{-30.4}$ & $0.30$ & \bronze & 21.8\% & $3040\pm160$ & ${1.3}^{+1.5}_{-0.7}\times10^9$\\
2940373266 & \civ & $2.146$ & $46.34$ & $1149.6^{+46.6}_{-60.7}$ & $365.4^{+14.8}_{-19.3}$ & $0.30$ & \bronze & 21.8\% & $2793\pm38$ & ${2.3}^{+2.5}_{-1.2}\times10^9$\\
2939408200 & \civ & $2.156$ & $45.83$ & $819.3^{+457.3}_{-350.9}$ & $259.6^{+144.9}_{-111.2}$ & $0.12$ & \bronze & 30.8\% & $5256\pm97$ & ${5.4}^{+7.6}_{-3.4}\times10^9$\\
2937414043 & \civ & $2.159$ & $44.74$ & $1437.6^{+24.0}_{-498.2}$ & $455.1^{+7.6}_{-157.7}$ & $0.02$ & \bronze & 32.0\% & $3363\pm249$ & ${4.1}^{+4.8}_{-2.2}\times10^9$\\
2925410674 & \civ & $2.226$ & $45.36$ & $495.2^{+763.7}_{-227.7}$ & $153.5^{+236.7}_{-70.6}$ & $0.28$ & \bronze & 21.8\% & - & -\\
2970717582 & \civ & $2.252$ & $46.16$ & $1154.1^{+99.4}_{-276.6}$ & $354.9^{+30.6}_{-85.1}$ & $0.16$ & \bronze & 29.5\% & $3664\pm57$ & ${3.8}^{+4.2}_{-2.0}\times10^9$\\
2938570770 & \civ & $2.253$ & $46.76$ & $746.4^{+179.5}_{-163.8}$ & $229.4^{+55.2}_{-50.4}$ & $0.33$ & \bronze & 20.6\% & $4696\pm25$ & ${4.0}^{+4.6}_{-2.2}\times10^9$\\
2925842332 & \civ & $2.286$ & $45.93$ & $637.8^{+347.7}_{-200.2}$ & $194.1^{+105.8}_{-60.9}$ & $0.05$ & \bronze & 32.0\% & $3707\pm125$ & ${2.0}^{+2.7}_{-1.2}\times10^9$\\
2940271483 & \civ & $2.289$ & $46.21$ & $1006.3^{+225.0}_{-102.7}$ & $306.0^{+68.4}_{-31.2}$ & $0.16$ & \bronze & 27.9\% & $4221\pm83$ & ${4.4}^{+4.9}_{-2.3}\times10^9$\\
2925792482 & \civ & $2.313$ & $46.04$ & $620.0^{+394.4}_{-257.1}$ & $187.2^{+119.1}_{-77.6}$ & $0.32$ & \bronze & 20.6\% & $3213\pm85$ & ${1.4}^{+2.1}_{-0.9}\times10^9$\\
2925856664 & \civ & $2.347$ & $46.09$ & $1087.3^{+215.2}_{-490.0}$ & $324.9^{+64.3}_{-146.4}$ & $0.32$ & \bronze & 20.6\% & $4067\pm49$ & ${4.2}^{+5.2}_{-2.4}\times10^9$\\
2925606098 & \civ & $2.355$ & $45.78$ & $508.2^{+588.4}_{-242.9}$ & $151.5^{+175.4}_{-72.4}$ & $0.23$ & \bronze & 21.8\% & $4472\pm80$ & ${2.2}^{+3.8}_{-1.8}\times10^9$\\
2938968217 & \civ & $2.449$ & $46.64$ & $1378.5^{+48.5}_{-348.5}$ & $399.7^{+14.1}_{-101.0}$ & $0.39$ & \bronze & 20.6\% & $4121\pm30$ & ${5.4}^{+6.1}_{-2.9}\times10^9$\\
2971090058 & \civ & $2.496$ & $45.98$ & $250.3^{+238.3}_{-56.9}$ & $71.6^{+68.2}_{-16.3}$ & $0.25$ & \bronze & 21.8\% & $3720\pm48$ & ${7.5}^{+11.0}_{-5.2}\times10^8$\\
2970730486 & \civ & $2.502$ & $45.93$ & $1132.6^{+12.5}_{-38.0}$ & $323.4^{+3.6}_{-10.8}$ & $0.56$ & \bronze & 20.6\% & $3633\pm52$ & ${3.5}^{+3.7}_{-1.8}\times10^9$\\
2971063757 & \civ & $2.520$ & $45.73$ & $1120.4^{+221.4}_{-502.6}$ & $318.3^{+62.9}_{-142.8}$ & $0.08$ & \bronze & 30.8\% & $2495\pm108$ & ${1.5}^{+1.9}_{-0.9}\times10^9$\\
2943195872 & \civ & $2.585$ & $46.12$ & $930.7^{+170.7}_{-412.7}$ & $259.6^{+47.6}_{-115.1}$ & $0.10$ & \bronze & 30.8\% & $4312\pm44$ & ${3.7}^{+4.6}_{-2.1}\times10^9$\\
2937772718 & \civ & $2.617$ & $45.62$ & $1315.2^{+64.0}_{-78.5}$ & $363.6^{+17.7}_{-21.7}$ & $0.16$ & \bronze & 27.9\% & $3922\pm157$ & ${4.5}^{+4.9}_{-2.4}\times10^9$\\
2939320724 & \civ & $2.622$ & $46.38$ & $250.2^{+25.3}_{-39.0}$ & $69.1^{+7.0}_{-10.8}$ & $0.42$ & \bronze & 20.6\% & $4212\pm89$ & ${9.9}^{+10.8}_{-5.1}\times10^8$\\
2938718841 & \civ & $2.743$ & $45.59$ & $361.2^{+40.0}_{-158.6}$ & $96.5^{+10.7}_{-42.4}$ & $0.33$ & \bronze & 20.6\% & $3643\pm116$ & ${10.0}^{+12.2}_{-5.6}\times10^8$\\
2939002332 & \civ & $2.793$ & $46.32$ & $1109.0^{+261.2}_{-285.6}$ & $292.4^{+68.9}_{-75.3}$ & $0.10$ & \bronze & 30.8\% & $3720\pm84$ & ${3.2}^{+3.7}_{-1.7}\times10^9$\\
2940162924 & \civ & $2.818$ & $45.75$ & $1313.7^{+53.3}_{-176.9}$ & $344.1^{+13.9}_{-46.3}$ & $0.02$ & \bronze & 32.3\% & - & -\\
2925645993 & \civ & $2.971$ & $46.49$ & $402.8^{+667.3}_{-148.3}$ & $101.4^{+168.1}_{-37.3}$ & $0.17$ & \bronze & 27.9\% & $5190\pm49$ & ${2.0}^{+3.8}_{-1.7}\times10^9$\\
2940645467 & \civ & $3.134$ & $45.55$ & $1011.0^{+196.8}_{-35.2}$ & $244.6^{+47.6}_{-8.5}$ & $0.27$ & \bronze & 21.8\% & $4972\pm128$ & ${4.9}^{+5.3}_{-2.5}\times10^9$\\
2940117301 & \civ & $3.175$ & $46.68$ & $635.3^{+446.3}_{-311.3}$ & $152.2^{+106.9}_{-74.6}$ & $0.09$ & \bronze & 30.8\% & $3686\pm33$ & ${1.5}^{+2.3}_{-1.0}\times10^9$\\
2971232645 & \civ & $3.375$ & $45.90$ & $548.2^{+187.6}_{-161.9}$ & $125.3^{+42.9}_{-37.0}$ & $0.21$ & \bronze & 25.8\% & $4430\pm209$ & ${1.9}^{+2.3}_{-1.1}\times10^9$\\
2925849629 & \civ & $3.427$ & $46.47$ & $682.7^{+429.5}_{-247.6}$ & $154.2^{+97.0}_{-55.9}$ & $0.19$ & \bronze & 26.1\% & $3777\pm58$ & ${1.6}^{+2.3}_{-1.0}\times10^9$\\
2925437937 & \civ & $3.450$ & $46.21$ & $412.8^{+490.9}_{-188.1}$ & $92.8^{+110.3}_{-42.3}$ & $0.09$ & \bronze & 30.8\% & $1444\pm213$ & ${1.3}^{+2.6}_{-1.1}\times10^8$\\
    \hline
    \caption{A collation of all of our published lags for all lines, along with monochromatic luminosities and estimates of mass. The \hbeta and \civ sources are for light curves from spectra co-added by run, except for those marked with a \textsuperscript{$*$} which are co-added by date. All results here are using the simple (no smoothing or jitter) DRW model, per our fiducial modelling choices outlined in Section~\ref{sec: selection_criteria}. Masses are estimated per Equation~\ref{eq: RM_mass} using the virial factor of $\log_{10}(f)=0.62 \pm 0.07 \pm 0.31$ from \citet{SDSS-Shen_2023} (the first and second uncertainties here being statistical error and population scatter). The velocity dispersions are as presented in \citet{McDougall_2025_LITMUS}. \\\textsuperscript{$\dagger$} Luminosities are measured at $5100 \angstrom$ for \hbeta, $3000 \angstrom$ for \mgii and $1350 \angstrom$ for \civ.}
    \label{tab: lags_all}
\end{longtable}


\end{document}